\documentclass[prd,groupedaddress,superscriptaddress,nofootinbib,floatfix,preprintnumbers,longbibliography,notitlepage,tightenlines,11pt,showpacs]{revtex4-2}

\usepackage[T1]{fontenc}
\usepackage[utf8]{inputenc}
\usepackage{amsmath,amssymb}
\usepackage{graphicx}
\usepackage{color,xcolor}
\usepackage{dsfont}
\usepackage{array}
\usepackage{natbib}
\usepackage{url}
\usepackage{mathtools}
\usepackage{float}
\usepackage{multirow}
\usepackage{cancel}
\usepackage{bm}
\usepackage[export]{adjustbox}
\usepackage[hypertexnames=false]{hyperref}
\hypersetup{colorlinks=true,linkcolor=blue,citecolor=blue,filecolor=blue,urlcolor=blue}

\renewcommand{\arraystretch}{1.6}

\def\1s0{^1 \hskip -0.03in S_0}
\def\3s1{^3 \hskip -0.025in S_1}

\begin{document}
\dimen\footins=6\baselineskip\relax

\preprint{\vbox{\hbox{ICCUB-20-020, UMD-PP-020-7, MIT-CTP/5238, INT-PUB-20-038}}}
\preprint{\vbox{\hbox{FERMILAB-PUB-20-498-T}}}

\begin{figure}
  \vskip -1.5cm
  \leftline{\includegraphics[width=0.18\textwidth]{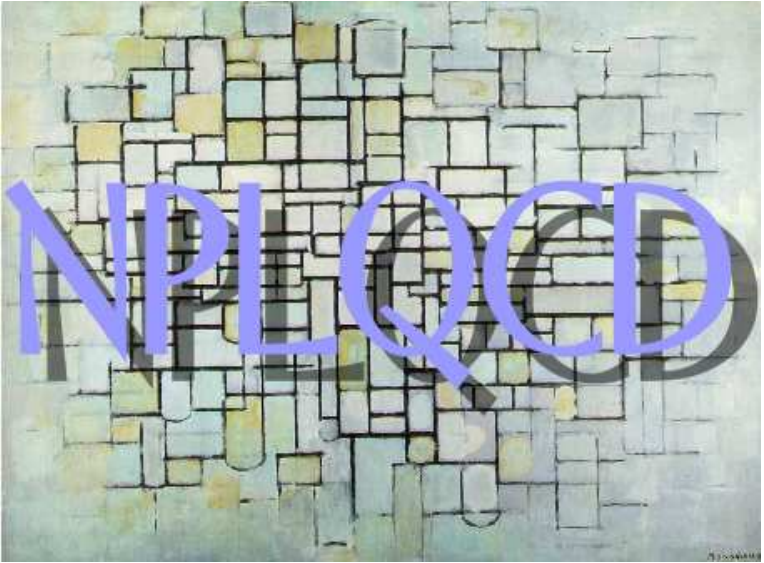}}
\end{figure}

\vspace{-1cm}

\title{Low-energy Scattering and Effective  Interactions of Two Baryons at \texorpdfstring{$m_{\pi}\sim 450$}{mpi=450}~MeV from Lattice Quantum Chromodynamics
}

\author{Marc~Illa}
\affiliation{Departament de F\'{\i}sica Qu\`{a}ntica i Astrof\'{\i}sica and Institut de Ci\`{e}ncies del Cosmos, Universitat de Barcelona, Mart\'{\i} Franqu\`es 1, E08028-Spain}

\author{Silas~R.~Beane}
\affiliation{Department of Physics, University of Washington, Seattle, WA 98195-1560, USA}

\author{Emmanuel~Chang}
\noaffiliation{}

\author{Zohreh~Davoudi}
\affiliation{Department of Physics and Maryland Center for Fundamental Physics, University of Maryland, College Park, MD 20742, USA}
\affiliation{RIKEN Center for Accelerator-based Sciences, Wako 351-0198, Japan}
 	
\author{William~Detmold}
\affiliation{Center for Theoretical Physics, Massachusetts Institute of Technology, Cambridge, MA 02139, USA}

\author{David~J.~Murphy}
\affiliation{Center for Theoretical Physics, Massachusetts Institute of Technology, Cambridge, MA 02139, USA}

\author{Kostas~Orginos}
\affiliation{Department of Physics, College of William and Mary, Williamsburg, VA 23187-8795, USA}
\affiliation{Jefferson Laboratory, 12000 Jefferson Avenue, Newport News, VA 23606, USA}

\author{Assumpta~Parre\~no}
\affiliation{Departament de F\'{\i}sica Qu\`{a}ntica i Astrof\'{\i}sica and Institut de Ci\`{e}ncies del Cosmos, Universitat de Barcelona, Mart\'{\i} Franqu\`es 1, E08028-Spain}

\author{Martin~J.~Savage}
\affiliation{Institute for Nuclear Theory, University of Washington, Seattle, WA 98195-1550, USA}

\author{Phiala~E.~Shanahan}
\affiliation{Center for Theoretical Physics, Massachusetts Institute of Technology, Cambridge, MA 02139, USA}

\author{Michael~L.~Wagman}
\affiliation{Fermi National Accelerator Laboratory, Batavia, IL 60510, USA}

\author{Frank~Winter}
\affiliation{Jefferson Laboratory, 12000 Jefferson Avenue, Newport News, VA 23606, USA}

\collaboration{NPLQCD Collaboration}

\date{\today}


\begin{abstract} 
The interactions between two octet baryons are studied at low energies using lattice Quantum Chromodynamics (LQCD) with larger-than-physical quark masses corresponding to a pion mass of $m_{\pi}\sim 450$ MeV and a kaon mass of $m_{K}\sim 596$ MeV. The two-baryon systems that are analyzed range from strangeness $S=0$ to $S=-4$ and include the spin-singlet and triplet $NN$, $\Sigma N$ ($I=3/2$), and $\Xi\Xi$ states, the spin-singlet $\Sigma \Sigma$ ($I=2$) and $\Xi \Sigma$ ($I=3/2$) states, and the spin-triplet $\Xi N$ ($I=0$) state. The corresponding $s$-wave scattering phase shifts, low-energy scattering parameters, and binding energies when applicable, are extracted using  L\"uscher's formalism. While the results are consistent with most of the systems being bound at this pion mass, the interactions in the spin-triplet $\Sigma N$ and $\Xi \Xi$ channels are found to be repulsive and do not support bound states. Using results from previous studies of these systems at a larger pion mass, an extrapolation of the binding energies to the physical point is performed and is compared with available experimental values and phenomenological predictions. The low-energy coefficients in pionless effective field theory (EFT) relevant for two-baryon interactions, including those responsible for $SU(3)$ flavor-symmetry breaking, are constrained. The $SU(3)$ flavor symmetry is observed to hold approximately at the chosen values of the quark masses, as well as the $SU(6)$ spin-flavor symmetry, predicted at large $N_c$. A remnant of an accidental $SU(16)$ symmetry found previously at a larger pion mass is further observed. The $SU(6)$-symmetric EFT constrained by these LQCD calculations is used to make predictions for two-baryon systems for which the low-energy scattering parameters could not be determined with LQCD directly in this study, and to constrain the coefficients of all leading $SU(3)$ flavor-symmetric interactions, demonstrating the predictive power of two-baryon EFTs matched to LQCD. 
\end{abstract}
\maketitle

%%%%%%%%%%%%%%%%%%%%%%%%%%%

%\tableofcontents

\section{Introduction \label{sec:Intro}}
\noindent
Hyperons ($Y$) are expected to appear in the interior of neutron stars~\cite{1960SvA4187A}, and unless the strong interactions between hyperons and nucleons ($N$) are sufficiently repulsive, the equation of state (EoS) of dense nuclear matter will be softer than for purely non-strange matter, leading to correspondingly lower maximum values for neutron star masses. 
While experimental data on scattering cross sections in the majority of the $YN$ channels are scarce, there are reasonably precise constraints on the interactions in the $\Lambda N$ channel from scattering and hypernuclear spectroscopy experiments~\cite{Feliciello:2015dua,Gal:2016boi}, and they indicate that the interactions in this channel are attractive. 
Given that the $\Lambda$ baryon is lighter than the other hyperons, it is likely the most abundant hyperon in the interior of neutron stars. 
However, models of the EoS including $\Lambda$ baryons and attractive $\Lambda N$ interactions~\cite{Lonardoni:2014bwa} predict a maximum neutron star mass that is below the maximum observed mass at $2M_{\odot}$~\cite{Champion:2008ge, Freire:2010tf,Demorest:2010bx,Antoniadis:2013pzd, Cromartie:2019kug}.\footnote{Very recently, the gravitational wave signal GW190814, originated from the merger of a $23 M_{\odot}$ black hole and a $2.6M_{\odot}$ compact object, was reported~\cite{Abbott:2020khf}, where the nature of the compact object is a subject of discussion. If this compact object was a neutron star, it would have been the most massive one known, imposing a mass-limit constraint very difficult to fulfill for the majority of existing nuclear EoS models.}
Several remedies have been suggested to solve this problem, known in the literature as the ``hyperon puzzle''~\cite{Chatterjee:2015pua, Vidana:2018bdi, Tolos:2020aln}. For example, if hyperons other than the $\Lambda$ baryon (such as $\Sigma$ baryons) are present in the interior of neutron stars and the interactions in the corresponding $YN$ and $YY$ channels are sufficiently repulsive, the EoS would become more stiff~\cite{Vidana:2010ip, Petschauer:2015nea}.
Another suggestion is that repulsive interactions in the $YNN$, $YYN$ and $YYY$ channels may render the EoS stiff enough to produce a $2M_{\odot}$ neutron star~\cite{Vidana:2010ip, Yamamoto:2013ada, Furumoto:2014ica, Lonardoni:2014bwa, Logoteta:2019utx}.
Repulsive density-dependent interactions in systems involving the $\Lambda$ and other hyperons have also been suggested, along with the possibility of a phase transition to quark matter in the interior of neutron stars; see Refs.~\cite{Chatterjee:2015pua,Vidana:2018bdi, Tolos:2020aln} for recent reviews. 
Given the scarcity or complete lack of experimental data on $YN$ and $YY$ scattering and all three-body interactions involving hyperons, $SU(3)$ flavor symmetry is used to constrain effective field theories (EFTs) and phenomenological meson-exchange models of hypernuclear interactions. In this way, quantities in channels for which experimental data exist can be related via symmetries to those in channels which lack such phenomenological constraints~\cite{Polinder:2006zh,Petschauer:2013uua}. 
For example, the lowest-order effective interactions in several channels with strangeness $S\in\{-2,-3,-4\}$ were constrained using experimental data on $pp$ phase shifts and the $\Sigma^+p$ cross section in the same $SU(3)$ representation in the framework of chiral EFT ($\chi$EFT) in Refs.~\cite{Haidenbauer:2015zqb,Haidenbauer:2009qn,Haidenbauer:2014rna}. However, only a few of the $SU(3)$-breaking low-energy coefficients (LECs) of the EFT could be constrained~\cite{Haidenbauer:2014rna}.\footnote{Note that $SU(3)_f$ $\chi$EFT is less convergent than for two flavors.} 
To date, the knowledge of these interactions in nature remains unsatisfactory, demanding more direct theoretical approaches.\footnote{Other observational means to constrain these interactions, such as radius measurements of neutron stars, their thermal and structural evolution, and the emission of gravitational waves in hot and rapidly rotating newly-born neutron stars, can be used to indirectly probe the strangeness content of dense matter and provide complementary constraints on models of hypernuclear interactions~\cite{Chatterjee:2015pua}.} 

Building upon our previous works~\cite{Beane:2006gf, Beane:2009gs, Beane:2009py, Beane:2010hg, Beane:2011iw, Beane:2012ey, Beane:2012vq, Wagman:2017tmp}, we present further studies which constrain hypernuclear forces in nature by direct calculations starting from the underlying theory of the strong interactions, Quantum Chromodynamics (QCD).
To this end, the numerical technique of lattice QCD (LQCD) is used to obtain information on the low-energy spectra and scattering on two-baryon systems, which can be used to constrain EFTs or phenomenological models of two-baryon interactions.
In recent years, LQCD has allowed a wealth of observables in nuclear physics, from hadronic spectra and structure to nuclear matrix elements~\cite{Detmold:2019ghl, Drischler:2019xuo, Davoudi:2020ngi}, to be calculated directly from interactions of quarks and gluons, albeit with uncertainties that are yet to be fully controlled. 
In the context of constraining hypernuclear interactions, LQCD is a powerful theoretical tool because the lowest-lying hyperons are stable when only strong interactions are included in the computation, circumventing the limitations faced by experiments on hyperons and hypernuclei. Nonetheless, LQCD studies in the multi-baryon sector require large computing resources as there is an inherent signal-to-noise degradation present in the correlation functions of baryons~\cite{Parisi:1983ae, Lepage:1989hd, Beane:2009kya, Beane:2009gs, Beane:2009py, Wagman:2016bam}, among other issues as discussed in a recent review~\cite{Davoudi:2020ngi}.
Consequently, most studies of two-baryon systems to date~\cite{Beane:2006mx, Beane:2006gf, Beane:2009py, Beane:2010hg, Beane:2011zpa, Beane:2011iw, Beane:2012ey, Beane:2012vq, Beane:2013br, Orginos:2015aya, Wagman:2017tmp, Fukugita:1994ve, Yamazaki:2011nd, Yamazaki:2012hi, Yamazaki:2015asa, Nemura:2008sp, Inoue:2010es, Berkowitz:2015eaa, Francis:2018qch, Junnarkar:2019equ} have used larger-than-physical quark masses to expedite computations, and only recently have results at the physical values of the quark masses emerged~\cite{Doi:2017zov, Gongyo:2017fjb, Iritani:2018sra, Aoki:2020bew}, making it possible to directly compare with experimental data~\cite{Acharya:2020asf}.
The existing studies are primarily based on two distinct approaches. In one approach, the low-lying spectra of two baryons in finite spatial volumes are determined from the time dependence of Euclidean correlation functions computed with LQCD, and are then converted to scattering amplitudes at the corresponding energies through the use of L\"uscher's formula~\cite{Luscher:1986pf, Luscher:1990ux} or its generalizations~\cite{Rummukainen:1995vs,Beane:2003da, Kim:2005gf, He:2005ey, Davoudi:2011md, Leskovec:2012gb, Hansen:2012tf, Briceno:2012yi, Gockeler:2012yj, Briceno:2013lba, Feng:2004ua, Lee:2017igf, Bedaque:2004kc, Luu:2011ep, Briceno:2013hya, Briceno:2013bda, Briceno:2017max}.
In another approach, non-local potentials are constructed based on the Bethe-Salpeter wavefunctions determined from LQCD correlation functions, and are subsequently used in the Lippmann-Schwinger equation to solve for scattering phase shifts~\cite{Aoki:2011ep, Aoki:2012tk, Aoki:2020bew}.
Given that L\"uscher's formalism is model-independent below inelastic thresholds, it is this approach that is used in the present study as the basis to constrain scattering amplitudes and their low-energy parametrizations in a number of two-(octet)baryon channels with strangeness $S\in\{0,-1,-2,-3,-4\}$.

While LQCD studies at unphysical values of the quark masses already shed light on the understanding of (hyper)nuclear and dense-matter physics, a full account of all systematic uncertainties, including precise extrapolations to the physical quark mass, is required to further impact phenomenology.
Additionally, LQCD results for scattering amplitudes can be used to better constrain the low-energy interactions within given phenomenological models and applicable EFTs. In the case of exact $SU(3)$ flavor symmetry and including only the lowest-lying octet baryons, there are six two-baryon interactions at leading order (LO) in pionless EFT~\cite{vanKolck:1998bw,Chen:1999tn} that can be constrained by the $s$-wave scattering lengths in two-baryon scattering~\cite{Savage:1995kv}.
LQCD has been used in Ref.~\cite{Wagman:2017tmp} to constrain the corresponding LECs of these interactions by computing the $s$-wave scattering parameters of two baryons at an $SU(3)$ flavor-symmetric point with $m_{\pi}\sim 806$ MeV. Strikingly, the first evidence of a long-predicted $SU(6)$ spin-flavor symmetry in nuclear and hypernuclear interactions in the limit of a large number of colors ($N_c$)~\cite{Kaplan:1995yg} was observed in that study, along with an accidental $SU(16)$ symmetry. This extended symmetry has been suggested in Ref.~\cite{Beane:2018oxh} to support the conjecture of entanglement suppression in nuclear and hypernuclear forces at low energies, pointing to intriguing aspects of strong interactions in nature. 

The objective of this paper is to extend our previous study to quark masses that are closer to their physical values, corresponding to a pion mass of $\sim 450$ MeV and a kaon mass of $\sim 596$ MeV, and further to study these systems in a setting with broken $SU(3)$ flavor symmetry as is the case in nature. The present study provides new constraints that allow preliminary extrapolations to physical quark masses to be performed, and complements previous independent LQCD studies at nearby quark masses~\cite{Beane:2006mx,Beane:2006gf,Beane:2009py,Beane:2010hg,Beane:2011iw,Beane:2012ey,Orginos:2015aya,Yamazaki:2012hi,Yamazaki:2015asa,Inoue:2011ai,Ishii:2013cta}.
In particular, predictions for the binding energies of ground states in a number of $YN$ and $YY$ channels based on the results of the current work and those of Ref.~\cite{Wagman:2017tmp} at larger quark masses are consistent with experiments and phenomenological results where they exist. 
Our LQCD results are used to constrain the leading $SU(3)$ symmetry-breaking coefficients in pionless EFT. This EFT matching enables the exploration of large-$N_c$ predictions, pointing to the validity of $SU(6)$ spin-flavor symmetry at this pion mass as well, and revealing a remnant of an accidental $SU(16)$ symmetry that was observed at a larger pion mass in Ref.~\cite{Wagman:2017tmp}. Strategies to make use of the QCD-constrained EFTs to advance the \emph{ab initio} many-body studies of larger hypernuclear isotopes and dense nuclear matter are beyond the scope of this work. Nevertheless, the methods applied in Refs.~\cite{Barnea:2013uqa,Contessi:2017rww,Bansal:2017pwn} to connect the results of LQCD calculations to higher-mass nuclei can also be applied in the hypernuclear sector using the results presented in this work.

This paper is organized as follows. Section~\ref{sec:metho} presents a summary of the computational details (Sec.~\ref{subsec:comp}), followed by the results for the lowest-lying energies of two-baryon systems from LQCD correlation functions, along with a description of the method used to obtain these spectra (Sec.~\ref{subsec:fitalg}), a determination of the $s$-wave scattering parameters in the two-baryon channels that are studied, along with the formalism used to extract the scattering amplitude (Sec.~\ref{subsec:luscher}), and finally the binding energies of the bound states that are identified in various channels, including an extrapolation to the physical point (Sec.~\ref{subsec:bind}).
Section~\ref{sec:pheno} discusses the constraints that these results impose on the low-energy coefficients of the next-to-leading order (NLO) pionless EFT Lagrangian, including some of the $SU(3)$ flavor-symmetry breaking terms (Sec.~\ref{subsec:lecs}). This is followed by a discussion of the predictions for the values of the coefficients that appear, in the limit of large $N_c$, in the $SU(6)$ spin-flavor Lagrangian at LO (Sec.~\ref{subsec:su6}).
The main results of this work are summarized in Sec.~\ref{sec:con}. In addition, several appendixes are presented to supplement the conclusions of this study.
Appendix~\ref{sec:appen-checks} contains an analysis of our results in view of the consistency checks of Refs.~\cite{Iritani:2017rlk,Iritani:2018talk}, demonstrating that the checks are unambiguously passed. Appendix~\ref{sec:appen-vs2015} presents an exhaustive comparison between the results obtained in this work and previous results presented in Ref.~\cite{Orginos:2015aya} for the two-nucleon channels using the same LQCD correlation functions, as well as with the predictions of the low-energy theorems analyzed in Ref.~\cite{Baru:2016evv}. Appendix~\ref{sec:appen-EFT} includes relations among the LECs of the three-flavor EFT Lagrangian of Ref.~\cite{Petschauer:2013uua} and the ones used in the present work, as well as a recipe to access the full set of leading symmetry-breaking coefficients from future studies of a more complete set of two-baryon systems. Last, Appendix~\ref{sec:appen-figtab} contains figures and tables that are omitted from the main body of the paper for clarity of presentation.
\section{Lowest-lying energies and low-energy scattering parameters \label{sec:metho}}
%%%%%%%%%%%%%%%%%%%%%%%%%%%%%%%%%%%%%%%%%%%
\subsection{Details of the LQCD computation \label{subsec:comp}}
This work continues, revisits, and expands upon the study of Ref.~\cite{Orginos:2015aya}. In particular, the same ensembles of QCD gauge-field configurations that were used in Ref.~\cite{Orginos:2015aya} to constrain the low-lying spectra and scattering amplitudes of spin-singlet and spin-triplet two-nucleon systems at a pion mass of $\sim 450$ MeV are used here. The same configurations have also been used to study properties of baryons and light nuclei at this pion mass, including the rate of the radiative capture process $np \to d \gamma$~\cite{Beane:2015yha}, the response of two-nucleon systems to large magnetic fields~\cite{Detmold:2015daa}, the magnetic moments of octet baryons~\cite{Parreno:2016fwu}, the gluonic structure of light nuclei~\cite{Winter:2017bfs}, and the gluon gravitational form factors of hadrons~\cite{Detmold:2017oqb,Shanahan:2018pib,Shanahan:2018nnv}.
\begin{table}[b!]
\caption{Parameters of the gauge-field ensembles used in this work. $L$ and $T$ are the spatial and temporal dimensions of the hypercubic lattice, $\beta$ is related to the strong coupling, $b$ is the lattice spacing, $m_{l(s)}$ is the bare light (strange) quark mass, $N_{\text{cfg}}$ is the number of configurations used and $N_{\text{src}}$ is the total number of sources computed. For more details, see Ref.~\cite{Orginos:2015aya}.}
\label{tab:gauge_param}
\begin{ruledtabular}
\renewcommand{\arraystretch}{1.2}
\begin{tabular}{ccccccccccc}
$L^3\times T$ & $\beta$ & $bm_l$ & $bm_s$ & $b$ [fm] & $L$ [fm] & $T$ [fm] & $m_{\pi}L$ & $m_{\pi}T$ & $N_{\text{cfg}}$ & $N_{\text{src}}$ \\ \hline 
$24^3\times 64$ & $6.1$ & $-0.2800$ & $-0.2450$ & $0.1167(16)$ & $2.8$ & $7.5$ & $6.4$ & $17.0$ & $4407$ & $1.16\times 10^6$ \\
$32^3\times 96$ & $6.1$ & $-0.2800$ & $-0.2450$ & $0.1167(16)$ & $3.7$ & $11.2$ & $8.5$ & $25.5$ & $4142$ & $3.95\times 10^5$ \\
$48^3\times 96$ & $6.1$ & $-0.2800$ & $-0.2450$ & $0.1167(16)$ & $5.6$ & $11.2$ & $12.8$ & $25.5$ & $1047$ & $6.8\times 10^4$
\end{tabular} 
\renewcommand{\arraystretch}{1}
\end{ruledtabular}
\end{table}
For completeness, a short summary of the technical details is presented here and a more detailed discussion can be found in Ref.~\citep{Orginos:2015aya}.

The LQCD calculations are performed with $n_f=2+1$ quark flavors, with the L\"uscher-Weisz gauge action~\cite{Luscher:1984xn} and a clover-improved quark action~\cite{Sheikholeslami:1985ij} with one level of stout smearing ($\rho=0.125$)~\cite{Morningstar:2003gk}. The lattice spacing is $b=0.1167(16)$ fm~\cite{Meinel:private}. The strange quark mass is tuned to its physical value, while the degenerate light (up and down)-quark masses produce a pion of mass $m_{\pi}=450(5)$ MeV and a kaon of mass $m_K=596(6)$ MeV.
Ensembles at these parameters with three different volumes are used. Using the two smallest volumes with dimensions $24^3\times 64$ and $32^3\times 96$, two different sets of correlation functions are produced, with sink interpolating operators that are either point-like or smeared with 80 steps of a gauge-invariant Gaussian profile with parameter $\rho=3.5$ at the quark level. In both cases, the source interpolating operators are smeared with the same parameters. These two types of correlation functions are labeled SP and SS, respectively. For the largest ensemble with dimensions $48^3\times 96$, only SP correlation functions are produced for computational expediency. Table~\ref{tab:gauge_param} summarizes the parameters of these ensembles.

Correlation functions are constructed by forming baryon blocks at the sink~\cite{Detmold:2012eu}:
\begin{equation}
\mathcal{B}^{ijk}_B(\bm{p},\tau;x_0)=\sum_{\bm{x}}e^{\mathrm{i}\bm{p}\cdot\bm{x}}\, S^{(f_1),i'}_i \hskip -0.03in (\bm{x},\tau;x_0)\; S^{(f_2),j'}_j \hskip -0.03in (\bm{x},\tau;x_0)\; S^{(f_3),k'}_k \hskip -0.03in (\bm{x},\tau;x_0) \; w^{B}_{i'j'k'}\, ,
\label{eq:blocks}
\end{equation}
where $S^{(f),n'}_n$ is a quark propagator with flavor $f\in\{u,d,s\}$ and with combined spin-color indices $n,n'\in\{1,\ldots,N_sN_c\}$, where $N_s=4$ is the number of spin components and $N_c=3$ is the number of colors. The weights $w^{B}_{i'j'k'}$ are tensors that antisymmetrize and collect the terms needed to have the quantum numbers of the baryons $B\in\{N,\Lambda, \Sigma, \Xi \}$. 
The interpolating operators for the single-baryon systems studied in this work are local, i.e., include no covariant derivatives. Explicitly,
\begin{align}
%p_{\uparrow}(x) &= \hat{N}_{121}(x)\, , \text{ with } \,
\hat{N}_{\mu_1\mu_2\mu_3}(x)&=\epsilon^{abc}\frac{1}{\sqrt{2}}[u^a_{\mu_1}(x)d^b_{\mu_2}(x)-d^a_{\mu_1}(x)u^b_{\mu_2}(x)]u^c_{\mu_3}(x),
\nonumber\\
%\Lambda_{\uparrow}(x) &= \hat{\Lambda}_{121}\,, \text{ with } \, 
\hat{\Lambda}_{\mu_1\mu_2\mu_3}(x)&=\epsilon^{abc}\frac{1}{\sqrt{2}}[u^a_{\mu_1}(x)d^b_{\mu_2}(x)-d^a_{\mu_1}(x)u^b_{\mu_2}(x)]s^c_{\mu_3}(x),
\nonumber\\
%\Sigma^+_{\uparrow}(x) &= \sqrt{\frac{2}{3}}[\hat{\Sigma}_{112}(x)-\hat{\Sigma}_{121}(x)]\,, \text{ with } \,
\hat{\Sigma}_{\mu_1\mu_2\mu_3}(x)&=\epsilon^{abc}u^a_{\mu_1}(x)u^b_{\mu_2}(x)s^c_{\mu_3}(x),
\nonumber\\
%\Xi^0_{\uparrow}(x) &= \sqrt{\frac{2}{3}}[\hat{\Xi}_{112}(x)-\hat{\Xi}_{121}(x)]\, , \text{ with } \,
\hat{\Xi}_{\mu_1\mu_2\mu_3}(x)&=\epsilon^{abc}s^a_{\mu_1}(x)s^b_{\mu_2}(x)u^c_{\mu_3}(x),
\end{align}
where $\mu_i$ denote spin indices and $a,b,c$ denote color indices~\cite{Basak:2005aq}. Only the upper-spin components in the Dirac spinor basis are used, requiring only specific $\mu_i$ indices: $p_{\uparrow}(x) = \hat{N}_{121}(x)$, $\Lambda_{\uparrow}(x) = \hat{\Lambda}_{121}$, $\Sigma^+_{\uparrow}(x) = \sqrt{\frac{2}{3}}[\hat{\Sigma}_{112}(x)-\hat{\Sigma}_{121}(x)]$, and $\Xi^0_{\uparrow}(x) = \sqrt{\frac{2}{3}}[\hat{\Xi}_{112}(x)-\hat{\Xi}_{121}(x)]$. The neutron, $\Sigma^-$, and $\Xi^-$ operators are obtained by simply interchanging $u\leftrightarrow d$ in the expressions above.
The sum over the sink position $\bm{x}$ in Eq.~(\ref{eq:blocks}) projects the baryon blocks to well-defined three-momentum $\bm{p}$.
In particular, two-baryon correlation functions were generated with total momentum $\bm{P}=\bm{p}_1+\bm{p}_2$, where $\bm{p}_i$ is the three-momentum of the $i$th baryon taking the values $\bm{p}_i=\tfrac{2\pi}{L}\bm{n}$ with $\bm{n}\in\{ (0,0,0),(0,0,\pm 1) \}$. Therefore, $\bm{P}=\tfrac{2\pi}{L}\bm{d}$, with $\bm{d}\in\{ (0,0,0),(0,0,\pm 2)\}$.\footnote{For the rest of the paper, $\bm{d}=(0,0,\pm 2)$ will be denoted as $\bm{d}=(0,0,2)$ for brevity.}
Additionally, two baryon correlation functions with back-to-back momenta were generated at the sink, with momenta $\bm{p}_1=-\bm{p}_2=\tfrac{2\pi}{L}\bm{n}$. This latter choice provides interpolating operators for the two-baryon system that primarily overlap with states that are unbound in the infinite-volume limit, providing a convenient means to identify excited states as well.
The construction of the correlation functions continues by forming a fully antisymmetrized local quark-level wavefunction at the location of the source, with quantum numbers of the two-baryon system of interest. Appropriate indices from the baryon blocks at the sink are then contracted with those at the source, in a way that is dictated by the quark-level wavefunction, see Refs.~\cite{Beane:2012vq,Detmold:2012eu} for more detail. The contraction codes used to produce the correlation functions in this study are the same as those used to perform the contractions for the larger class of interpolating operators used in our previous studies of the $SU(3)$ flavor-symmetric spectra of nuclei and hypernuclei up to $A= 5$~\cite{Beane:2012vq}, and two-baryon scattering~\cite{Beane:2013br,Orginos:2015aya,Wagman:2017tmp}.\footnote{The same code was generalized to enable studies of $np\rightarrow d\gamma$~\cite{Beane:2015yha}, proton-proton fusion~\cite{Savage:2016kon}, and other electroweak processes, as reviewed in Ref.~\cite{Davoudi:2020ngi}.
%For the $n_f=2+1$ computations of this work, appropriately modified production scripts are used, as first presented in Ref.~\cite{Orginos:2015aya}
}

In this study, correlation functions for nine different two-baryon systems have been computed, ranging from strangeness $S=0$ to $-4$. Using the notation $(^{2s+1} L_J,\, I)$, where $s$ is the total spin, $L$ is the orbital momentum, $J$ is the total angular momentum, and $I$ is the isospin, the systems are:
\begin{align*}
 S=\phantom{-}0\; &: \; NN \;(\1s0,\, I=1), \ NN \;(\3s1,\, I=0), \\
 S=-1\; &: \; \Sigma N \;(\1s0,\, I=\tfrac{3}{2}),\ \Sigma N (\3s1,\, I=\tfrac{3}{2}),\\
 S=-2\; &: \; \Sigma \Sigma \;(\1s0,\, I=2),\ \Xi N \;(\3s1,\, I=0), \\
 S=-3\; &: \; \Xi \Sigma \;(\1s0,\, I=\tfrac{3}{2}),\\
 S=-4\; &: \; \Xi\Xi \;(\1s0,\, I=1), \ \Xi\Xi \;(\3s1,\, I=0).
\end{align*}
Under strong interactions, these channels do not mix with other two-baryon channels or other hadronic states below three-particle inelastic thresholds. In the limit of exact $SU(3)$ flavor symmetry, the states belong to irreducible representations (irreps) of $SU(3)$: $\mathbf{27}$ (all the singlet states), $\overline{\mathbf{10}}$ (triplet $NN$), $\mathbf{10}$ (triplet $\Sigma N$ and $\Xi \Xi$), and $\mathbf{8}_a$ (triplet $\Xi N$). In the rest of this work, the isospin label will be dropped for simplicity.

%%%%%%%%%%%%%%%%%%%%%%%%%%%%%%%%%%%%%%%%%%%
\subsection{Low-lying finite-volume spectra of two baryons \label{subsec:fitalg}}
The two-point correlation functions constructed in the previous section have spectral representations in Euclidean spacetime. Explicitly, the correlation function $C_{\hat{\mathcal{O}},\hat{\mathcal{O}}'}(\tau;\bm{d})$ formed using the source (sink) interpolating operators $\hat{\mathcal{O}}$ ($\hat{\mathcal{O}}'$) can be written as
\begin{equation}
C_{\hat{\mathcal{O}},\hat{\mathcal{O}}'}(\tau;\bm{d}) = \sum_{\bm{x}}e^{2\pi i \bm{d}\cdot \bm{x}/L}\langle \hat{\mathcal{O}}'(\bm{x},\tau)\hat{\mathcal{O}}^\dagger(\bm{0},0) \rangle = \sum_i \mathcal{Z}'_i\mathcal{Z}^*_i e^{-E^{(i)}\tau},
\label{eq:correlator}
\end{equation}
where all quantities are expressed in lattice units. $E^{(i)}$ is the energy of the $i$th eigenstate $| E^{(i)} \rangle$, $\mathcal{Z}_i$ ($\mathcal{Z}'_i$) is an overlap factor defined as $\mathcal{Z}_i= \sqrt{V} \langle 0| \hat{\mathcal{O}}(\bm{0},0) |E^{(i)}\rangle$ ($\mathcal{Z}'_i= \sqrt{V} \langle 0| \hat{\mathcal{O}}'(\bm{0},0) |E^{(i)}\rangle$), and $V=L^3$. The boost-vector dependence of the energies, states, and overlap factors is implicit.
The lowest-lying energies of the one- and two-baryon systems required for the subsequent analyses can be extracted by fitting the correlation functions to this form. To reliably discern the first few exponents given the discrete $\tau$ values and the finite statistical precision of the computations is a challenging task.
In particular, a well-known problem in the study of baryons with LQCD is the exponential degradation of the signal-to-noise ratio in the correlation function as the source-sink separation time increases---an issue that worsens as the masses of the light quarks approach their physical values. First highlighted by Parisi~\cite{Parisi:1983ae} and Lepage~\cite{Lepage:1989hd}, and studied in detail for light nuclei in Refs.~\cite{Beane:2009gs, Beane:2009py}, it was later shown that this problem is related to the behavior of the complex phase of the correlation functions~\cite{Wagman:2016bam,Wagman:2017xfh}.
Another problem that complicates the study of multi-baryon systems is the small excitation gaps in the finite-volume spectrum that lead to significant excited-state contributions to correlation functions. To overcome these issues, sophisticated methods have been developed to analyze the correlation functions, such as Matrix Prony~\cite{Beane:2009kya} and the generalized pencil-of-function~\cite{Aubin:2010jc} techniques, as well as signal-to-noise optimization techniques~\cite{Detmold:2014hla}.
Ultimately, a large set of single- and multi-baryon interpolating operators with the desired quantum numbers must be constructed to provide a reliable variational basis to isolate the lowest-lying energy eigenvalues via solving a generalized eigenvalue problem~\cite{Blossier:2009kd}, as is done in the mesonic sector~\cite{Briceno:2017max}. 
Such an approach is not yet widely applied to the study of two-baryon correlation functions, given its computational-resource requirement, but progress is being made. In Ref.~\cite{Francis:2018qch}, a partial set of two-baryon scattering interpolating operators were used to study the two-nucleon and $H$-dibaryon channels with results that generally disagreed with previous works~\cite{Beane:2012vq, Berkowitz:2015eaa,Beane:2010hg}. Investigations continue to understand and resolve the observed discrepancies~\cite{Iritani:2017rlk, Beane:2017edf, Wagman:2017tmp, Davoudi:2017ddj, Yamazaki:2017jfh, Drischler:2019xuo, Davoudi:2020ngi}.
For the present study, in which only up to two types of interpolating operators were computed, no variational analysis could be performed. Instead, we have developed a robust automated fitting methodology to sample and combine fit range and model selection choices for uncertainty quantification. 

Given that the correlation functions are only evaluated at a finite number of times and with finite precision, to fit Eq.~\eqref{eq:correlator} the spectral representation is truncated to a relatively small number of exponentials and fitted in a time range $\{\tau_{\text{min}},\tau_{\text{max}}\}$, where $\tau_{\text{max}}$ is set by a threshold value determined by examining the signal-to-noise ratio, and $\tau_{\text{min}}$ is chosen to take values in the interval $[2,\tau_{\text{max}}-\tau_{\text{plateau}}]$.
Here, $\tau_{\text{plateau}}=5$ is chosen to be the minimum length of the fitting window (numbers are expressed in units of the lattice spacing). A scan over all possible fitting windows is treated as a means to quantify the associated systematic uncertainty.
With a fixed window, a correlated $\chi^2$-function is minimized to obtain the fit parameters $\mathcal{Z}'_i\mathcal{Z}_i^*$ and $E_i$ for $i\in\{0,1,\ldots,e\}$, where $e+1$ is the number of exponentials in a given fit form. Variable projection techniques~\cite{Golub_2003,Olearly_2013} are used to obtain the value of the overlap factors for a given energy, since they appear linearly in $C_{\hat{\mathcal{O}},\hat{\mathcal{O}}'}(\tau;\bm{d})$.
Furthermore, given the finite statistical sampling of correlation functions, shrinkage techniques~\cite{Rinaldi:2019thf} are used to better estimate the covariance matrix. The number of excited states included in the fit is decided via the Akaike information criterion~\cite{AkaikeAIC}.
The confidence intervals of the parameters are estimated via the bootstrap resampling method. For a fit to be included in the set of accepted fits (later used to extract the fit parameters and assess the resulting uncertainties), several checks must be passed, including $\chi^2/N_{\text{dof}}$ being smaller than 2, and different optimization algorithms leading to consistent results for the parameters (within a tolerance)---see Ref.~\cite{Beane:2020ycc} for further details on this and other checks. The accepted fits are then combined to give the final result for the mean value of the energy,
\begin{equation}
\overline{E}=\sum_f \omega_f E_f\, ,
\end{equation}
with weights $\omega_f$ that are chosen to be the following combination of the $p$-value, $p_f$, and the uncertainty of each fit, $\delta E_{f}$:
\begin{equation}
\omega_f=\frac{p_f/\delta E_f^{2}}{\sum_{f'}p_{f'}/\delta E_{f'}^{2}}\, ,
\label{eq:weights}
\end{equation}
see Ref.~\cite{Jay:2020jkz} for a Bayesian framework. Here, the indices $f, f'$ run over all the accepted fits.
The statistical uncertainty is defined as that of the fit with the highest weight, while the systematic uncertainty is defined as the average difference between the weighted mean value and each of the accepted fits:
\begin{equation}
\delta \overline{E}_{\text{stat}}=\delta E^{f:\text{max}[\{w_f\}]}\, ,\qquad \delta \overline{E}_{\text{sys}}=\sqrt{\sum_f w_f \left(E^f-\overline{E}\right)^2}\, .
\label{eq:weights_fitting}
\end{equation}
It should be noted that instead of fitting to the correlation function, the effective energy function can be employed, derived from the logarithm of the ratio of correlation functions at displaced times,
\begin{equation}
\mathcal{C}_{\hat{\mathcal{O}},\hat{\mathcal{O}}'}(\tau;\bm{d},\tau_J)=\frac{1}{\tau_J}\log\left[\frac{C_{\hat{\mathcal{O}},\hat{\mathcal{O}}'}(\tau;\bm{d})}{C_{\hat{\mathcal{O}},\hat{\mathcal{O}}'}(\tau+\tau_J;\bm{d})}\right]\xrightarrow[]{\tau\rightarrow\infty}E^{(0)}\, ,
\label{eq:effmassfun}
\end{equation}
where $\tau_J$ is a non-zero integer that is introduced to improve the extraction of $E^{(0)}$ (for a detailed study, see Ref.~\cite{Beane:2009kya}). Consistent results are obtained when either correlation functions or the effective energy functions are used as input.

In order to identify the shift in the finite-volume energies of two baryons compared with non-interacting baryons, the following ratio of two-baryon and single-baryon correlation functions can be formed
\begin{equation}
R_{\scriptscriptstyle B_1B_2}(\tau;\bm{d})=\frac{C_{\hat{\mathcal{O}}_{B_1B_2},\hat{\mathcal{O}}'_{B_1B_2}}(\tau;\bm{d})}{C_{\hat{\mathcal{O}}_{B_1},\hat{\mathcal{O}}'_{B_1}}(\tau;\bm{d}) C_{\hat{\mathcal{O}}_{B_2},\hat{\mathcal{O}}'_{B_2}}(\tau;\bm{d})}\, ,
\label{eq:energyshiftfun}
\end{equation}
with an associated effective energy-shift function,
\begin{equation}
\mathcal{R}_{\scriptscriptstyle B_1B_2} (\tau;\bm{d},\tau_J) = \frac{1}{\tau_J}\log\left[\frac{R_{\scriptscriptstyle B_1B_2}(\tau;\bm{d})}{R_{\scriptscriptstyle B_1B_2}(\tau+\tau_J;\bm{d})}\right]\, .
\label{eq:effenergyshiftfun}
\end{equation}
Constant fits to ratios of correlation functions can be used to obtain energy shifts $\Delta E^{(0)}=E^{(0)}-m_1-m_2$ (where $m_1$ and $m_2$ are the masses of baryons $B_1$ and $B_2$, respectively), requiring time ranges such that both the two-baryon and the single-baryon correlation functions are described by a single-state fit.
However, if both correlation functions are not in their ground states, cancellations may occur between excited states (including the finite-volume states that would correspond to elastic scattering states in the infinite volume), either in correlation function or in ratios of correlation functions, producing a ``mirage plateau''~\cite{Iritani:2017rlk}. Despite this issue, as demonstrated in Ref.~\cite{Beane:2017edf}, our previous results, such as those in Refs.~\cite{Beane:2012vq,Wagman:2017tmp} are argued to be free of this potential issue (similar discussions can be found in Ref.~\cite{Yamazaki:2017jfh}).
To determine $\Delta E^{(0)}$ in this work, the two-baryon and single-baryon correlation functions are fit to multi-exponential forms (which account for excited states) within the same fitting range, and afterwards the energy shifts are computed at the bootstrap level, in such a way that the correlations between the different correlation functions are taken into account.
The use of correlated differences of multi-state fit results is convenient in particular for automated fit range sampling, since the number of excited states can be varied independently for one- and two-baryon correlation functions, unlike fits to the ratio in Eq.~\eqref{eq:energyshiftfun}. Consistent results were obtained via fitting the ratio in Eq.~\eqref{eq:energyshiftfun} in the allowed time regions. 

The effective mass plots (EMPs) for the single-baryon correlation functions, and for each of the ensembles studied in the present work, are displayed in Fig.~\ref{fig:B1_EMP} of Appendix~\ref{sec:appen-figtab}. The bands shown in the figures indicate the baryon mass which results from the fitting strategy explained above, with the statistical and systematic uncertainties included, and the corresponding numerical values listed in Table~\ref{tab:baryon_mass}. The table also shows the baryon masses extrapolated to infinite volume, obtained by fitting the masses in the three different volumes to the following form:
\begin{equation}
M^{(V)}_{B}(m_{\pi}L)=M^{(\infty)}_B+c_B\frac{e^{-m_{\pi}L}}{m_{\pi}L}\, ,
\label{eq:mass_extrap}
\end{equation}
where $M^{(\infty)}_B$ and $c_B$ are the two fit parameters. This form incorporates LO volume corrections to the baryon masses in heavy-baryon chiral perturbation theory (HB$\chi$PT)~\cite{Beane:2011pc}. As evident from the $m_\pi L$ values listed in Table~\ref{tab:gauge_param}, the volumes used are large enough to ensure small volume dependence in the single-baryon masses.\footnote{For the smallest volume $e^{-m_{\pi}L}/m_{\pi}L \sim 10^{-4}$ and $c_B$ are of $\mathcal{O}(1)$ but consistent with zero within uncertainties, e.g., for the nucleon, $c_B=3(4)(7)$ l.u.}
This is supported by the observation that the value $M^{(\infty)}_B$ obtained for each baryon is compatible with all the finite-volume results $M^{(V)}_{B}$.
In physical units, $M_N\sim 1226$ MeV, $M_{\Lambda}\sim 1313$ MeV, $M_{\Sigma}\sim 1346$ MeV and $M_{\Xi}\sim 1414$ MeV. While the $\Lambda$ baryon is not relevant to subsequent analysis of the two-baryon systems studied in this work, the centroid of the four octet-baryon masses is used to define appropriate units for the EFT LECs, hence $M_{\Lambda}$ is reported for completeness.

\begin{table}[t!]
\caption{The values of the masses of the octet baryons. The first uncertainty is statistical, while the second is systematic. Quantities are expressed in lattice units (l.u.).}
\label{tab:baryon_mass}
\begin{ruledtabular}
\begin{tabular}{ccccc}
Ensemble & $M_N$ [l.u.] & $M_{\Lambda}$ [l.u.] & $M_{\Sigma}$ [l.u.] & $M_{\Xi}$ [l.u.] \\
\hline
$24^3\times 64$ & $0.7261(08)(15)$ & $0.7766(07)(13)$ & $0.7959(07)(10)$ & $0.8364(07)(08)$ \\
$32^3\times 96$ & $0.7258(05)(08)$ & $0.7765(05)(06)$ & $0.7963(05)(06)$ & $0.8362(05)(05)$ \\
$48^3\times 96$ & $0.7250(06)(12)$ & $0.7761(05)(09)$ & $0.7955(06)(07)$ & $0.8359(08)(08)$ \\ \hline
$\infty$ & $0.7253(04)(08)$ & $0.7763(04)(06)$ & $0.7959(04)(05)$ & $0.8360(05)(05)$ \\
\end{tabular}
\end{ruledtabular}
\end{table}
\begin{figure}[t!]
\includegraphics[width=\textwidth]{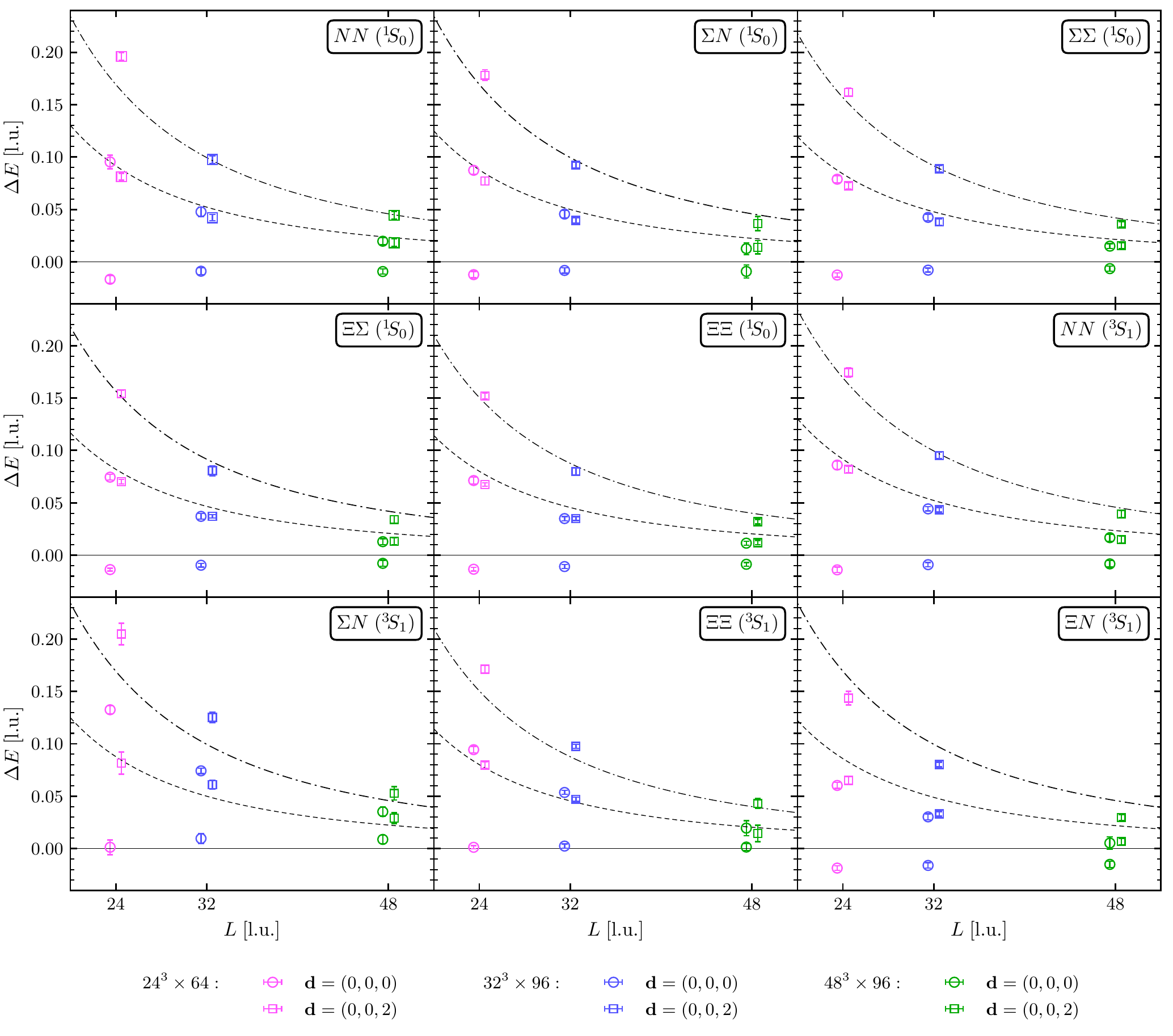}
\caption{Summary of the energy shifts extracted from LQCD correlation functions for all two-baryon systems studied in this work, together with the non-interacting energy shifts defined as $\Delta E = \sqrt{m_1^2+|\bm{p}_1|^2}+\sqrt{m_2^2+|\bm{p}_2|^2}-m_1-m_2$, where $|\bm{p}_1|^2=|\bm{p}_2|^2=0$ corresponds to systems that are at rest (continuous line), $|\bm{p}_1|^2=|\bm{p}_2|^2=(\frac{2\pi}{L})^2$ corresponds to systems which are either boosted or are unboosted but have back-to-back momenta (dashed line), and $|\bm{p}_1|^2=0$ and $|\bm{p}_2|^2=(\frac{4\pi}{L})^2$ corresponds to boosted systems where only one baryon has non-zero momentum (dash-dotted line). The points with no boost have been shifted slightly to the left, and the ones with boosts have been shifted to the right for clarity. Quantities are expressed in lattice units.}
\label{fig:energy-levels}
\end{figure}

The results for the two-baryon energy shifts are shown in Fig.~\ref{fig:energy-levels}.\footnote{The channels within the figures/tables are sorted according to the $SU(3)$ irrep they belong to in the limit of exact flavor symmetry, ordered as $\mathbf{27},~\overline{\mathbf{10}},~\mathbf{10},~\text{and}~\mathbf{8}_a$, and within each irrep according to their strangeness, from the largest to the smallest.}
For display purposes, the effective energy-shift functions, defined in Eq.~\eqref{eq:effenergyshiftfun}, are shown in Figs.~\ref{fig:NN1s0_EMP}-\ref{fig:XN3s1_EMP} of Appendix~\ref{sec:appen-figtab}, along with the corresponding two-baryon effective-energy functions, defined in Eq.~\eqref{eq:effmassfun}. The associated numerical values are listed in Tables~\ref{tab:eshift_ini}-\ref{tab:eshift_fin} of the same appendix.\footnote{Future studies with a range of values of the lattice spacing will be needed to extrapolate the results of two-baryon studies to the continuum limit. Nonetheless, the use of an improved lattice action in this study suggests that the discretization effects may be mild, and the associated systematic uncertainty, which has not been reported in the values in this paper, may not be significant at the present level of precision.}
In each subfigure of Figs.~\ref{fig:NN1s0_EMP}-\ref{fig:XN3s1_EMP}, two correlation functions are displayed: the one yielding the lowest energy (labeled as $n=1$ in Tables~\ref{tab:eshift_ini}-\ref{tab:eshift_fin}) corresponds to having both baryons at rest or, if boosted, with the same value of the momentum, and the one yielding a higher energy (labeled as $n=2$ in the tables) corresponds to the two baryons having different momenta, e.g., having back-to-back momenta or one baryon at rest and the other with non-zero momentum. While the first case ($n=1$) couples primarily to the ground state, the latter ($n=2$) is found to have small overlap onto the ground state, and gives access to the first excited state directly. 

As a final remark, it should be noted that the single-baryon masses and the energies extracted for the two-nucleon states within the present analysis are consistent within $1\sigma$ with the results of Ref.~\cite{Orginos:2015aya}, obtained with the same set of data but using different fitting strategies. Despite this overall consistency, the uncertainties of the two-nucleon energies in the present work are generally larger compared with those reported in Ref.~\cite{Orginos:2015aya} for the channels where results are available in that work. The reason lies in a slightly more conservative systematic uncertainty analysis employed here. The comparison between the results of this work and that of Ref.~\cite{Orginos:2015aya} is discussed extensively in Appendix~\ref{sec:appen-vs2015}. 

%%%%%%%%%%%%%%%%%%%%%%%%%%%%%%%%%%%%%%%%%%%
\subsection{Low-energy scattering phase shifts and effective-range parameters\label{subsec:luscher}}
Below three-particle inelastic thresholds, L{\"u}scher's quantization condition~\cite{Luscher:1986pf, Luscher:1990ux} provides a means to extract the infinite-volume two-baryon scattering amplitudes from the energy eigenvalues of two-baryon systems obtained from LQCD calculations, e.g., those presented in Sec.~\ref{subsec:fitalg}.
This condition holds if the range of interactions is smaller than (half of) the spatial extent of the cubic volume, $L$, and the corrections to this condition scale as $e^{-m_\pi L}$ for the two-baryon systems. Such corrections are expected to be small in the present work given the $m_\pi L$ values in Table~\ref{tab:gauge_param}. The quantization conditions are those used in Refs.~\cite{Orginos:2015aya,Wagman:2017tmp}: in the case of spin-singlet states, only the $s$-wave limit of the full quantization condition is considered. For coupled $\3s1 - ^3\hskip -0.06 in D_1$ states, in which the Blatt-Biedenharn parametrization~\cite{Blatt:1952zza} of the scattering matrix can be used, only the $\alpha$-wave approximation of the quantization condition is considered~\cite{Briceno:2013bda}.
In both cases, and denoting the ($s$-wave or $\alpha$-wave) phase shift by $\delta$, the condition can be written as~\cite{Davoudi:2011md}
\begin{equation}
k^* \cot\delta= 4\pi c^{\bm{d}}_{00}(k^{*2};L)\, ,
\label{eq:QC}
\end{equation}
where $k^*$ is the center-of-mass (c.m.\@) relative momentum of each baryon, $\bm{d}$ is the total c.m.\ momentum in units of $2\pi/L$, and $c^{\bm{d}}_{lm}$ is a kinematic function related to L\"uscher's $\mathcal{Z}$-function, $\mathcal{Z}^{\bm{d}}_{lm}$:
\begin{equation}
c^{\bm{d}}_{lm}(k^{*2};L)=\frac{\sqrt{4\pi}}{\gamma L^3}\left(\frac{2\pi}{L}\right)^{l-2} \mathcal{Z}^{\bm{d}}_{lm}[1;(k^*L/2\pi)^2]\, ,
\label{eq:kcotextract}
\end{equation}
with $\gamma=E/E^*$ being the relativistic gamma factor. Here, $E$ and $E^*$ are the energies of the system in the laboratory and c.m.\ frames, respectively. The three-dimensional zeta-function is defined as
\begin{equation}
\mathcal{Z}^{\bm{d}}_{lm}[s;x^2]=\sum_{\bm{n}}\frac{|\bm{r}|^l Y_{lm}(\bm{r})}{(\bm{r}^2-x^2)^s}\, ,
\label{eq:zfun}
\end{equation}
where $\bm{r}=\hat{\gamma}^{-1}(\bm{n}-\alpha\bm{d})$ and $\alpha=\frac{1}{2}\left[1+(m_1^2-m_2^2)/E^{*2}\right]$,\footnote{Here $\alpha$ should not be confused with $\alpha$-wave mentioned above.} with $m_1$ and $m_2$ being the masses of the two baryons.
The factor $\hat{\gamma}^{-1}$ acting on a vector $\bm{u}$ modifies the parallel component with respect to $\bm{d}$, while leaving the perpendicular component invariant, i.e., $\hat{\gamma}^{-1}\bm{u}=\gamma^{-1}\bm{u}_{\parallel}+\bm{u}_{\perp}$.
Convenient expressions have been derived to exponentially accelerate the numerical evaluation of the function in Eq.~\eqref{eq:zfun}~\cite{Yamazaki:2004qb, Beane:2011sc, Fu:2011xz, Leskovec:2012gb}, and the following expression is used in the present analysis:
\begin{equation}
\mathcal{Z}^{\bm{d}}_{00}[1;x^2]=-\gamma \pi e^{x^2}+\frac{e^{x^2}}{\sqrt{4\pi}}\sum_{\bm{n}}\frac{e^{-|\bm{r}|^2}}{|\bm{r}|^2-x^2}+\gamma \frac{\pi}{2}\int^1_0 dt \, \frac{e^{tx^2}}{t^{3/2}}\left[\sum_{\bm{m}\neq \bm{0}}\cos(2\pi \alpha {\bm{m}\cdot\bm{d}})e^{\frac{-\pi^2|\hat{\gamma}\bm{m}|^2}{t}}+2tx^2\right].
\end{equation}
\begin{figure}[t!]
\includegraphics[width=\textwidth]{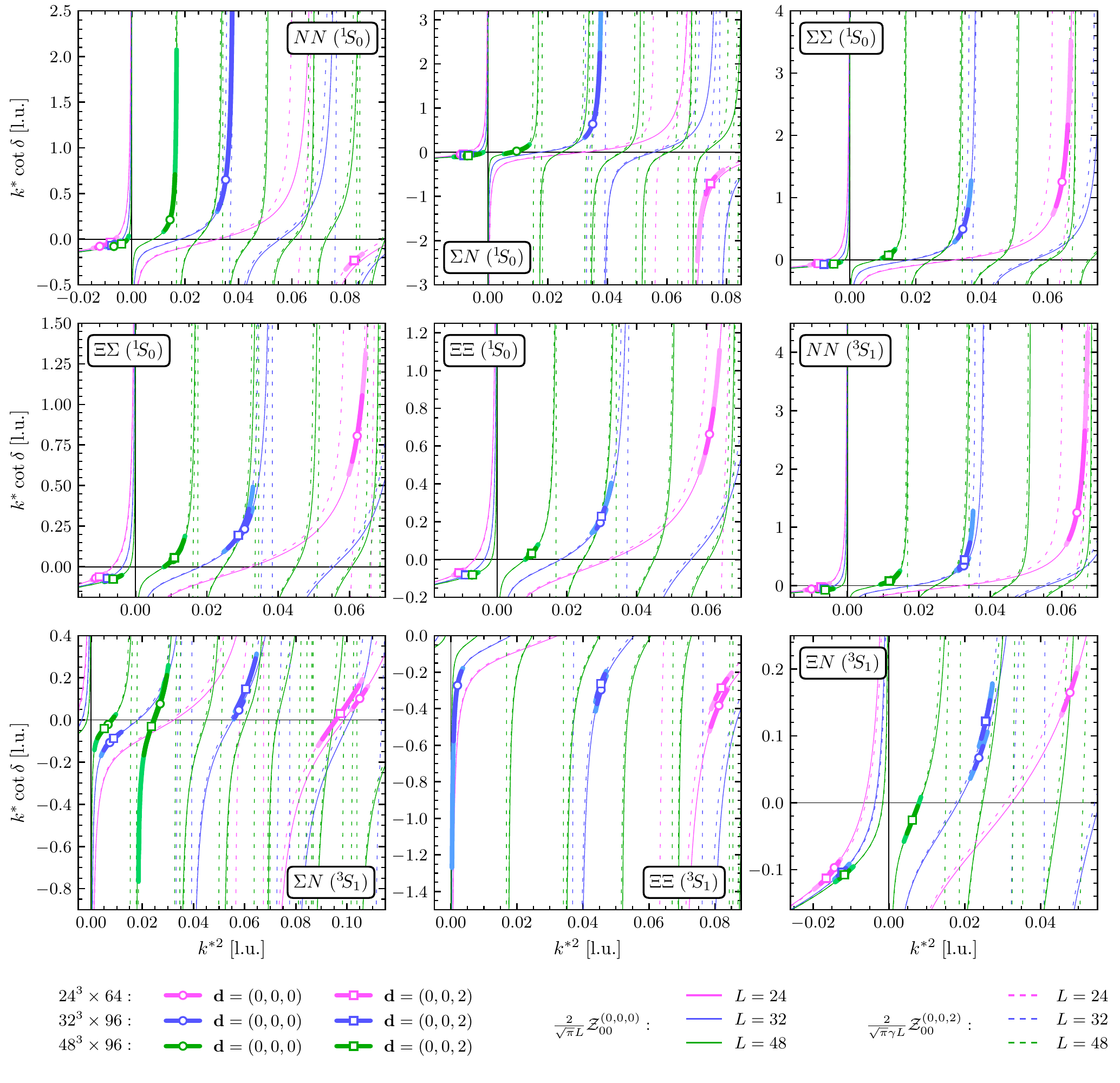}
\caption{$k^*\cot \delta$ values as a function of the squared c.m.\ momentum $k^{*2}$ for all two-baryon systems studied in this work. The darker uncertainty bands are statistical, while the lighter bands show the statistical and systematic uncertainties combined in quadrature. The kinematic functions $c^{\bm{d}}_{lm}(k^{*2};L)$, given by Eq.~\eqref{eq:kcotextract}, are also shown as continuous and dashed lines. Quantities are expressed in lattice units.}
\label{fig:kcotdelta}
\end{figure}

The values of $k^* \cot\delta$ at given $k^{*2}$ values are shown for all two-baryon systems in Fig.~\ref{fig:kcotdelta}, and the associated numerical values are listed in Tables~\ref{tab:eshift_ini}-\ref{tab:eshift_fin} of Appendix~\ref{sec:appen-figtab}. 
The validity of L\"uscher's quantization condition must be verified in each channel, in particular in those that have exhibited anomalously large ranges, such as $\Sigma N\; (\3s1)$, in previous calculations.
The consistency between solutions to L\"uscher's condition and the finite-volume Hamiltonian eigenvalue equation using a LO EFT potential was established in Ref.~\cite{Beane:2012ey} for the $\Sigma N\; (\3s1)$ channel and at values of the quark masses ($m_{\pi}\sim 389$ MeV) close to those of the current analysis. The conclusion of Ref.~\cite{Beane:2012ey}, therefore, justifies the use of L\"uscher's quantization condition in the current work for this channel.

The energy dependence of $k^* \cot\delta$ can be parametrized by an effective range expansion (ERE) below the $t$-channel cut~\cite{Schwinger:notes,Blatt:1949zz,Bethe:1949yr},\footnote{Since the pion is the lightest hadron that can be exchanged between any of the two baryons considered in the present study, $k^*_{t\text{-cut}}=m_{\pi}/2$.}
\begin{equation}
k^*\cot\delta = -\frac{1}{a}+\frac{1}{2}rk^{*2}+Pk^{*4}+\mathcal{O}(k^{*6})\, ,
\label{eq:ERE}
\end{equation}
where $a$ is the scattering length, $r$ is the effective range, and $P$ is the leading shape parameter. These parameters can be constrained by fitting $k^* \cot\delta$ values obtained from the use of L\"uscher's quantization condition as a function of $k^{*2}$. To this end, one could use a one-dimensional choice of the $\chi^2$ function, minimizing the vertical distance between the fitted point and the function,
\begin{equation}
\chi^2(a^{-1},r,P)=\sum_i \frac{[(k^*\cot\delta)_i-f(a^{-1},r,P,k_i^{*2})]^2}{\sigma_i^2}\, ,
\label{eq:chisquared1D}
\end{equation}
where\footnote{The inverse scattering length can be constrained far more precisely compared with the scattering length itself given that $a^{-1}$ samples can cross zero in the channels considered. As a result, in the following all dependencies on $a$ enter via $a^{-1}$.} $f(a^{-1},r,P,k^{*2})$ corresponds to the ERE parametrization given by the right-hand side of Eq.~\eqref{eq:ERE}, and the sum runs over all extracted pairs of $\{k_i^{*2}, (k^*\cot\delta)_i \}$, where the compound index $i$ counts data points for different boosts, $n$ values of the level, and different volumes.
Each contribution is weighted by an effective variance that results from the combination of the uncertainty in both $k_i^{*2}$ and $(k^*\cot\delta)_i$, $\sigma_i^2 =[\delta (k^*\cot\delta)_i]^2+[\delta k_i^{*2}]^2\,$, with $\delta x$ being the mid-68\% confidence interval of the quantity $x$. The uncertainty on the $\{k_i^{*2}, (k^*\cot\delta)_i \}$ pair can be understood by recalling that each pair is a member of a bootstrap ensemble with the distribution obtained in the previous step of the analysis.
To generate the distribution of the scattering parameters, pairs of $\{k_i^{*2}, (k^*\cot\delta)_i \}$ are randomly selected from each bootstrap ensemble and are used in Eq.~\eqref{eq:chisquared1D} to obtain a new set of $\{a^{-1},r,P\}$ parameters. This procedure is repeated $N$ times, where $N$ is chosen to be equal to the number of bootstrap ensembles for $\{k_i^{*2}, (k^*\cot\delta)_i \}$. This produces an ensemble of $N$ values of fit parameters $\{a^{-1},r,P\}$, from which the central value and the associated uncertainty in the parameters can be determined (median and mid-$68\%$ intervals are used for this purpose). 

Alternatively, one can use a two-dimensional choice of the $\chi^2$ function.\footnote{We thank Sinya Aoki for suggesting that we further explore this choice of $\chi^2$.}
Knowing that $k^*\cot\delta$ values must lie along the $\mathcal{Z}$-function, as can be seen from Eq.~\eqref{eq:QC} and Fig.~\ref{fig:kcotdelta}, one could take the distance between the data point and the point where the ERE crosses the $\mathcal{Z}$-function along this function (arc length) in the definition of $\chi^2$. Explicitly,
\begin{equation}
\chi^2(a^{-1},r,P)=\sum_i \frac{D_{\mathcal{Z}}[\{k_i^{*2},(k^*\cot\delta)_i\},\{K_i^{*2},f(a^{-1},r,P,K_i^{*2})\}]^2}{\sigma_i^2}\, ,
\label{eq:newchi}
\end{equation}
where $\sigma_i^2$ is now defined as
\begin{equation}
\sigma_i^2 =[\delta (k^*\cot\delta)_i]^2+\left(\left.\frac{\partial (k^*\cot\delta)_i}{\partial k^{*2}}\right|_{k^{*2}=k_i^{*2}}\right)^2 [\delta k_i^{*2}]^2\, ,
\label{eq:weights_chi2}
\end{equation}
and $D_{\mathcal{Z}}[\{x_1,y_1\},\{x_2,y_2\}]$ denotes the distance between the two points $\{x_1,y_1\}$ and $\{x_2,y_2\}$ along the $\mathcal{Z}$-function.
The quantity $K^{*2}$ is the point where the ERE ($f$ in Eq.~\eqref{eq:newchi}) crosses the $\mathcal{Z}$-function. To obtain this point, and given the large number of discontinuities present in the $\mathcal{Z}$-function, Householder's third order method can be used as a reliable root-finding algorithm~\cite{Householder:231838}:
\begin{equation}
4\pi c^{\bm{d}}_{00}(K^{*2},L)-f(a^{-1},r,P,K^{*2})\equiv F(K^{*2})=0 \; : \; K_{m+1}^{*2}=K_{m}^{*2}+3\left . \frac{(1/F)''}{(1/F)'''}\right |_{K_{m}^{*2}}\, ,
\end{equation}
where the starting point is set to be $K^{*2}_0=k_i^{*2}$ and the number of primes over $1/F$ indicates the order of the derivative computed at the point $K_{m}^{*2}$. The stopping criterion is defined as $|K_{m+1}^{*2}-K_{m}^{*2}|<10^{-6}$, which occurs for $m\sim\mathcal{O}(10)$.
Since the extraction of this point requires knowledge of scattering parameters, the minimization must be implemented iteratively. This second choice of $\chi^2$ function has been used in the main analysis of this work, however, the use of the one-dimensional $\chi^2$ function is shown to yield statistically consistent results (within $1\sigma$) for scattering parameters, as demonstrated in Appendix~\ref{sec:appen-vs2015}.

\begin{figure}[t!]
\includegraphics[width=\textwidth]{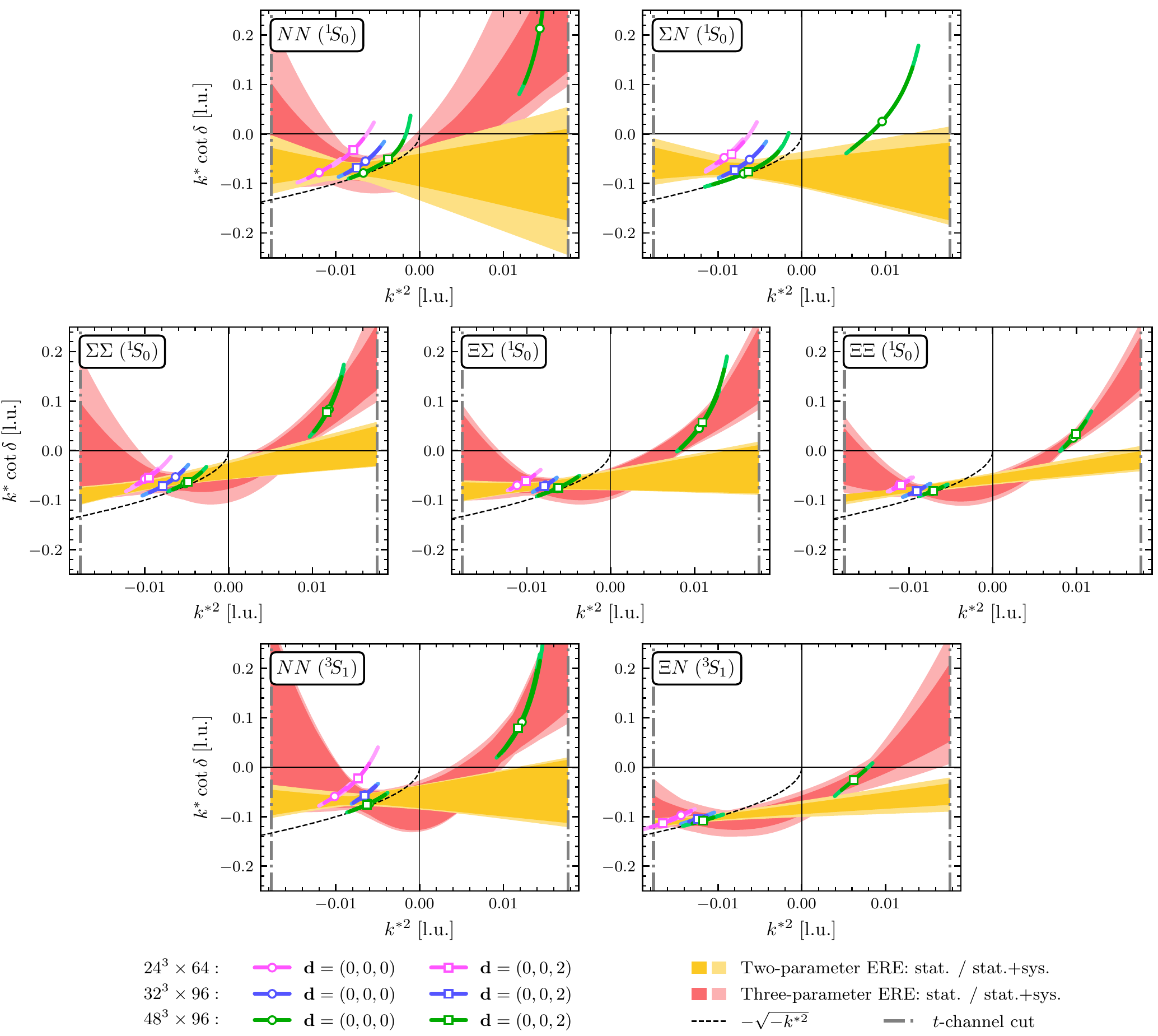}
\caption{$k^*\cot \delta$ values as a function of the c.m.\ momenta $k^{*2}$, along with the band representing the two- (yellow) and three-parameter (red) ERE for the two-baryon channels shown. The bands denote the 68\% confidence regions of the fits. Quantities are expressed in lattice units.}
\label{fig:ere-fit}
\end{figure}
\begin{figure}[t!]
\includegraphics[width=\textwidth]{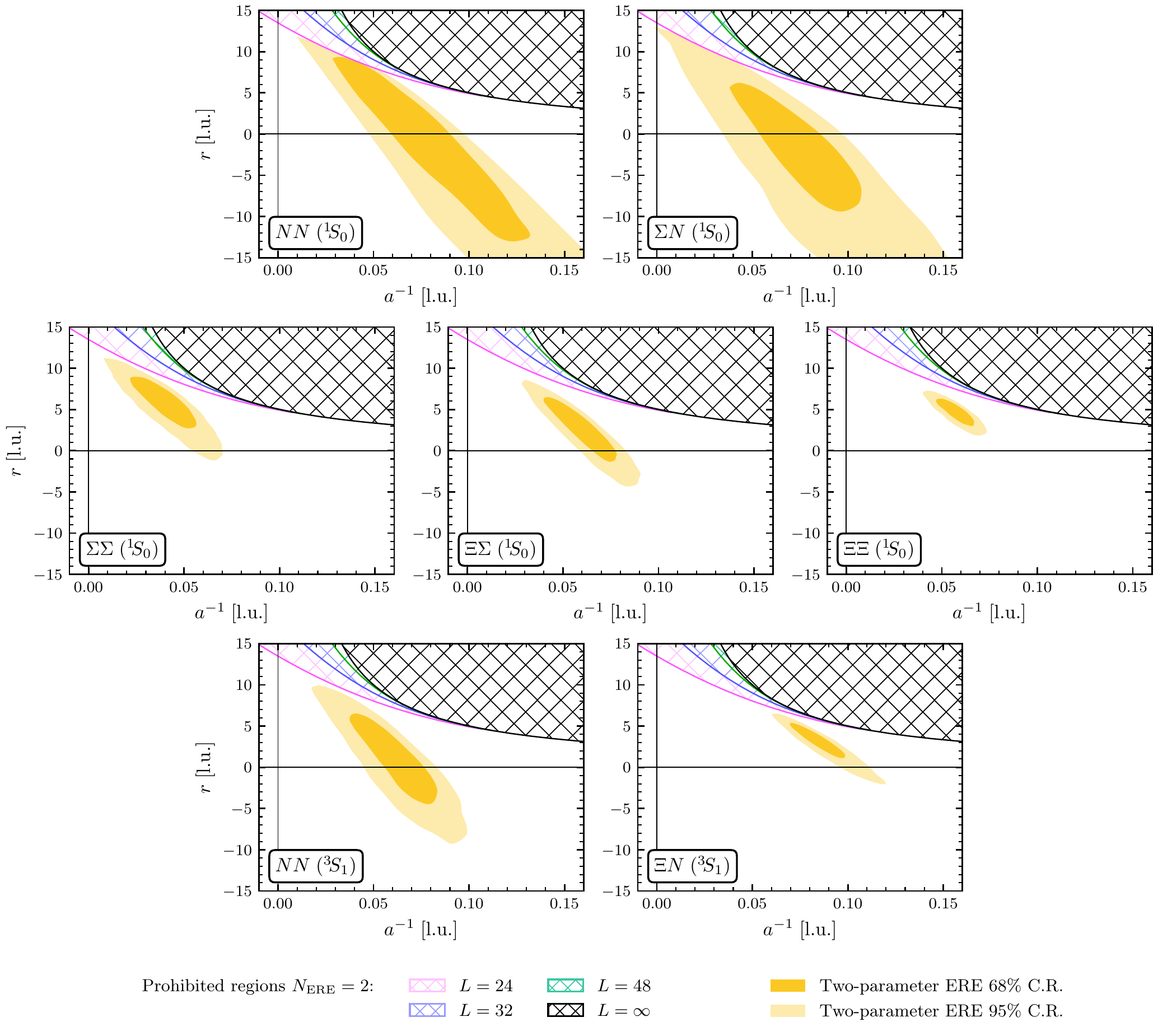}
\caption{The two-dimensional 68\% and 95\% confidence regions (C.R.) corresponding to the combined statistical and systematic uncertainty on the scattering parameters $a^{-1}$ and $r$ for all two-baryon systems that exhibit bound states, obtained from two-parameter ERE fits. The prohibited regions where the two-parameter ERE does not cross the $\mathcal{Z}$-function for given volumes (as well as the infinite-volume case) are denoted by hashed areas. Quantities are expressed in lattice units.}
\label{fig:ere-parameters}
\end{figure}
For a precise extraction of the ERE parameters, a sufficient number of points below the $t$-channel cut must be available, for positive or negative $k^{*2}$.
In general, for the channels studied throughout this work, there are only a few points in the positive $k^{*2}$ region below the $t$-channel cut (starting at $k_{t\text{-cut}}^{*2}\sim 0.018 \text{ l.u.}$). 
For a non-interacting system, states above scattering threshold have c.m.\ energies $\sqrt{m_1^2+k^{*2}}+\sqrt{m_2^2+k^{*2}}$, with the c.m.\ momenta roughly scaling with the volume as $k^{*2}\sim(2\pi |\bm{n}|/L)^2$. With the minimum value of $|\bm{n}|^2$ used in this work ($|\bm{n}|^2=1$), only the states from the ensemble with $L=48$ are expected to lie below the $t$-channel cut ($4\pi^2/48^2<k_{t\text{-cut}}^{*2}$).
This behavior is consistent with the data. Comparing with the results of the analysis at $m_\pi \sim 806$ MeV in Ref.~\cite{Wagman:2017tmp}, where lattice configurations of comparable size (in lattice units) were used, the larger value of the pion mass resulted in the position of the $t$-channel cut being moved further away from zero, and the majority of the lowest-lying states extracted in that study remained inside the region where the ERE parametrization could be used.
Therefore, with only ground-state energies available for the analysis of the ERE in the ensembles with $L\in\{24,32\}$, the precision in the extraction of scattering parameters is noticeably reduced compared with the study at $m_{\pi} \sim 806$ MeV in Ref.~\cite{Wagman:2017tmp}.
Inclusion of the shape parameter, $P$, does not improve the fits, and although the scattering lengths remain consistent with those obtained with a two-parameter fit, the effective ranges are larger in magnitude, and the uncertainties in the scattering parameters are increased. Moreover, the central values of the extracted shape parameters are rather large, bringing into question the assumption that the contribution of each order in the ERE should be smaller than the previous order. However, uncertainties on the shape parameters are sufficiently large that no conclusive statement can be made regarding the convergence of EREs.
In one case, i.e., the $\Sigma N\; (\1s0)$ channel, the three-parameter ERE fit is not performed given the large uncertainties. For these reasons, while the scattering parameters are reported for both the two- and three-parameter fits in this section, only those of the two-parameter fits will be used in the EFT study in the next section.

\begin{table}[t!]
\caption{The values of the inverse scattering length $a^{-1}$, effective range $r$, and shape parameter $P$ determined from the two- and three-parameter ERE fits to $k^*\cot\delta$ versus $k^{*2}$ for various two-baryon channels (see Fig.~\ref{fig:ere-fit}). Quantities are expressed in lattice units.}
\label{tab:scattPar}
\begin{ruledtabular}
\begin{tabular}{c|cr|ccr}
 & \multicolumn{2}{c|}{Two-parameter ERE fit} & \multicolumn{3}{c}{Three-parameter ERE fit} \\
 & $a^{-1}$ [l.u.] & \multicolumn{1}{c|}{$r$ [l.u.]} & $a^{-1}$ [l.u.] & $r$ [l.u.] & \multicolumn{1}{c}{$P$ [l.u.]} \\
\hline
$NN\;(\1s0)$	&	$0.084_{(-42)(-35)}^{(+20)(+44)}$	&	$-2.4_{(-5.5)(-9.0)}^{(+8.4)(+8.3)}$	&	$0.053_{(-29)(-52)}^{(+33)(+43)}$	&	$15.4_{(-6.2)(-5.7)}^{(+6.5)(+20.8)}$	&	$803_{(-570)(-190)}^{(+46)(+510)}$	\\
$\Sigma N\;(\1s0)$	&	$0.079_{(-27)(-31)}^{(+25)(+14)}$	&	$-2.8_{(-5.3)(-4.0)}^{(+6.7)(+6.0)}$	&	-	&	-	&	\multicolumn{1}{c}{-}	\\
$\Sigma\Sigma\;(\1s0)$	&	$0.040_{(-13)(-14)}^{(+15)(+06)}$	&	$5.8_{(-2.9)(-0.9)}^{(+2.8)(+1.5)}$	&	$0.059_{(-28)(-18)}^{(+17)(+41)}$	&	$10.0_{(-2.4)(-4.1)}^{(+3.8)(+3.8)}$	&	$563_{(-330)(-260)}^{(+200)(+490)}$	\\
$\Xi\Sigma\;(\1s0)$	&	$0.061_{(-17)(-12)}^{(+16)(+06)}$	&	$2.4_{(-3.4)(-1.6)}^{(+3.6)(+1.8)}$	&	$0.062_{(-22)(-11)}^{(+28)(+21)}$	&	$10.6_{(-2.1)(-0.9)}^{(+2.5)(+1.8)}$	&	$469_{(-280)(-140)}^{(+310)(+210)}$	\\
$\Xi\Xi\;(\1s0)$	&	$0.058_{(-07)(-08)}^{(+07)(+07)}$	&	$4.6_{(-1.4)(-0.8)}^{(+0.8)(+1.5)}$	&	$0.075_{(-22)(-16)}^{(+16)(+19)}$	&	$10.9_{(-1.0)(-1.0)}^{(+0.9)(+1.0)}$	&	$538_{(-250)(-180)}^{(+190)(+200)}$	\\
$NN\;(\3s1)$	&	$0.063_{(-24)(-09)}^{(+18)(+10)}$	&	$0.5_{(-4.1)(-2.9)}^{(+5.5)(+2.4)}$	&	$0.082_{(-47)(-26)}^{(+42)(+18)}$	&	$8.0_{(-5.1)(-1.9)}^{(+5.0)(+4.8)}$	&	$812_{(-560)(-340)}^{(+570)(+300)}$	\\
$\Xi N\;(\3s1)$	&	$0.086_{(-10)(-13)}^{(+07)(+11)}$	&	$3.0_{(-0.9)(-1.6)}^{(+1.7)(+1.7)}$	&	$0.080_{(-21)(-22)}^{(+14)(+23)}$	&	$12.2_{(-3.0)(-4.5)}^{(+3.5)(+4.8)}$	&	$307_{(-190)(-170)}^{(+220)(+310)}$	\\
\end{tabular}
\end{ruledtabular}
\end{table}
\begin{figure}[t!]
\includegraphics[width=\textwidth]{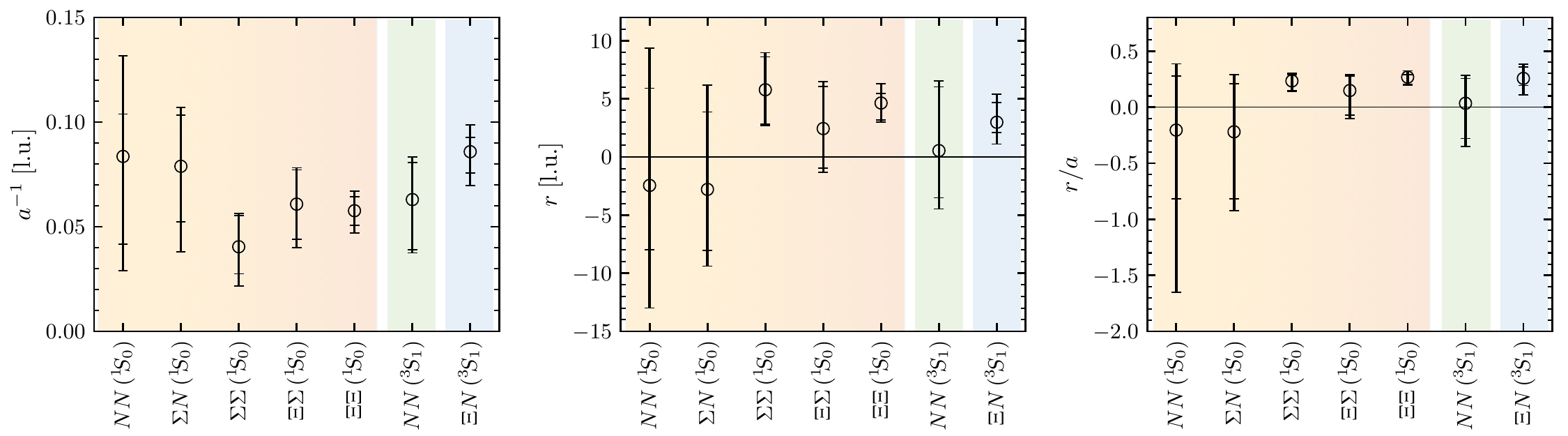}
\caption{Summary of the inverse scattering length $a^{-1}$ (left panel), effective range $r$ (middle panel), and ratio $r/a$ (right panel) determined from the two-parameter ERE fit for the two-baryon systems analyzed. The background color groups the channels by the $SU(3)$ irreps they would belong to if $SU(3)$ flavor symmetry were exact (orange for $\mathbf{27}$, green for $\overline{\mathbf{10}}$, and blue for $\mathbf{8}_a$). Quantities are expressed in lattice units.}
\label{fig:scattPar}
\end{figure}
\begin{table}[t!]
\caption{The values of the ratio of the effective range and scattering length, $r/a$, determined from the two-parameter ERE fit to $k^*\cot\delta$ values in each channel.}
\label{tab:scattPar-ratio}
\begin{ruledtabular}
\begin{tabular}{ccccccc}
\multicolumn{7}{c}{$r/a$} \\
$NN\;(\1s0)$	&	$\Sigma N\;(\1s0)$	&	$\Sigma\Sigma\;(\1s0)$	&	$\Xi\Sigma\;(\1s0)$	&	$\Xi\Xi\;(\1s0)$	&	$NN\;(\3s1)$	&	$\Xi N\;(\3s1)$	\\ \hline
$-0.2_{(-0.6)(-1.3)}^{(+0.5)(+0.3)}$	&	$-0.22_{(-60)(-37)}^{(+43)(+27)}$	&	$0.23_{(-08)(-04)}^{(+06)(+04)}$	&	$0.15_{(-22)(-11)}^{(+13)(+05)}$	&	$0.27_{(-07)(-03)}^{(+02)(+05)}$	&	$0.03_{(-31)(-23)}^{(+22)(+12)}$	&	$0.26_{(-06)(-13)}^{(+10)(+07)}$	\\
\end{tabular}
\end{ruledtabular}
\end{table}
Fits to $k^*\cot{\delta}$ as a function of $k^{*2}$ in various two-baryon channels are shown in Fig.~\ref{fig:ere-fit}, along with the correlation between the inverse scattering length and the effective range in each channel depicted in Fig.~\ref{fig:ere-parameters} using the 68\% and 95\% confidence regions of the parameters.
The areas in the parameter space that are prohibited by the constraints imposed by Eq.~\eqref{eq:QC} are also shown in Fig.~\ref{fig:ere-parameters}, highlighting the fact that the two-parameter ERE must cross the $\mathcal{Z}$-functions for each volume in the negative-$k^{*2}$ region. For fits including higher-order parameters, these constraints are more complicated and are not shown.
For the $\Sigma N \; (\3s1)$ and $\Xi \Xi \; (\3s1)$ channels, the ground-state energy is positively shifted, i.e., $\Delta E \gtrsim 0$, and only the values of $k^{*2}$ associated with the ground states are inside the range of validity of the ERE. As a result, no extraction of the ERE parameters is possible in these channels given the number of data points.
Results for the scattering parameters obtained using two- and three-parameter ERE fits in the other seven channels are summarized in Table~\ref{tab:scattPar}, and are shown in Fig.~\ref{fig:scattPar} for better comparison in the case of two-parameter fits.

The inverse scattering lengths extracted for all systems are compatible with each other (albeit within rather large uncertainties), signaling that there may exist enhanced flavor symmetries at this pion mass at low energies, a feature that will be thoroughly examined in Sec.~\ref{sec:pheno}.
The effective range in most systems is compatible with zero.\footnote{In Appendix~\ref{sec:appen-vs2015}, results for $NN$ channels are compared with the previous scattering parameters obtained in Ref.~\cite{Orginos:2015aya} using the same correlation functions, as well as with the predictions obtained from low-energy theorems in Ref.~\cite{Baru:2016evv}. Through a thorough investigation, the various tensions are discussed and resolved.}
Furthermore, the ratio $r/a$ can be used as an indicator of the naturalness of the interactions; for natural interactions, $r/a\sim 1$, while for unnatural interactions $r/a \ll 1$.
At the physical point, both $NN$ channels are unnatural and exhibit large scattering lengths, with $r/a$ being close to $0.1$ for the spin-singlet channel, and $0.3$ for the spin-triplet channel.
From Table~\ref{tab:scattPar-ratio}, the most constrained ratios are obtained for the $\Sigma \Sigma\; (\1s0)$, $\Xi\Xi\; (\1s0)$, and $\Xi N\; (\3s1)$ channels, for which $r/a\sim 0.2-0.3$, indicating unnatural interactions at low energies.
For other channels, the larger uncertainty in this ratio precludes drawing conclusions about naturalness.
Alternatively, naturalness can be assessed by considering the ratio of the binding momentum to the pion mass, as this quantity is better constrained in this study, see Table~\ref{tab:binding_momenta} in the next subsection. The values for $\kappa^{(\infty)}/m_{\pi}$ in each of the bound two-baryon channels are between $0.2$ and $0.4$, indicating that the range of interactions mediated by the pion exchange is not the only characteristic scale in the system, suggesting unnaturalness.
However, at larger-than-physical quark masses, pion exchange may not be the only significant contribution to the long-range component of the nuclear force, as is discussed in Ref.~\cite{Beane:2013br}.
For these reasons, both natural and unnatural interactions are considered in the next section when adopting a power-counting scheme in constraining the LECs of the EFT.\footnote{For a detailed discussion of naturalness in EFTs, see Ref.~\cite{vanKolck:2020plz}.}
\begin{table}[t]
\caption{The values of the parameters $\tilde{a}^{-1}, \tilde{r},\tilde{P}$ from a two- or three-parameter polynomial fit for two-baryon channels that exhibit smooth and monotonic behavior in $k^*\cot\delta$ as a function of $k^{*2}$ beyond the $t$-channel cut. Quantities are expressed in lattice units.}
\label{tab:scattPar-beyond}
\begin{ruledtabular}
\begin{tabular}{c|cr|ccr}
 & \multicolumn{2}{c|}{Two-parameter polynomial fit} & \multicolumn{3}{c}{Three-parameter polynomial fit} \\
 & $\tilde{a}^{-1}$ [l.u.] & \multicolumn{1}{c|}{$\tilde{r}$ [l.u.]} & $\tilde{a}^{-1}$ [l.u.] & $\tilde{r}$ [l.u.] & \multicolumn{1}{c}{$\tilde{P}$ [l.u.]} \\
\hline
$\Sigma\Sigma\;(\1s0)$	&	$0.038_{(-16)(-05)}^{(+12)(+09)}$	&	$6.2_{(-2.6)(-0.7)}^{(+2.7)(+0.8)}$	&	$0.044_{(-12)(-07)}^{(+08)(+11)}$	&	$10.9_{(-2.6)(-0.6)}^{(+1.3)(+2.5)}$	&	$331_{(-120)(-80)}^{(+100)(+98)}$	\\
$\Xi\Sigma\;(\1s0)$	&	$0.043_{(-10)(-05)}^{(+08)(+07)}$	&	$6.3_{(-1.1)(-1.3)}^{(+1.7)(+0.5)}$	&	$0.052_{(-11)(-07)}^{(+11)(+07)}$	&	$7.7_{(-2.4)(-1.0)}^{(+1.6)(+1.8)}$	&	$173_{(-46)(-37)}^{(+43)(+25)}$	\\
$\Xi\Xi\;(\1s0)$	&	$0.047_{(-07)(-03)}^{(+03)(+08)}$	&	$6.9_{(-0.3)(-0.9)}^{(+0.9)(+0.5)}$	&	$0.053_{(-06)(-04)}^{(+03)(+06)}$	&	$8.9_{(-0.9)(-1.0)}^{(+0.7)(+1.3)}$	&	$149_{(-23)(-28)}^{(+23)(+31)}$	\\
$NN\;(\3s1)$	&	$0.038_{(-16)(-07)}^{(+12)(+07)}$	&	$7.2_{(-1.9)(-1.3)}^{(+2.3)(+1.0)}$	&	$0.051_{(-13)(-08)}^{(+12)(+09)}$	&	$8.3_{(-3.1)(-2.4)}^{(+2.2)(+2.0)}$	&	$265_{(-66)(-72)}^{(+89)(+62)}$	\\
$\Sigma N\;(\3s1)$	&	$0.073_{(-20)(-21)}^{(+22)(+16)}$	&	$3.5_{(-1.2)(-0.8)}^{(+1.2)(+0.9)}$	&	$0.085_{(-39)(-19)}^{(+23)(+31)}$	&	$5.2_{(-5.6)(-3.2)}^{(+2.9)(+4.3)}$	&	$-8_{(-14)(-15)}^{(+27)(+16)}$	\\
$\Xi\Xi\;(\3s1)$	&	$0.20_{(-09)(-18)}^{(+17)(+14)}$	&	$-2.6_{(-2.9)(-6.1)}^{(+4.3)(+1.0)}$	&	$0.25_{(-14)(-13)}^{(+22)(+29)}$	&	$1_{(-15)(-10)}^{(+14)(+22)}$	&	$-19_{(-54)(-88)}^{(+87)(+62)}$	\\
$\Xi N\;(\3s1)$	&	$0.059_{(-01)(-05)}^{(+05)(+02)}$	&	$6.9_{(-0.3)(-0.3)}^{(+0.2)(+0.4)}$	&	$0.066_{(-04)(-03)}^{(+02)(+04)}$	&	$7.1_{(-0.3)(-0.5)}^{(+0.5)(+0.3)}$	&	$36_{(-11)(-04)}^{(+08)(+08)}$	\\
\end{tabular}
\end{ruledtabular}
\end{table}
\begin{figure}[t]
\includegraphics[width=\textwidth]{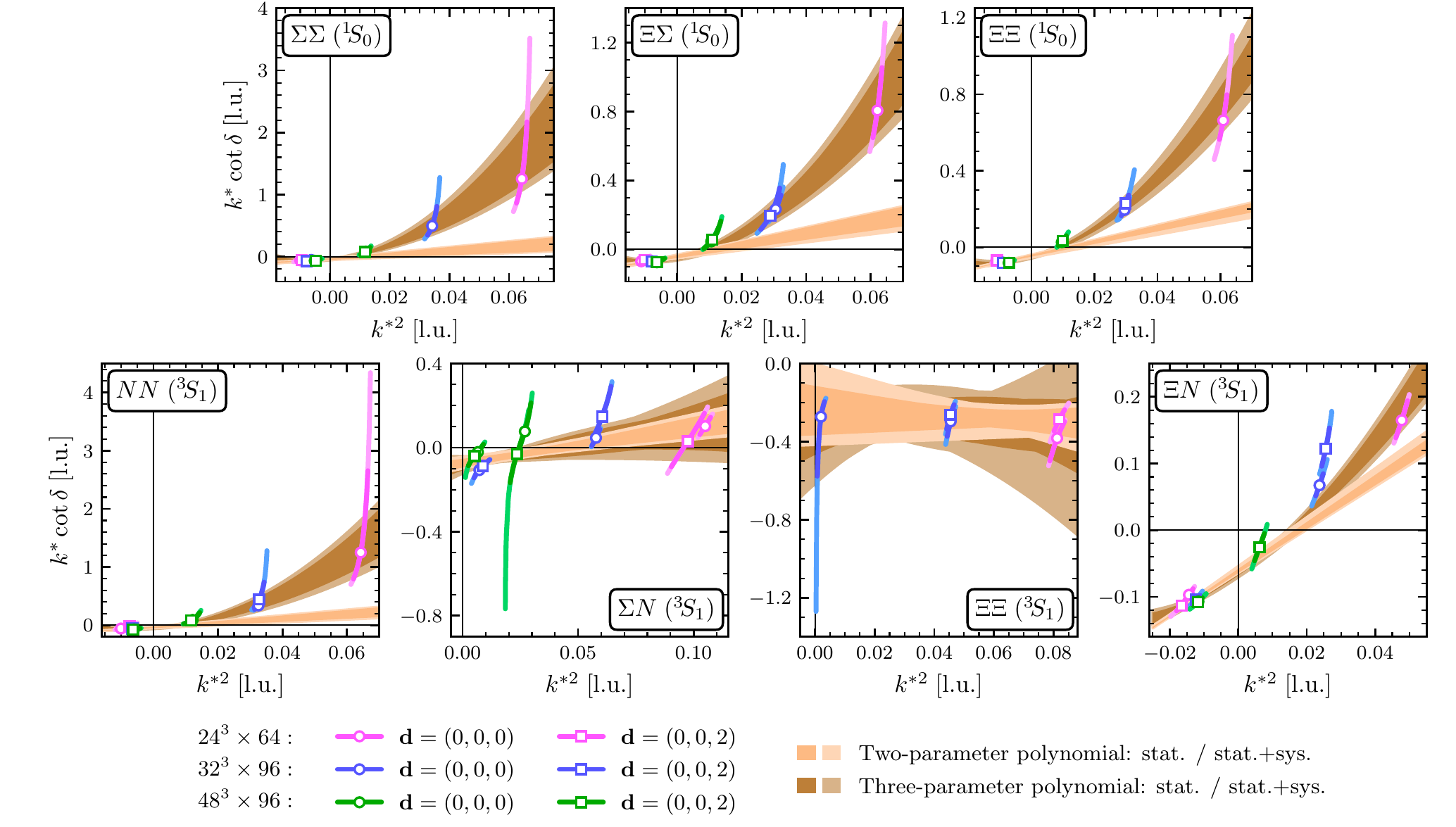}
\caption{$k^*\cot \delta$ values as a function of the c.m.\ momenta $k^{*2}$, along with the bands representing the two- and three-parameter polynomial fits for two-baryon systems under the assumption that there is a smooth and monotonic behavior in $k^*\cot\delta$ as a function of $k^{*2}$ beyond the $t$-channel cut. Quantities are expressed in lattice units.}
\label{fig:ere-fit-beyond}
\end{figure}
Although the ERE is only valid below the $t$-channel cut, one may still fit the $k^*\cot\delta$ values beyond this threshold using a similar polynomial form as the ERE in Eq.~\eqref{eq:ERE}.
To distinguish the ``model'' fit parameters from those obtained from the ERE, two- and three-parameter polynomials are characterized by two $\{\tilde{a}^{-1},\tilde{r}\}$ or three $\{\tilde{a}^{-1},\tilde{r},\tilde{P}\}$ parameters. Such forms are motivated by the fact that in most channels, $k^*\cot\delta$ values as a function of $k^{*2}$ exhibit smooth and monotonic behavior beyond the $t$-channel cut, as is seen in Fig.~\ref{fig:kcotdelta}. The only exceptions are the spin-singlet $NN$ and $\Sigma N$ channels, for which such a polynomial fit will not be performed.
The results of this fit, using the same strategy as described above for ERE fits, are shown in Table~\ref{tab:scattPar-beyond} and Fig.~\ref{fig:ere-fit-beyond}. In the next section, the EFTs and approximate symmetries of the interactions will be utilized to make predictions for the inverse scattering length in channels for which ERE fits could not be performed, i.e., $\Sigma N\;(\3s1)$ and $\Xi\Xi\;(\3s1)$ channels, and in those cases, the scattering length is found consistent with the $\tilde{a}^{-1}$ values obtained from this model analysis. It should be emphasized that such a polynomial fit beyond the $t$-channel cut is only one out of many applicable parametrizations of the amplitude, and a systematic uncertainty associated with multiple model choices and model-selection criteria needs to be assigned to reliably constrain the energy dependence of the amplitude at higher energies.\footnote{More precise LQCD results may be required to identify non-polynomial behavior in $k^{*2}$. This is analogous to the efforts to uniquely identify non-analytic terms in chiral expansions, such as in $\pi$-$\pi$ scattering, where very high-precision calculations are required to reveal the logarithmic dependence on $m_\pi$, see e.g., Ref.~\cite{Beane:2007xs}.}
%
%%%%%%%%%%%%%%%%%%%%%%%%%%%%%%%%%%%%%%%%%%%
\subsection{Binding energies \label{subsec:bind}}

A negative shift in the energy of two baryons in a finite volume compared with that of the non-interacting baryons may signal the presence of a bound state in the infinite-volume limit. However, to conclusively discern a bound state from a scattering state, a careful inspection of the volume dependence of the energies is required.
L\"uscher's quantization condition can be used to identify the volume dependence of bound-state energies.\footnote{Alternatively, LQCD eigenenergies in a finite volume can be matched to an EFT description of the system in the same volume to constrain the interactions. The constrained EFT can then be used to obtain the infinite-volume binding energy, see e.g., Ref.~\cite{Eliyahu:2019nkz}. This approach is more easily applicable to the multi-baryon sector, however it relies on the validity of the EFT that is used.}
Explicitly, the infinite-volume binding momentum $\kappa^{(\infty)}$ can be determined by expanding Eq.~\eqref{eq:QC} in the negative-$k^{*2}$ region~\cite{Davoudi:2011md}:
\begin{equation}
|k^*|=\kappa^{(\infty)}+\frac{Z^2}{L}\left[\sum_{\bm{m}}\frac{1}{|\hat{\gamma}\bm{m}|}e^{i2\pi \alpha \bm{m}\cdot\bm{d}}e^{-|\hat{\gamma}\bm{m}|\kappa^{(\infty)}L}\right] , 
\label{eq:kinf}
\end{equation}
where $Z^2$ is the residue of the scattering amplitude at the bound-state pole. In this study, the boost vectors are $\bm{d}=(0,0,0)$ and $\bm{d}=(0,0, 2)$, and the values of $\gamma$ deviate from one at the percent level.\footnote{The largest value of $\gamma$ is found in the $NN \, (\1s0)$ system with $L=24$, where $\gamma \sim 1.015$.}
Therefore, all systems considered are non-relativistic to a good approximation. Only the first few terms in the sum in Eq.~\eqref{eq:kinf}, corresponding to $|\bm{m}|\in\{0,1,\sqrt{2}\}$, are considered in the volume extrapolation performed below, with corrections that scale as $\mathcal{O}(e^{-2\kappa^{(\infty)}L})$. 

Alternatively, one can compute $\kappa^{(\infty)}$ by finding the pole location in the $s$-wave scattering amplitude:
\begin{equation}
\left. k^* \cot\delta \right|_{k^*=i \kappa^{(\infty)}}+\kappa^{(\infty)}=0\, .
\label{eq:kinf3}
\end{equation}
To obtain $\kappa^{(\infty)}$, the scattering amplitude has to first be constrained using L\"uscher's quantization condition as discussed in the previous subsection, and then be expressed in terms of an ERE expansion.
This approach, therefore, requires an intermediate step compared with the first method, but does not require a truncation of the sum in Eq.~\eqref{eq:kinf}.

Results for the infinite-volume binding momenta $\kappa^{(\infty)}$ are shown in Table~\ref{tab:binding_momenta}. The columns labeled as $\bm{d}=(0,0,0)$ and $\bm{d}=(0,0,2)$ correspond, respectively, to fitting separately the values of $k^*$ with no boost, or with boost $\bm{d}=(0,0,2)$, using Eq.~\eqref{eq:kinf}.
The column labeled as $\bm{d}=\{(0,0,0),(0,0,2)\}$ is the result of fitting both sets of $k^*$ values simultaneously, i.e., imposing the same value for $\kappa^{(\infty)}$ and $Z^2$ in both fits.
The last column shows the $\kappa^{(\infty)}$ values obtained using Eq.~\eqref{eq:kinf3}, with the parameters listed in Table~\ref{tab:scattPar} as obtained with a two-parameter ERE fit to $k^*\cot\delta$.
The results obtained with the different extractions of $\kappa^{\infty}$ are seen to be consistent with each other within uncertainties. The largest difference observed is in the $\Xi\Xi\;(\1s0)$ channel, with a difference between the volume-extrapolation and pole-location results of around $1.5\sigma$. The agreement between the two approaches suggests that the higher-order terms neglected in the sum in Eq.~\eqref{eq:kinf} are not significant.

\begin{table}[t!]
\caption{The infinite-volume binding momenta $\kappa^{(\infty)}$ for bound states obtained either by using the extrapolation in Eq.~\eqref{eq:kinf} or from the pole location of the scattering amplitude as in Eq.~\eqref{eq:kinf3}. Quantities are expressed in lattice units.}
\label{tab:binding_momenta}
\begin{ruledtabular}
\begin{tabular}{ccccc}
 & \multicolumn{4}{c}{$\kappa^{(\infty)}$ [l.u.]} \\
 & $\bm{d}=(0,0,0)$ & $\bm{d}=(0,0,2)$ & $\bm{d}=\{(0,0,0),(0,0,2)\}$ & $\left.-k^*\cot\delta\right|_{k^*=i\kappa^{(\infty)}}$ \\
\hline
$NN\;(\1s0)$	&	$0.077_{(-11)(-04)}^{(+08)(+06)}$	&	$0.072_{(-14)(-16)}^{(+10)(+08)}$	&	$0.075_{(-10)(-01)}^{(+05)(+06)}$	&	$0.076_{(-28)(-32)}^{(+06)(+12)}$	\\
$\Sigma N\;(\1s0)$	&	$0.073_{(-16)(-13)}^{(+13)(+05)}$	&	$0.083_{(-09)(-13)}^{(+09)(+06)}$	&	$0.080_{(-09)(-09)}^{(+08)(+02)}$	&	$0.072_{(-14)(-24)}^{(+12)(+09)}$	\\
$\Sigma\Sigma\;(\1s0)$	&	$0.068_{(-10)(-11)}^{(+08)(+08)}$	&	$0.072_{(-10)(-16)}^{(+11)(+07)}$	&	$0.069_{(-07)(-09)}^{(+07)(+06)}$	&	$0.047_{(-15)(-17)}^{(+15)(+07)}$	\\
$\Xi\Sigma\;(\1s0)$	&	$0.078_{(-09)(-09)}^{(+08)(+06)}$	&	$0.080_{(-08)(-11)}^{(+08)(+05)}$	&	$0.079_{(-05)(-07)}^{(+06)(+04)}$	&	$0.066_{(-14)(-14)}^{(+10)(+05)}$	\\
$\Xi\Xi\;(\1s0)$	&	$0.086_{(-05)(-06)}^{(+05)(+05)}$	&	$0.086_{(-05)(-09)}^{(+06)(+06)}$	&	$0.086_{(-03)(-05)}^{(+04)(+04)}$	&	$0.069_{(-08)(-09)}^{(+05)(+08)}$	\\
$NN\;(\3s1)$	&	$0.072_{(-11)(-08)}^{(+08)(+06)}$	&	$0.076_{(-09)(-08)}^{(+08)(+03)}$	&	$0.074_{(-07)(-05)}^{(+08)(+04)}$	&	$0.064_{(-20)(-08)}^{(+10)(+08)}$	\\
$\Xi N\;(\3s1)$	&	$0.108_{(-04)(-08)}^{(+04)(+06)}$	&	$0.106_{(-04)(-08)}^{(+05)(+06)}$	&	$0.107_{(-03)(-05)}^{(+03)(+05)}$	&	$0.101_{(-05)(-09)}^{(+05)(+06)}$	\\
\end{tabular}
\end{ruledtabular}
\end{table}
The binding energy, $B$, is defined in terms of the infinite-volume baryon masses and binding momenta as
\begin{equation}
B=M^{(\infty)}_1+M^{(\infty)}_2-\sqrt{M^{(\infty)2}_1-\kappa^{(\infty)2}}-\sqrt{M^{(\infty)2}_2-\kappa^{(\infty)2}}\, ,
\label{eq:bind_en}
\end{equation}
where $M_i^{(\infty)}$ is the infinite-volume mass of baryon $i$ obtained from Eq.~\eqref{eq:mass_extrap}. This quantity is computed for all systems that exhibit a negative c.m.\ momentum squared in the infinite-volume limit, i.e., those listed in Table~\ref{tab:binding_momenta}.
The binding energies in physical units are listed for these systems in Table~\ref{tab:binding_energy}. The binding energies of the two-nucleon systems computed here are consistent within $1 \sigma$ with the values published previously in Ref.~\cite{Orginos:2015aya} using the same LQCD correlation functions.
The same two-baryon systems studied here were also studied at $m_{\pi}\sim 806$ MeV in Ref.~\cite{Wagman:2017tmp}, and were found to be bound albeit with larger binding energies. While the results at $m_{\pi} \sim 806$ MeV were inconclusive regarding the presence of bound states in the ${\bf 10}$ irrep, the $\Sigma N \; (\3s1)$ and $\Xi \Xi \; (\3s1)$ systems are found to be unbound at this pion mass.
The results obtained in the present work can be combined with those of Ref.~\cite{Wagman:2017tmp} obtained at $m_{\pi} \sim 806$ MeV to perform a preliminary extrapolation of the binding energies to the physical pion mass.\footnote{The results in the literature for the binding energies of two-baryon systems obtained at larger-than-physical quark masses must be compared with the results of the current work with caution, as the use of different scale setting schemes makes a comparison in physical units meaningless, unless the physical limit of the quantities are taken. In the two-baryon sector, no continuum extrapolation has been performed in any of the previous studies.}
This enables a postdiction of binding energies in nature in cases where there are experimental data, and a prediction for the presence of bound states and their binding in cases where no experimental information is available.
\begin{table}[t!]
\caption{Binding energies for bound states in MeV. The values are obtained using $\kappa^{(\infty)}$ from the volume-extrapolation method with a combined fit to $\bm{d}=(0,0,0)$ and $\bm{d}=(0,0,2)$ data. The uncertainty from scale setting is an order of magnitude smaller than the statistical and systematic uncertainties quoted.}
\label{tab:binding_energy}
\begin{ruledtabular}
\begin{tabular}{ccccccc}
\multicolumn{7}{c}{$B$ [MeV]} \\
$NN\;(\1s0)$	&	$\Sigma N\;(\1s0)$	&	$\Sigma\Sigma\;(\1s0)$	&	$\Xi\Sigma\;(\1s0)$	&	$\Xi\Xi\;(\1s0)$	&	$NN\;(\3s1)$	&	$\Xi N\;(\3s1)$	\\ \hline
$13.1_{(-3.1)(-0.4)}^{(+2.0)(+2.3)}$	&	$14.3_{(-3.0)(-2.8)}^{(+3.1)(+0.9)}$	&	$10.2_{(-1.9)(-2.3)}^{(+2.1)(+2.0)}$	&	$12.8_{(-1.6)(-2.2)}^{(+2.1)(+1.6)}$	&	$14.9_{(-1.0)(-1.8)}^{(+1.5)(+1.4)}$	&	$12.7_{(-2.4)(-1.7)}^{(+2.4)(+1.5)}$	&	$25.3_{(-1.5)(-2.2)}^{(+1.5)(+2.2)}$	\\
\end{tabular}
\end{ruledtabular}
\end{table}

For systems with non-zero strangeness, experimental knowledge is notably limited in comparison to the nucleon-nucleon sector, and almost all phenomenological predictions are based on $SU(3)$ flavor-symmetry assumptions as discussed in the Introduction.
There is a significant body of work devoted to building phenomenological models of two-baryon interactions based on one-boson-exchange potentials, such as the Nijmegen hard-core~\cite{Nagels:1973rq,Nagels:1976xq,Nagels:1978sc}, soft-core (NSC)~\cite{Maessen:1989sx,Rijken:1998yy,Stoks:1999bz} and extended-soft-core (ESC)~\cite{Rijken:2005md,Rijken:2006ep,Rijken:2006kg,Rijken:2010zzb,Rijken:2013wxa,Nagels08II:2015,Nagels08III:2015,Nagels:2015lfa,doi:10.1063/1.5118371} models, as well as the Jülich~\cite{Holzenkamp:1989tq,Reuber:1993ip,Haidenbauer:2005zh} and Ehime~\cite{Tominaga:1998iy,Yamaguchi:2001ip} models. EFTs~\cite{Korpa:2001au,Polinder:2006zh,Polinder:2007mp,Haidenbauer:2009qn,Haidenbauer:2013oca,Haidenbauer:2014rna,Haidenbauer:2015zqb,Haidenbauer:2018gvg,Haidenbauer:2019boi,Li:2018tbt} and quark models~\cite{Fujiwara:1996qj,Fujiwara:2006yh,Valcarce:2010kz} have also been used to construct two-baryon potentials.
A short summary of the results in the literature for the relevant channels with non-zero strangeness is as follows:
\begin{itemize}
\item The $\1s0$ and $\3s1$ $\Sigma N$ channels do not exhibit bound states in any of the models listed above. The spin-singlet state behaves in a similar way to $NN\; (\1s0)$, and the interactions are slightly attractive, while those in the spin-triplet channel are found to be repulsive.
\item For the $\Xi N\;(\3s1)$ system, almost all the models find that the interactions are slightly attractive, but only a few exhibit a bound state.\footnote{Since the binding energies are not explicitly computed in these references and only the $s$-wave scattering parameters are reported, binding energies are computed here using Eqs.~\eqref{eq:kinf3} and \eqref{eq:bind_en}, assuming a two-parameter ERE for $k^*\cot\delta$. These are marked with the symbol $^{\ddagger}$.} %``ESC04d''~\cite{Rijken:2006kg} gives $B=0.002$ MeV$^{\ddagger}$,
Among the most recent results are ``ESC08a''~\cite{Rijken:2010zzb} which gives $B=0.9$ MeV$^{\ddagger}$, and ``ESC08c1''~\cite{Rijken:2013wxa} which gives $B=0.5$ MeV$^{\ddagger}$. There is one LQCD calculation of this system near the physical values of the quark masses performed by the HAL QCD collaboration~\cite{Sasaki:2019qnh} using a different method than the current work, and no bound state is observed.
\item The ``NSC97'' model~\cite{Stoks:1999bz} finds a bound state for the $\Sigma \Sigma \;(\1s0)$ channel, with binding energies ranging from $1.53$ to $3.17$ MeV. $\chi$EFT at NLO~\cite{Haidenbauer:2014rna} finds a binding energy between $0$ and $0.01$ MeV (no bound state is found with ESC or quark models in this channel).
\item The $\Xi \Sigma \;(\1s0)$ system is found to be bound in the ``NSC97'' model~\cite{Stoks:1999bz}, with a binding energy between $3.02$ and $16.5$ MeV, and by $\chi$EFT~\cite{Haidenbauer:2009qn,Haidenbauer:2014rna}, with a binding energy between $2.23$ and $6.18$ MeV at LO and $0.19$ and $0.58$ MeV at NLO. With the quark model ``fss2''~\cite{Fujiwara:2006yh}, although the interaction in this system is found to be attractive, no bound state is predicted (similar to the ``ESC08c1'' model~\cite{Rijken:2013wxa}).
\item Using one-boson-exchange potentials, with ``NSC97''~\cite{Stoks:1999bz} the $\Xi \Xi \;(\1s0)$ state is bound with a binding energy between $0.1$ and $15.8$ MeV, and with ``Ehime''~\cite{Yamaguchi:2001ip} between $0.23$ and $0.71$ MeV (no bound state is found with ``ESC08c1''~\cite{Rijken:2013wxa}). $\chi$EFT~\cite{Haidenbauer:2009qn,Haidenbauer:2014rna} also finds this state to be bound with a binding energy of $2.56-7.27$ MeV at LO and $0.40-1.00$ MeV at NLO. The quark model ``fss2''~\cite{Fujiwara:2006yh} does not find a bound state. In the $\Xi \Xi \;(\3s1)$ channel, no bound state is found with one-boson-exchange potentials, except for ``Ehime'' that finds a deeply bound state with a binding energy of $9-15$ MeV. ``fss2''~\cite{Fujiwara:2006yh} finds this channel to be repulsive.
\end{itemize}

The quark-mass dependences of multi-baryon spectra have not been studied extensively in the literature. For the octet-baryon masses, it was found that LQCD calculations performed with 2+1 dynamical fermions are consistent with a linear dependence on the pion mass at unphysical values of the quark masses, compared to the HB$\chi$PT prediction of quadratic dependence at LO~\cite{WalkerLoud:2008bp,WalkerLoud:2008pj,Walker-Loud:2014iea}.
Nonetheless, recent precision studies near the physical values of the quark masses appear to be more consistent with chiral predictions~\cite{Borsanyi:2020bpd}.
In the two-baryon sector the situation is more complicated. On the theoretical side, $\chi$EFT was used in Ref.~\cite{Haidenbauer:2011za} to extrapolate LQCD results to the physical point, assuming no dependence on the light quark masses for the LECs of the EFT (at a fixed order).
The same premise was taken in Ref.~\cite{Beane:2012ey} to determine the $I=3/2$ $\Sigma N$ interaction at LO, which was used to address the possible appearance of $\Sigma^-$ hyperons in dense nuclear matter.
In the absence of a conclusive form for the quark-mass extrapolation of two-baryon binding energies, two naive expressions with linear and quadratic $m_{\pi}$ dependence were used in Ref.~\cite{Beane:2011zpa} to extrapolate the binding energy of $H$-dibaryon to its physical value.
In Refs.~\cite{Shanahan:2011su, Thomas:2011cg, Shanahan:2013yta}, under the assumption that the $H$-dibaryon is a compact six valence-quark state (and not a two-baryon molecule), $\chi$EFT was used to extrapolate the binding energies, resulting in an unbound state.

\begin{figure}[t]
\includegraphics[width=\textwidth]{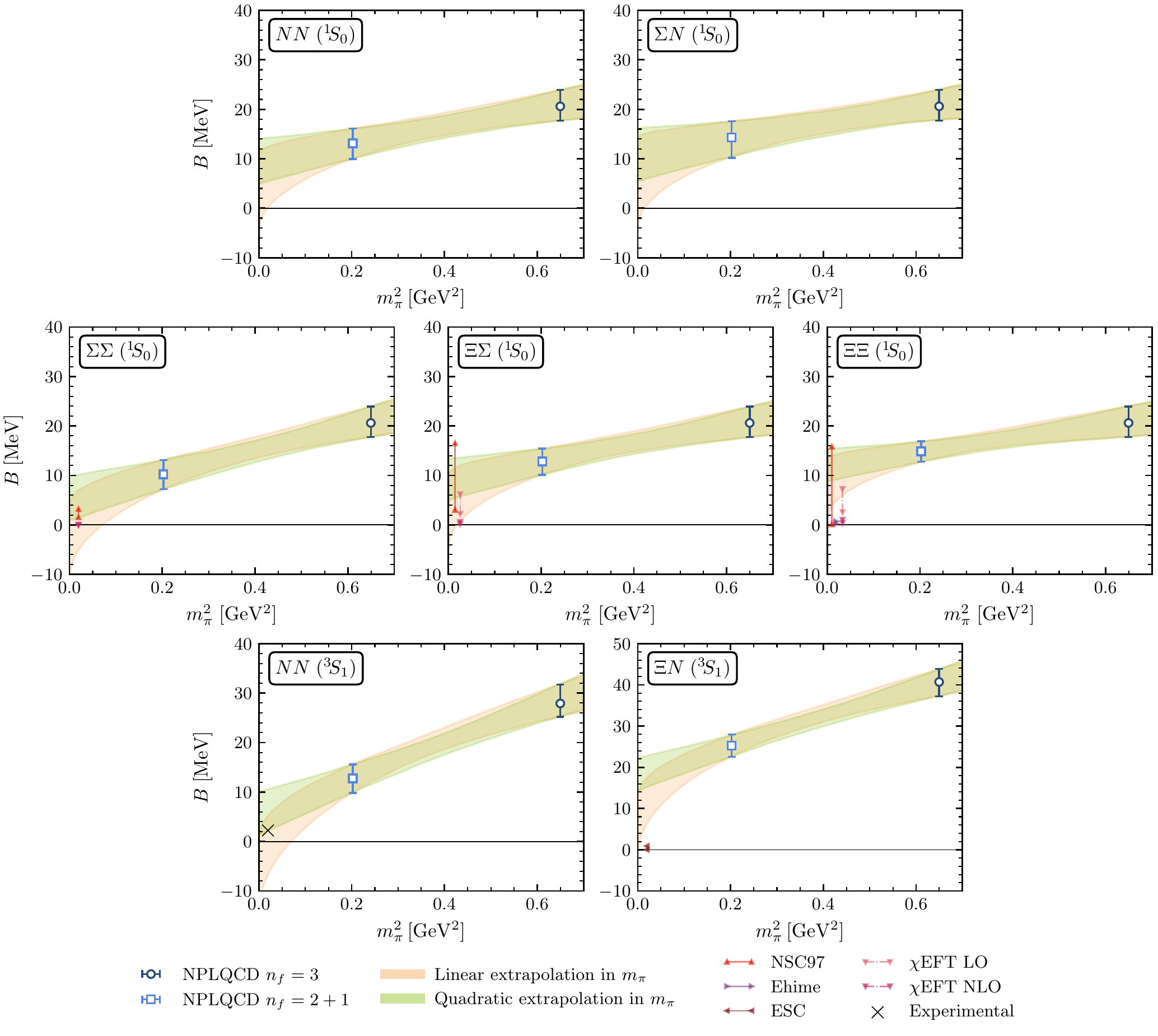}
\caption{Extrapolation of the binding energies of different two-baryon systems, using the results obtained in this work and those at $m_{\pi}\sim 806$ MeV from Ref.~\cite{Wagman:2017tmp}. For comparison, the results with values obtained using one-boson-exchange models or $\chi$EFTs are also shown (and where needed, are shifted slightly in the horizontal direction for clarity).}
\label{fig:extrap_bind}
\end{figure}

Two analytical forms with different $m_{\pi}$ dependence are used here to obtain the binding energies at the physical light-quark masses, using the results presented in Ref.~\cite{Wagman:2017tmp} at $m_\pi \sim 806$ MeV and those listed in Table~\ref{tab:binding_energy} for $m_\pi \sim 450$ MeV,
\begin{align}
B_{\rm lin}(m_{\pi})&=B^{(0)}_{\rm lin}+B^{(1)}_{\rm lin}\, m_{\pi}\, , \label{eq:extrapLin}\\
B_{\rm quad}(m_{\pi})&=B^{(0)}_{\rm quad}+B^{(1)}_{\rm quad} \, m^2_{\pi}\, ,\label{eq:extrapQuad}
\end{align}
where $B^{(0)}_{\rm lin}$, $B^{(1)}_{\rm lin}$, $B^{(0)}_{\rm quad}$, and $B^{(1)}_{\rm quad}$ are parameters to be constrained by fits to data. These fits are shown in Fig.~\ref{fig:extrap_bind}, along with the experimental value and predictions at the physical point. The binding energies extrapolated to the physical point, i.e., $B_{\rm lin}(m_{\pi}^{\rm phys})$ and $B_{\rm quad}(m_{\pi}^{\rm phys})$, are summarized in Table~\ref{tab:extrap_bind}.\footnote{Performing fits to dimensionless ratios of the binding energies to the baryon masses (to minimize the effects of non-zero lattice spacing) do not change the qualitative conclusions presented in the text.} 
It should be emphasized that given the lack of knowledge of the quark-mass dependence of binding energies, the preliminary extrapolations performed here are only to point out an emerging trend in the binding energies toward the physical point, and they do not provide conclusive predictions.

These extrapolations highlight some interesting features. The values obtained at the physical point are consistent with the experimental values for the $NN$ channels. The rest of the binding predictions are at the same level of precision as the phenomenological results.
The $\Xi \Xi \; (\1s0)$ and $\Xi N \; (\3s1)$ channels are more consistent with being bound than the other channels, using both extrapolation functions. Moreover, the $\Sigma N \; (\3s1)$ channel was found not to support a bound state in this study, a conclusion that is in agreement with phenomenological models.
The same conclusion holds for $\Xi \Xi \; (\3s1)$, noting that only in one model, namely ``Ehime'', a different conclusion is reached~\cite{Yamaguchi:2001ip}. The spread of results and some contradictory conclusions in the models motivate the need for LQCD studies of these states at near-physical values of the quark masses in the upcoming years. 
\begin{table}[t!]
\caption{Extrapolated binding energies at the physical quark masses for bound states in MeV using two different forms, linear and quadratic in $m_{\pi}$.}
\label{tab:extrap_bind}
\begin{ruledtabular}
\begin{tabular}{crrrrrrr}
&	\multicolumn{1}{c}{$NN\;(\1s0)$}	&	\multicolumn{1}{c}{$\Sigma N\;(\1s0)$}	&	\multicolumn{1}{c}{$\Sigma\Sigma\;(\1s0)$}	&	\multicolumn{1}{c}{$\Xi\Sigma\;(\1s0)$}	&	\multicolumn{1}{c}{$\Xi\Xi\;(\1s0)$}	&	\multicolumn{1}{c}{$NN\;(\3s1)$}	&	\multicolumn{1}{c}{$\Xi N\;(\3s1)$}	\\ \hline
$B_{\rm lin}(m_{\pi}^{\rm phys})$	&	$6.4_{(-6.5)}^{(+6.3)}$	&	$8.4_{(-6.6)}^{(+7.8)}$	&	$1.0_{(-6.1)}^{(+6.1)}$	&	$5.9_{(-5.8)}^{(+5.7)}$	&	$9.6_{(-4.7)}^{(+4.5)}$	&	$-0.9_{(-6.1)}^{(+6.1)}$	&	$11.7_{(-6.2)}^{(+5.4)}$	\\
$B_{\rm quad}(m_{\pi}^{\rm phys})$	&	$9.9_{(-4.5)}^{(+4.6)}$	&	$11.5_{(-4.8)}^{(+5.7)}$	&	$5.8_{(-4.3)}^{(+4.2)}$	&	$9.5_{(-4.0)}^{(+3.8)}$	&	$12.4_{(-3.1)}^{(+3.0)}$	&	$6.3_{(-4.4)}^{(+4.3)}$	&	$18.9_{(-4.1)}^{(+3.8)}$	\\
\end{tabular}
\end{ruledtabular}
\end{table}

\section{Effective low-energy interactions of two baryons \label{sec:pheno}}
%%%%%%%%%%%%%%%%%%%%%%%%%%%%%%%%%%%%%%%%%%%
\subsection{Leading and next-to-leading order interactions in the EFT \label{subsec:lecs}}
Even though $SU(3)$ flavor symmetry is explicitly broken in this study by the different values of the light- and strange-quark masses, it is still useful to classify the different two-(octet)baryon channels according to the $SU(3)$ irrep that they belong to.
In the spin-flavor decomposition of the product of two octet baryons with $J^P=\frac{1}{2}^+$, the 64 existing channels can be grouped into:
\begin{equation}
\mathbf{8}\otimes \mathbf{8} = \mathbf{27}\oplus \mathbf{8}_s \oplus \mathbf{1} \oplus \overline{\mathbf{10}} \oplus \mathbf{10} \oplus \mathbf{8}_a\, .
\end{equation}
The states belonging to the $\{\mathbf{27}, \mathbf{8}_s, \mathbf{1}\}$ irreps are symmetric with respect to the exchange of two baryons, and by the Pauli exclusion principle must have total spin $J=0$.
The $\{\overline{\mathbf{10}}, \mathbf{10}, \mathbf{8}_a \}$ irreps, with an antisymmetric flavor wavefunction, have $J=1$. Each of the systems studied in this work belongs to only one single irrep: all of the singlet states belong to the $\mathbf{27}$ irrep, $NN \; (\3s1)$ to $\overline{\mathbf{10}}$, the triplet states $\Sigma N$ and $\Xi \Xi$ to the $\mathbf{10}$ irrep, and $\Xi N \; (\3s1)$ to the $\mathbf{8}_a$ irrep.
However, since $m_u=m_d\neq m_s$, which explicitly breaks $SU(3)$ symmetry, mixing among the irreps will appear. Note that the structure of the LQCD interpolating operators used in this study, i.e., single-point quark-level wavefunctions at the source, does not allow accessing channels in the $\mathbf{8}_s$ irrep.
Moreover, the state in the $\mathbf{1}$ irrep is a coupled flavor channel, which is excluded from this study given that a large number of kinematic inputs are required to constrain the corresponding coupled-channel scattering amplitudes.\footnote{The ground state of the flavor channels belonging to the $\mathbf{1}$ irrep has been determined in previous LQCD studies to be bound at larger-than-physical values of the quark masses, corresponding to the long-sought-for $H$-dibaryon state, see Refs.~\cite{Beane:2009py, Beane:2010hg, Inoue:2010es, Beane:2011zpa, Francis:2018qch}.}

The Lagrangian for the low-energy interactions of two octet baryons was first constructed in Ref.~\cite{Savage:1995kv} using the HB$\chi$EFT formalism, and consists of two-baryon contact operators at LO.
These interactions have also been studied in chiral perturbation theory ($\chi$PT) in Refs.~\cite{Polinder:2006zh,Haidenbauer:2007ra}, where in addition to the momentum-independent operators at LO, the pseudoscalar-meson exchanges are included in the interacting potential. At LO, all terms in both HB$\chi$EFT and $\chi$PT are $SU(3)$ symmetric.
At NLO, there are two types of contributions: the $SU(3)$-symmetric interactions, obtained by the addition of derivative terms to the LO Lagrangian, and the $SU(3)$ symmetry-breaking interactions, denoted by $\cancel{SU(3)}$ in the following, that arise from the inclusion of the quark-mass matrix. The NLO extension of the two-baryon potential within $\chi$PT was first presented in Refs.~\cite{Petschauer:2013uua,Haidenbauer:2013oca} and includes interactions in higher partial waves.

In this paper, two-baryon systems are analyzed at low energies; therefore only $s$-wave interactions are considered. The LO Lagrangian of Ref.~\cite{Savage:1995kv} is used, and the NLO contributions are formed to follow the organization of the LO terms.
In other words, the same spin-flavor operator structure is preserved in the NLO Lagrangian, up to the inclusion of derivative operators and the quark-mass matrix. The EFT considered is therefore a pionless EFT~\cite{vanKolck:1998bw,Chen:1999tn} in the hypernuclear sector.
The LO coefficients are known as Savage-Wise coefficients in the literature. This organization is different from that of Petschauer and Kaiser in Ref.~\cite{Petschauer:2013uua}, and while the notation used here to label the NLO LECs is the same as in Ref.~\cite{Petschauer:2013uua}, their meaning is different. The differences between the two organizations and the relations between both sets of $\cancel{SU(3)}$ coefficients are presented in Appendix~\ref{sec:appen-EFT}. 
The full pionless EFT Lagrangian, up to NLO, is written as
\begin{equation}
\mathcal{L}_{BB}=\, \mathcal{L}^{(0),\, SU(3)}_{BB} + \mathcal{L}^{(2),\, SU(3)}_{BB} + \mathcal{L}^{(2),\, \cancel{SU(3)}}_{BB}\, ,
\label{eq:LEFT}
\end{equation}
with
\begin{align}
\mathcal{L}^{(0),\, SU(3)}_{BB}=&-c_1 \text{Tr}(B^{\dag}_i B_i B^{\dag}_j B_j)-c_2 \text{Tr}(B^{\dag}_i B_j B^{\dag}_j B_i)-c_3 \text{Tr}(B^{\dag}_i B^{\dag}_j B_i B_j)\nonumber \\
&-c_4 \text{Tr}(B^{\dag}_i B^{\dag}_j B_j B_i)-c_5\text{Tr}(B^{\dag}_i B_i )\text{Tr}(B^{\dag}_j B_j)-c_6\text{Tr}(B^{\dag}_i B_j )\text{Tr}(B^{\dag}_j B_i)\, , \label{eq:LagLO}\\
\mathcal{L}^{(2),\, SU(3)}_{BB} = &-\tilde{c}_{1} \text{Tr}( B^{\dag}_i \nabla^2 B_i B^{\dag}_j B_j+\text{h.c.})
-\tilde{c}_{2}\text{Tr}(B^{\dag}_i \nabla^2 B_j B^{\dag}_j B_i+\text{h.c.}) \nonumber\\
&-\tilde{c}_{3}\text{Tr}(B^{\dag}_i B^{\dag}_j \nabla^2B_i B_j+\text{h.c.})
-\tilde{c}_{4}\text{Tr}(B^{\dag}_i B^{\dag}_j \nabla^2 B_j B_i+\text{h.c.})\nonumber\\
&-\tilde{c}_{5}[\text{Tr}(B^{\dag}_i \nabla^2 B_i)\text{Tr}(B^{\dag}_j B_j)+\text{h.c.}]
-\tilde{c}_{6}[\text{Tr}(B^{\dag}_i \nabla^2 B_j)\text{Tr}(B^{\dag}_j B_i)+\text{h.c.}]\, , \label{eq:LagNLO1}\\
\mathcal{L}^{(2),\, \cancel{SU(3)}}_{BB} = &-c_1^{\chi}\text{Tr}(B^{\dag}_i \chi B_i B^{\dag}_j B_j)-c_2^{\chi}\text{Tr}(B^{\dag}_i \chi B_j B^{\dag}_j B_i)-c_3^{\chi}\text{Tr}(B^{\dag}_i B_i \chi B^{\dag}_j B_j) \nonumber\\
&-c_4^{\chi}\text{Tr}(B^{\dag}_i B_j \chi B^{\dag}_j B_i)-c_5^{\chi}\text{Tr}(B^{\dag}_i \chi B^{\dag}_j B_i B_j +\text{h.c.})-c_6^{\chi}\text{Tr}(B^{\dag}_i \chi B^{\dag}_j B_j B_i+\text{h.c.}) \nonumber\\
&-c_7^{\chi}\text{Tr}(B^{\dag}_i B^{\dag}_j \chi B_i B_j )-c_8^{\chi}\text{Tr}(B^{\dag}_i B^{\dag}_j \chi B_j B_i)-c_9^{\chi}\text{Tr}(B^{\dag}_i B^{\dag}_j B_i B_j \chi) \nonumber\\
&-c_{10}^{\chi}\text{Tr}(B^{\dag}_i B^{\dag}_j B_j B_i \chi )-c_{11}^{\chi}\text{Tr}(B^{\dag}_i \chi B_i ) \text{Tr}(B^{\dag}_j B_j )-c_{12}^{\chi}\text{Tr}(B^{\dag}_i \chi B_j) \text{Tr}( B^{\dag}_j B_i)\, , 
\label{eq:LagNLO2}
\end{align}
where only terms that contribute to $s$-wave interactions are included in the NLO Lagrangian $\mathcal{L}^{(2),\, SU(3)}_{BB}$. The indices $i$ and $j$ denote spin indices, $B$ is the octet-baryon flavor matrix,
\begin{equation}
B=\begin{bmatrix}
\frac{\Sigma^0}{\sqrt{2}}+\frac{\Lambda}{\sqrt{6}} & \Sigma^+ & p \\
\Sigma^- & -\frac{\Sigma^0}{\sqrt{2}}+\frac{\Lambda}{\sqrt{6}} & n \\
\Xi^- & \Xi^0 & -\sqrt{\frac{2}{3}}\Lambda
\end{bmatrix},
\end{equation}
and $\chi$ is the quark-mass matrix, which can be written in terms of the meson masses using the Gell-Mann–Oakes–Renner relation~\cite{GellMann:1968rz}:
\begin{equation}
\chi=2B_0\begin{bmatrix}
m_{u} & 0 & 0 \\
0 & m_{d} & 0 \\
0 & 0 & m_s
\end{bmatrix}\propto \begin{bmatrix}
m^2_{\pi} & 0 & 0 \\
0 & m^2_{\pi} & 0 \\
0 & 0 & 2m^2_{K}-m^2_{\pi}
\end{bmatrix},
\end{equation}
where the constant $B_0$ is proportional to the quark condensate. 

In order to constrain the values of the LECs $c_i$, $\tilde{c}_i$, and $c^{\chi}_i$, the LO and NLO EREs of the inverse scattering amplitudes in the $s$-wave can be used.
It is known that if the interactions between octet baryons are unnatural, that is $r/a \ll 1$, a better justified power-counting scheme in the EFT is the KSW-vK (Kaplan, Savage and Wise~\cite{Kaplan:1998tg, Kaplan:1998we} and van Kolck~\cite{vanKolck:1998bw}) scheme, where at LO in the scattering amplitude, the contributions from LO momentum-independent operators are summed to all orders.
With natural interactions, a power-counting scheme based on naive dimensional analysis is used and the expansion of the amplitude remains perturbative in the interaction strength, including for the LO interaction.
As mentioned in Sec.~\ref{subsec:luscher}, given the large uncertainties in the scattering parameters (in particular in the effective range), the ratio $r/a$ shown in Table~\ref{tab:scattPar-ratio} is not well constrained, and does not conclusively prove unnaturalness in all channels.
Since in at least two channels the interactions seem unnatural, in the following both the natural and the unnatural cases will be considered in expressing relations between LECs and the scattering parameters.
These relations for each two-baryon channel can be separated into those that are momentum-independent, with contributions from LO and NLO $\cancel{SU(3)}$ terms in the Lagrangian, and momentum-dependent, with only contributions from NLO $SU(3)$ terms:
\begin{align}
\left[-\frac{1}{a_{\scriptscriptstyle B_1B_2}}+\mu\right]^{-1}&= \frac{\overline{M}_{\scriptscriptstyle B_1B_2}}{2\pi}(c^{(\text{irrep})}+\bm{c}^{\chi}_{\scriptscriptstyle B_1B_2})\, , \label{eq:scattparam1}\\
\frac{r_{\scriptscriptstyle B_1B_2}}{2}\left[-\frac{1}{a_{\scriptscriptstyle B_1B_2}}+\mu\right]^{-2}&=\frac{\overline{M}_{\scriptscriptstyle B_1B_2}}{2\pi}\tilde{c}^{(\text{irrep})} \, , \label{eq:scattparam2}
\end{align}
where $c^{(\text{irrep})}$ stands for the appropriate linear combinations of the $c_i$ LECs defined in the Lagrangian in Eq.~\eqref{eq:LEFT}. These relations are given in Table~\ref{tab:LECtab} for each two baryon channel consisting of baryons $B_1$ and $B_2$, where the LECs corresponding to given $SU(3)$ irreps in this table are related to the $c_i$ LECs by:\footnote{While the relations for the $\mathbf{1}$ and $\mathbf{8}_s$ irreps are not used here, they will be needed in Sec.~\ref{subsec:su6} in connection to the $SU(6)$ spin-flavor symmetry relations.}
\begin{align}
c^{(27)}&= 2(c_1-c_2+c_5-c_6)\, ,& c^{(\overline{10})}&= 2(c_1+c_2+c_5+c_6)\, ,\nonumber\\
c^{(8_s)}&= \frac{1}{3}(-4c_1+4c_2-5c_3+5c_4+6c_5-6c_6),& c^{(10)}&= 2(-c_1-c_2+c_5+c_6)\, ,\nonumber\\
c^{(1)}&= \frac{2}{3}(-c_1+c_2-8c_3+8c_4+3c_5-3c_6)\, ,& c^{(8_a)}&= 3c_3+3c_4+2c_5+2c_6\, . \label{eq:SU3eq}
\end{align}
The same relations hold for $\tilde{c}^{(\text{irrep})}$, replacing $c_i$ with $\tilde{c}_i$. Similarly, $\bm{c}^{\chi}_{\scriptscriptstyle B_1B_2}\equiv c^{\chi}_{\scriptscriptstyle B_1B_2}(m^2_K-m^2_{\pi})$ and $c^{\chi}_{\scriptscriptstyle B_1B_2}$ are linear combinations of the $c_i^\chi$ LECs as given in Table~\ref{tab:LECtab}.
The variables $a_{\scriptscriptstyle B_1B_2}$ and $r_{\scriptscriptstyle B_1B_2}$ are the scattering length and effective range of the channel $B_1 B_2$, and $\overline{M}_{\scriptscriptstyle B_1B_2}$ is the reduced mass of that system.
The renormalization scale $\mu$ depends on the naturalness of the interactions. For the natural case $\mu=0$, and Eqs.~\eqref{eq:scattparam1} and~\eqref{eq:scattparam2} correspond to a tree-level expansion of the scattering amplitude. For the unnatural case, the expansion does not converge for momenta larger than $a^{-1}$, and in the KSW-vK scheme $\mu$ is introduced as a renormalization scale for the $s$-channel two-baryon loops appearing in the all-orders expansion of the amplitude with LO interactions.
Since a pionless EFT is used, a convenient choice is $\mu=m_{\pi}$ (where $m_{\pi}\sim 450$ MeV is the mass of the pion obtained with the quark masses used in the LQCD study).
\begin{table}[t!]
\caption{The LECs of the LO and NLO pionless EFT that contribute to the scattering amplitude of the various two-baryon channel. The first three columns are total angular momentum ($J$), strangeness ($S$), and isospin ($I$).}
\label{tab:LECtab}
\begin{ruledtabular}
\begin{tabular}{ccccccc}
$J$ & $S$ & $I$ & Channel & $SU(3)$ LO & $SU(3)$ NLO & $\cancel{SU(3)}$ NLO \\ \hline
$0$ & $\phantom{-}0$ & $1$ & $NN$ & $c^{(27)}$ & $\tilde{c}^{(27)}$ & $4(c_3^{\chi}-c_4^{\chi})$\\
 & $-1$ & $\frac{3}{2}$ & $\Sigma N$ & $c^{(27)}$ & $\tilde{c}^{(27)}$ & $2(c_3^{\chi}-c_4^{\chi})$\\ 
 & $-2$ & $2$ & $\Sigma\Sigma$ & $c^{(27)}$ & $\tilde{c}^{(27)}$ & $0$\\ 
 & $-3$ & $\frac{3}{2}$ & $\Xi \Sigma$ & $c^{(27)}$ & $\tilde{c}^{(27)}$ & $2(c_1^{\chi}-c_2^{\chi}+c_{11}^{\chi}-c_{12}^{\chi})$\\ 
 & $-4$ & $1$ & $\Xi \Xi$ & $c^{(27)}$ & $\tilde{c}^{(27)}$ & $4(c_1^{\chi}-c_2^{\chi}+c_{11}^{\chi}-c_{12}^{\chi})$\\ 
$1$ & $\phantom{-}0$ & $0$ & $NN$ & $c^{(\overline{10})}$ & $\tilde{c}^{(\overline{10})}$ & $4(c_3^{\chi}+c_4^{\chi})$\\ 
 & $-1$ & $\frac{3}{2}$ & $\Sigma N$ & $c^{(10)}$ & $\tilde{c}^{(10)}$ & $-2(c_3^{\chi}+c_4^{\chi})$\\ 
 & $-4$ & $0$ & $\Xi \Xi$ & $c^{(10)}$ & $\tilde{c}^{(10)}$ & $-4(c_1^{\chi}+c_2^{\chi}-c_{11}^{\chi}-c_{12}^{\chi})$\\ 
 & $-2$ & $0$ & $\Xi N$ & $c^{(8_a)}$ & $\tilde{c}^{(8_a)}$ & $2(2c_5^{\chi}+2c_6^{\chi}+2c_7^{\chi}+2c_8^{\chi}+2c_9^{\chi}+2c_{10}^{\chi}+c_{11}^{\chi}+c_{12}^{\chi})$\\ 
\end{tabular}
\end{ruledtabular} 
\end{table}

\begin{figure}[t!]
\includegraphics[width=0.9\textwidth]{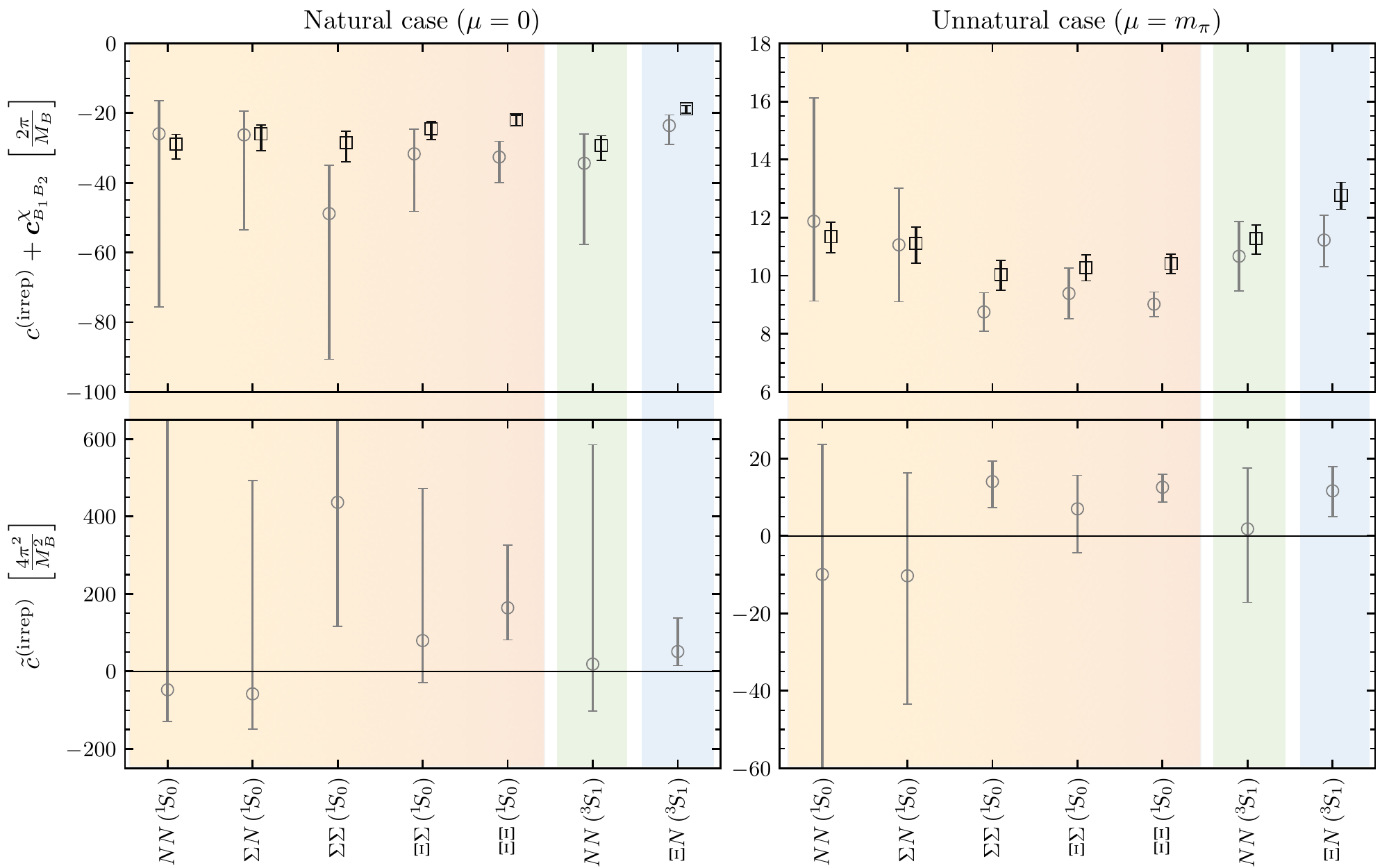}
\caption{LECs obtained by solving Eqs.~\eqref{eq:scattparam1} (upper panels) and~\eqref{eq:scattparam2} (lower panels) under the assumption of natural (left panels) and unnatural (right panels) interactions. The LECs of momentum-independent operators are in units of $[\frac{2\pi}{M_{B}}]$ and those of the momentum-dependent operators are in units of $[\frac{4\pi^2}{M^2_{B}}]$, where $M_B$ is the centroid of the octet-baryon masses. The gray-circle markers denote quantities that are extracted using the ERE parameters (method I), while black-square markers are those obtained from scattering lengths that are computed from binding momenta (method II).}
\label{fig:EFT_LONLO}
\end{figure}

Two sets of inputs can be used to constrain the numerical values for the LECs: 1) the scattering parameters $\{a^{-1},r\}$ obtained from two-parameter ERE fits in Sec.~\ref{subsec:luscher}, tabulated in Table~\ref{tab:scattPar}, can be used to compute LECs of both momentum-independent and momentum-dependent operators (method~I), and 2) the binding momenta from Sec.~\ref{subsec:bind} can be used to compute the corresponding scattering length, related at LO by \mbox{$-a^{-1}+\kappa^{(\infty)}=0$}, and this single parameter can be used to constrain the LECs of momentum-independent operators (method~II). 
This second method is motivated by the fact that $\kappa^{(\infty)}$ is extracted with higher precision than the parameters from the ERE fits, therefore enabling tighter constraints on the LECs of momentum-independent operators. The results for both types of LECs are presented in Table~\ref{tab:LECtab1}, and are depicted in Fig.~\ref{fig:EFT_LONLO}.
Results are presented in units of $2\pi/M_{B}$ for the momentum-independent operators and $4\pi^2/M^2_{B}$ for the momentum-dependent operators, where $M_B$ is the centroid of the octet-baryon masses, $M_B=\tfrac{1}{4}M_N+\tfrac{1}{8}M_{\Lambda}+\tfrac{3}{8}M_{\Sigma}+\tfrac{1}{4}M_{\Xi}=0.78583(23)(30)$ l.u.
\begin{table}[b!]
\caption{LECs of the momentum-independent and momentum-dependent operators as they appear in Table~\ref{tab:LECtab} for the two-baryon channels, obtained by solving Eq.~\eqref{eq:scattparam1} in units of $[\frac{2\pi}{M_{B}}]$ for the momentum-independent operators, and Eq.~\eqref{eq:scattparam2} in units of $[\frac{4\pi^2}{M^2_{B}}]$ for the momentum-dependent operators, where $M_B$ is the centroid of the octet-baryon masses. $\tilde{c}^{(\text{irrep})}$ are only determined using method I.}
\label{tab:LECtab1}
\begin{ruledtabular}
\begin{tabular}{cccccccccc}
LECs	&	$\mu$	&	Method	&	$NN\;(\1s0)$	&	$\Sigma N\;(\1s0)$	&	$\Sigma\Sigma\;(\1s0)$	&	$\Xi\Sigma\;(\1s0)$	&	$\Xi\Xi\;(\1s0)$	&	$NN\;(\3s1)$	&	$\Xi N\;(\3s1)$	\\\hline
\multirow{4}{*}{$c^{(\text{irrep})}+\bm{c}^{\chi}_{\scriptscriptstyle B_1B_2}$}	&	\multirow{2}{*}{$0$}	&	I	&	$-26_{(-50)}^{(+9)}$	&	$-26_{(-27)}^{(+7)}$	&	$-49_{(-42)}^{(+14)}$	&	$-32_{(-17)}^{(+7)}$	&	$-33_{(-7)}^{(+5)}$	&	$-34_{(-23)}^{(+8)}$	&	$-24_{(-5)}^{(+3)}$	\\
	&		&	II	&	$-29_{(-4)}^{(+3)}$	&	$-26_{(-5)}^{(+3)}$	&	$-28_{(-5)}^{(+3)}$	&	$-25_{(-3)}^{(+2)}$	&	$-22_{(-2)}^{(+1)}$	&	$-29_{(-4)}^{(+3)}$	&	$-19_{(-1)}^{(+1)}$	\\
	&	\multirow{2}{*}{$m_{\pi}$}	&	I	&	$11.9_{(-2.7)}^{(+4.2)}$	&	$11.1_{(-2.0)}^{(+2.0)}$	&	$8.8_{(-0.7)}^{(+0.7)}$	&	$9.4_{(-0.9)}^{(+0.9)}$	&	$9.0_{(-0.4)}^{(+0.4)}$	&	$10.7_{(-1.2)}^{(+1.2)}$	&	$11.2_{(-0.9)}^{(+0.9)}$	\\
	&		&	II	&	$11.3_{(-0.5)}^{(+0.5)}$	&	$11.1_{(-0.7)}^{(+0.6)}$	&	$10.0_{(-0.5)}^{(+0.5)}$	&	$10.3_{(-0.5)}^{(+0.4)}$	&	$10.4_{(-0.3)}^{(+0.3)}$	&	$11.3_{(-0.5)}^{(+0.5)}$	&	$12.8_{(-0.5)}^{(+0.5)}$	\\ \hline
\multirow{2}{*}{$\tilde{c}^{(\text{irrep})}$}	&	$0$	&	I	&	$-47_{(-82)}^{(+1600)}$	&	$-58_{(-91)}^{(+550)}$	&	$437_{(-320)}^{(+1800)}$	&	$80_{(-110)}^{(+390)}$	&	$164_{(-83)}^{(+160)}$	&	$19_{(-120)}^{(+570)}$	&	$51_{(-36)}^{(+86)}$	\\
	&	$m_{\pi}$	&	I	&	$-10_{(-84)}^{(+34)}$	&	$-10_{(-33)}^{(+27)}$	&	$14_{(-7)}^{(+5)}$	&	$7_{(-11)}^{(+9)}$	&	$13_{(-4)}^{(+3)}$	&	$2_{(-19)}^{(+16)}$	&	$12_{(-7)}^{(+6)}$	\\
\end{tabular}
\end{ruledtabular}
\end{table}

As can be seen from the values of the LECs that are obtained, the NLO $SU(3)$ coefficients have large uncertainties, and are mostly consistent with zero, because the effective ranges used to constrain them have rather large uncertainties.
Another feature of the results is that assuming the interactions to be unnatural leads to better-constrained parameters in general, as a non-zero scale $\mu$ in the left-hand side of Eqs.~\eqref{eq:scattparam1} and \eqref{eq:scattparam2} reduces the effect of uncertainties on the scattering lengths (this was also observed in Ref.~\cite{Wagman:2017tmp} for systems at $m_{\pi} \sim 806$ MeV).
Furthermore, as is expected, the values obtained with method II have smaller uncertainties than the ones obtained from method I, given the more precise scattering lengths, although the method is limited to LO predictions.
Another anticipated feature is that in the cases where the effective range is resolved from zero within uncertainties (e.g., in the $\Xi\Xi\; (\1s0)$ channel), the values from method II are slightly different from those obtained from method I, indicating the non-negligible effect of the NLO effective-range contributions that are neglected with this method.

\begin{table}[b!]
\caption{The values of the momentum-independent $SU(3)$ coefficient $c^{(27)}$ and specific linear combinations of the $\cancel{SU(3)}$ coefficients $\bm{c}^{\chi}_i$. Quantities are expressed in units of $[\frac{2\pi}{M_{B}}]$, where $M_B$ is the centroid of the octet-baryon masses.}
\label{tab:LECtab2}
\begin{ruledtabular}
\begin{tabular}{ccccccc}
$\mu$	&	Method	&	$c^{(27)}$ $\{NN, \Sigma N\}$	&	$c^{(27)}$ $\{\Sigma\Sigma\}$	&	$c^{(27)}$ $\{\Xi\Sigma, \Xi\Xi\}$	&	$\bm{c}^{\chi}_3-\bm{c}^{\chi}_4$	&	$\bm{c}^{\chi}_1-\bm{c}^{\chi}_2+\bm{c}^{\chi}_{11}-\bm{c}^{\chi}_{12}$	\\ \hline
\multirow{2}{*}{$0$}	&	I	&	$-27_{(-62)}^{(+58)}$	&	$-49_{(-42)}^{(+14)}$	&	$-31_{(-36)}^{(+21)}$	&	$0_{(-26)}^{(+18)}$	&	$0_{(-7)}^{(+10)}$	\\
	&	II	&	$-23_{(-12)}^{(+9)}$	&	$-28_{(-5)}^{(+3)}$	&	$-27_{(-7)}^{(+6)}$	&	$-1_{(-3)}^{(+4)}$	&	$1_{(-2)}^{(+2)}$	\\
\multirow{2}{*}{$m_{\pi}$}	&	I	&	$10.3_{(-7.7)}^{(+6.3)}$	&	$8.8_{(-0.7)}^{(+0.7)}$	&	$9.8_{(-2.1)}^{(+2.1)}$	&	$0.4_{(-2.2)}^{(+2.9)}$	&	$-0.2_{(-0.6)}^{(+0.6)}$	\\
	&	II	&	$10.9_{(-1.9)}^{(+1.6)}$	&	$10.0_{(-0.5)}^{(+0.5)}$	&	$10.1_{(-1.2)}^{(+1.2)}$	&	$0.1_{(-0.5)}^{(+0.6)}$	&	$0.1_{(-0.4)}^{(+0.4)}$	\\
\end{tabular}
\end{ruledtabular}
\end{table}
\begin{figure}[tb]
\includegraphics[width=0.8\textwidth]{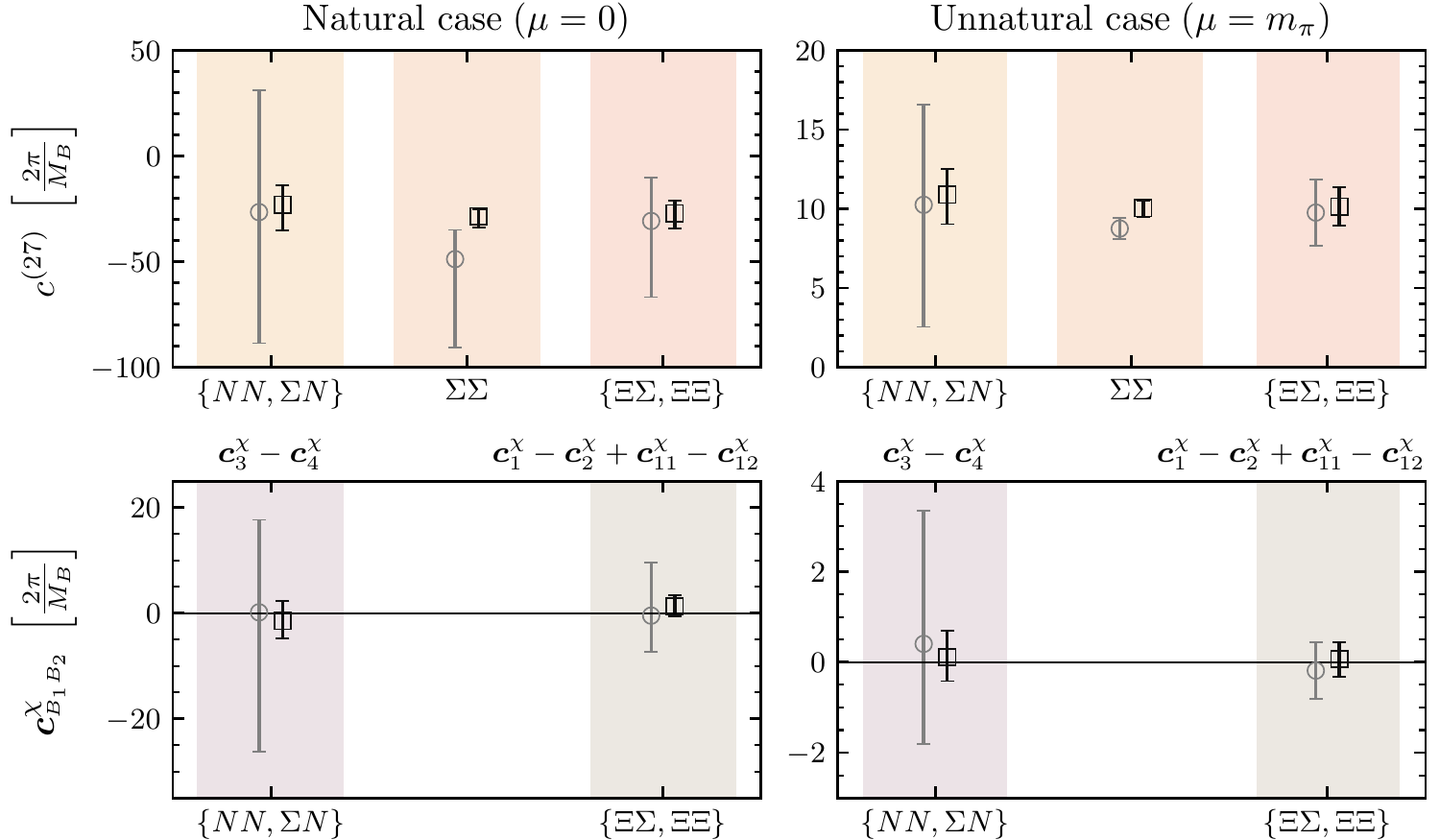}
\caption{The LO $SU(3)$ LEC $c^{(27)}$ (upper panels) and NLO $\cancel{SU(3)}$ LECs $\bm{c}^{\chi}_{\scriptscriptstyle B_1B_2}$ (lower panels) under the assumption of natural (left panels) and unnatural (right panels) interactions, in units of $[\frac{2\pi}{M_{B}}]$, where $M_B$ is the centroid of the octet-baryon masses. The gray-circle markers denote quantities that are extracted using method I, while black-square markers show results obtained from method II. See the text for further details.}
\label{fig:EFT_c27cchi}
\end{figure}
It should be noted that the input for scattering parameters is not sufficient to disentangle the LO $SU(3)$ and NLO $\cancel{SU(3)}$ coefficients in general, hence the $c^{(\text{irrep})}+\bm{c}^{\chi}_{\scriptscriptstyle B_1B_2}$ entry in Table~\ref{tab:LECtab1} and Fig.~\ref{fig:EFT_LONLO}.
For the systems that belong to the ${\bf 27}$ irrep, since the spin-singlet pairs $\{NN,\Sigma N\}$ and $\{\Xi \Sigma, \Xi \Xi\}$ depend on the same $SU(3)$ LO and $\cancel{SU(3)}$ NLO LECs but with different linear combinations of the coefficients, a system of equations can be formed to separate each contribution.
The results are shown in Table~\ref{tab:LECtab2} and Fig.~\ref{fig:EFT_c27cchi}, along with the result for the $\Sigma \Sigma$ channel for comparison purposes, as there is no contribution from $\cancel{SU(3)}$ interactions for this channel at this order.
From these results, it can be seen that the values of the symmetry-breaking coefficients $\bm{c}^{\chi}_3-\bm{c}^{\chi}_4$ and $\bm{c}^{\chi}_1-\bm{c}^{\chi}_2+\bm{c}^{\chi}_{11}-\bm{c}^{\chi}_{12}$ are compatible with zero. Together with the observation that the scattering lengths and binding energies in all of the systems are similar within uncertainties, it appears that the $SU(3)$ flavor symmetry remains an approximate symmetry at the quark masses used in this study.
These observations in the two-baryon sector are consistent with those in the single-baryon sector as presented in Ref.~\cite{Orginos:2015aya} at the same quark masses. There, it was found that the quantity $\delta_{\text{GMO}}=\frac{1}{M_B}(M_{\Lambda}+\frac{1}{3}M_{\Sigma}-\frac{2}{3}M_N-\frac{2}{3}M_{\Xi})$, which is a measure of $SU(3)$ flavor-symmetry breaking, is an order of magnitude smaller than its experimental value.\footnote{The violation of the Gell-Mann-Okubo mass relation~\cite{GellMann:1962xb,Okubo:1961jc} results from $SU(3)$ breaking transforming in the $\mathbf{27}$ irrep of $SU(3)$ flavor symmetry, which can only arise from insertions of the light-quark mass matrix or from nonanalytic meson-mass dependence induced by loops in $\chi$PT.}

In Appendix~\ref{sec:appen-EFT}, the full list of relations needed to independently constrain all 24 different LECs that appear at LO and NLO are shown, demonstrating that the proper combinations of 18 two-baryon flavor channels are sufficient to extract all these LECs. These channels will be the subject of upcoming LQCD studies toward the physical values of the quark masses.

%%%%%%%%%%%%%%%%%%%%%%%%%%%%%%%%%%%%%%%%%%%
\subsection{Compatibility with large-\texorpdfstring{$N_c$}{Nc} predictions \label{subsec:su6}}
In the limit of $SU(3)$ flavor symmetry and large $N_c$, two-baryon interactions are predicted to be invariant under an $SU(6)$ spin-flavor symmetry, with corrections that generally scale as $1/N_c$~\cite{Kaplan:1995yg}.
In the two-nucleon sector, this encompasses the $SU(4)$ spin-flavor Wigner symmetry~\cite{Wigner:1936dx, Wigner:1937zz, Wigner:1939zz}, with corrections that scale as $1/N^2_c$.
Under $SU(6)$ group transformations, the baryons transform as a three-index symmetric tensor $\Psi^{\mu\nu\rho}$, where each $SU(6)$ index is a pair of spin and flavor indices $(i \alpha)$. At LO, only two independent terms contribute to the interacting Lagrangian of two-baryon systems:
\begin{equation}
\mathcal{L}^{(0),SU(6)}_{BB}=-a(\Psi^{\dag}_{\mu\nu\rho}\Psi^{\mu\nu\rho})^2-b\Psi^{\dag}_{\mu\nu\sigma}\Psi^{\mu\nu\tau}\Psi^{\dag}_{\rho\delta\tau}\Psi^{\rho\delta\sigma}\, ,
\label{eq:su6lag}
\end{equation}
where the baryon tensor can be expressed as a function of the octet-baryon matrices, $B$:\footnote{Those components of the field $\Psi$ that correspond to decuplet baryons~\cite{Kaplan:1995yg} have been neglected as they are not relevant to the low-energy scattering of two octet baryons.}
\begin{equation}
\Psi^{\mu\nu\rho}=\Psi^{(i \alpha)(j \beta)(k \gamma)}=\frac{1}{\sqrt{18}}\left(B^{\alpha}_{\omega,i}\epsilon^{\omega \beta \gamma}\epsilon_{jk}+B^{\beta}_{\omega, j}\epsilon^{\omega \gamma \alpha}\epsilon_{ik}+B^{\gamma}_{\omega,k}\epsilon^{\omega \alpha \beta}\epsilon_{ij}\right) .
\end{equation}
Here, $\alpha,\beta,\gamma,\omega$ are flavor indices, $i,j,k$ are spin indices, and the Levi-Civita tensor $\epsilon$ is in either flavor or spin space depending on the type and number of indices.
A priori, the relative size of the Kaplan-Savage coefficients, $a$ and $b$, is unknown, and only experimental data or LQCD input may constrain these LECs.
As is seen in Eqs.~\eqref{eq:SU6eq} below, the contribution from the $b$ coefficient to the LO amplitude is parametrically suppressed compared with that of the coefficient $a$. As a result, if $b$ in Eq.~\eqref{eq:su6lag} is comparable to or smaller than $a$, there remains only one type of interaction that contributes significantly to the scattering amplitude, a situation that would realize an accidental $SU(16)$ symmetry of the nuclear and hypernuclear forces.
The first evidence for $SU(16)$ symmetry in the two-(octet)baryon sector was observed in a LQCD study at a pion mass of $\sim 806$ MeV~\cite{Wagman:2017tmp}, and the goal of the present study is to examine these predictions at smaller values of the light-quark masses.
Such a symmetry is suggested in Ref.~\cite{Beane:2018oxh} to be consistent with the conjecture of maximum entanglement suppression of the low-energy sector of QCD.
\begin{table}[b!]
\caption{The leading $SU(6)$ LECs, $a$ and $b/3$, obtained by solving a given pair of equations in Eqs.~\eqref{eq:SU6eq}. The last column shows the results of a constant fit to the LECs obtained in each case as described in Eqs.~\eqref{eq:averging}. The spin specifications are dropped from channel labels for brevity, but one clarification is necessary: in the first pair of two-baryon channels, $NN$ refers to the spin-singlet case, while in the last pair, it denotes the spin-triplet case. Quantities are expressed in units of $[\frac{2\pi}{M_{B}}]$, where $M_B$ is the centroid of the octet-baryon masses.}
\label{tab:SU6tab}
\begin{ruledtabular}
\begin{tabular}{ccccccccc|c}
LEC	&	$\mu$	&	Method	&	$\{NN, \Xi N\}$	&	$\{\Sigma N, \Xi N\}$	&	$\{\Sigma \Sigma, \Xi N\}$	&	$\{\Xi \Sigma, \Xi N\}$	&	$\{\Xi \Xi, \Xi N\}$	&	$\{NN, \Xi N\}$	&	Combined	\\ \hline
\multirow{4}{*}{$a$}	&	\multirow{2}{*}{$0$}	&	I	&	$-12_{(-13)}^{(+3)}$	&	$-12_{(-8)}^{(+2)}$	&	$-18_{(-11)}^{(+4)}$	&	$-14_{(-5)}^{(+2)}$	&	$-14_{(-3)}^{(+2)}$	&	$-14_{(-7)}^{(+3)}$	&	$-15_{(-4)}^{(+4)}$	\\
	&		&	II	&	$-11.9_{(-1.3)}^{(+0.9)}$	&	$-11.2_{(-1.4)}^{(+0.9)}$	&	$-11.8_{(-1.6)}^{(+1.0)}$	&	$-10.8_{(-1.0)}^{(+0.8)}$	&	$-10.2_{(-0.7)}^{(+0.6)}$	&	$-12.0_{(-1.4)}^{(+0.9)}$	&	$-11.0_{(-1.0)}^{(+1.0)}$	\\
	&	\multirow{2}{*}{$m_{\pi}$}	&	I	&	$5.8_{(-0.9)}^{(+1.2)}$	&	$5.6_{(-0.7)}^{(+0.7)}$	&	$5.0_{(-0.4)}^{(+0.4)}$	&	$5.2_{(-0.4)}^{(+0.4)}$	&	$5.1_{(-0.3)}^{(+0.3)}$	&	$5.5_{(-0.5)}^{(+0.5)}$	&	$5.2_{(-0.4)}^{(+0.4)}$	\\
	&		&	II	&	$6.0_{(-0.3)}^{(+0.2)}$	&	$6.0_{(-0.3)}^{(+0.3)}$	&	$5.7_{(-0.3)}^{(+0.2)}$	&	$5.8_{(-0.2)}^{(+0.2)}$	&	$5.8_{(-0.2)}^{(+0.2)}$	&	$6.0_{(-0.2)}^{(+0.2)}$	&	$5.9_{(-0.2)}^{(+0.2)}$	\\ \hline
\multirow{4}{*}{$\dfrac{b}{3}$}	&	\multirow{2}{*}{$0$}	&	I	&	$5_{(-31)}^{(+110)}$	&	$6_{(-26)}^{(+65)}$	&	$57_{(-42)}^{(+99)}$	&	$18_{(-27)}^{(+41)}$	&	$20_{(-22)}^{(+23)}$	&	$24_{(-29)}^{(+57)}$	&	$24_{(-36)}^{(+36)}$	\\
	&		&	II	&	$23_{(-9)}^{(+11)}$	&	$16_{(-8)}^{(+12)}$	&	$22_{(-10)}^{(+14)}$	&	$13_{(-7)}^{(+9)}$	&	$7_{(-6)}^{(+6)}$	&	$24_{(-9)}^{(+12)}$	&	$14_{(-9)}^{(+9)}$	\\
	&	\multirow{2}{*}{$m_{\pi}$}	&	I	&	$-1_{(-11)}^{(+7)}$	&	$0_{(-6)}^{(+6)}$	&	$6_{(-3)}^{(+3)}$	&	$4_{(-4)}^{(+4)}$	&	$5_{(-3)}^{(+3)}$	&	$1_{(-4)}^{(+4)}$	&	$4_{(-4)}^{(+4)}$	\\
	&		&	II	&	$3_{(-2)}^{(+2)}$	&	$4_{(-2)}^{(+2)}$	&	$6_{(-2)}^{(+2)}$	&	$6_{(-2)}^{(+2)}$	&	$5_{(-2)}^{(+2)}$	&	$3_{(-2)}^{(+2)}$	&	$5_{(-2)}^{(+2)}$	\\
\end{tabular}
\end{ruledtabular}
\end{table}

\begin{figure}[t!]
\includegraphics[width=0.9\textwidth]{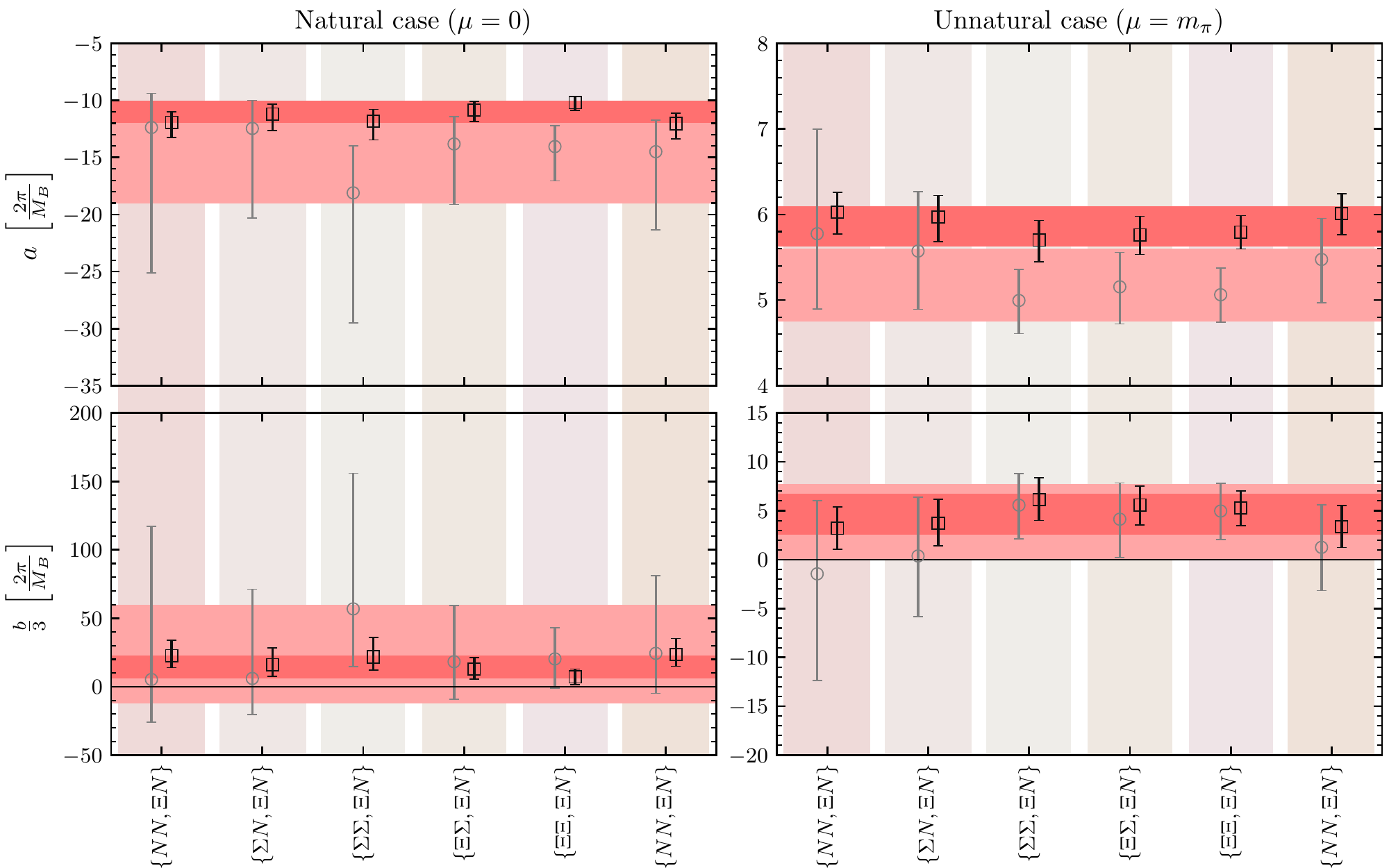}
\caption{The leading $SU(6)$ LECs, $a$ (upper panels) and $b/3$ (lower panels), under the assumption of natural (left panels) and unnatural (right panels) interactions, in units of $[\frac{2\pi}{M_{B}}]$, where $M_B$ is the centroid of the octet-baryon masses. The gray-circle markers denote quantities extracted using the ERE parameters (method I), with the light pink band showing the averaged value, while black-square markers show results obtained from scattering lengths that are constrained by binding momenta (method II), with the dark pink band showing the averaged value.}
\label{fig:SU6coeff}
\end{figure}

As in Sec.~\ref{subsec:lecs}, the $a$ and $b$ coefficients can be matched to scattering amplitudes in a momentum expansion at LO. Since at least some of the $SU(3)$ symmetry-breaking LECs $\bm{c}^{\chi}_i$ were found to be consistent with zero in this study, one can assume an approximate $SU(3)$ symmetry in general, and relate the $SU(6)$ LECs $a$ and $b$ directly to the LECs of the LO $SU(3)$-symmetric Lagrangian for given irreps:
\begin{align}
c^{(27)}&=2a-\frac{2b}{27}+\mathcal{O}\left(\frac{1}{N^2_c}\right), & c^{(\overline{10})}&=2a-\frac{2b}{27}+\mathcal{O}\left(\frac{1}{N^2_c}\right), \nonumber \\
c^{(8_s)}&=2a+\frac{2b}{3}+\mathcal{O}\left(\frac{1}{N_c}\right), & c^{(10)}&=2a+\frac{14b}{27}+\mathcal{O}\left(\frac{1}{N_c}\right), \nonumber \\
c^{(1)}&=2a-\frac{2b}{3}+\mathcal{O}\left(\frac{1}{N_c}\right), & c^{(8_a)}&=2a+\frac{2b}{27}+\mathcal{O}\left(\frac{1}{N_c}\right).
\label{eq:SU6eq}
\end{align}
In order to extract $a$ and $b$, states in the $\mathbf{27}$ and $\overline{\mathbf{10}}$ irreps can be combined with those in the $\mathbf{8}_a$ irrep, allowing for six possible extractions.\footnote{Note that the ERE parameters were obtained in the previous section only for two-baryon channels belonging to the $\{\mathbf{27},\overline{\mathbf{10}},\mathbf{8}_a\}$ irreps.}
The results are shown in Table~\ref{tab:SU6tab} and Fig.~\ref{fig:SU6coeff}. As seen in Eqs.~\eqref{eq:SU6eq}, the contributions from the $b$ coefficient are suppressed by at least a factor of 3 compared with those from the $a$ coefficient, and thus the rescaled coefficient $b/3$ is considered.
Considering that the results presented should be valid only up to corrections that scale as $1/N_c$, individual values of the coefficients $a$ and $b/3$ obtained from different pairs of channels exhibit remarkable agreement, indicating that the $SU(6)$ spin-flavor symmetry is a good approximation at these values of the quark masses.
A correlated weighted average\footnote{The average of a series of values $\{x_i\}$ with uncertainties $\{\sigma_i\}$ is computed as
\begin{equation}
x_{\text{average}}=\sum_i x_i w_i\, , \quad w_i=\frac{\sigma^{-2}_i}{\sum_j\sigma^{-2}_j}\, , \quad \sigma^2_{\text{average}}=\sum_{ij} w_i w_j C_{ij}\, , \quad C_{ij}=\sigma_i\sigma_j\, ,\label{eq:averging}
\end{equation}
where, since the different values of $x_i$ (and their uncertainties) are correlated, a 100\% correlation is assumed when computing $\sigma^2_{\text{average}}$. For asymmetric uncertainties in $x_i$, the following procedure is used to symmetrize them: a value $x_i=c^{(+u)}_{(-l)}$ is modified to $c+(u-l)/4$ with uncertainty $\sigma=\text{max}[(u+3l)/4,(3u+l)/4]$.} of the results is obtained, following the procedure introduced by Schmelling~\cite{Schmelling:1994pz} and used by the FLAG collaboration~\cite{Aoki:2013ldr}, and is shown as the pink bands in Fig.~\ref{fig:SU6coeff}.
Given the uncertainty in $b/3$, no conclusion can be drawn about the relative importance of $a$ and $b/3$. We will return to the question of the presence of an accidental $SU(16)$ symmetry shortly.

\begin{figure}[hbt!]
\includegraphics[width=0.9\textwidth]{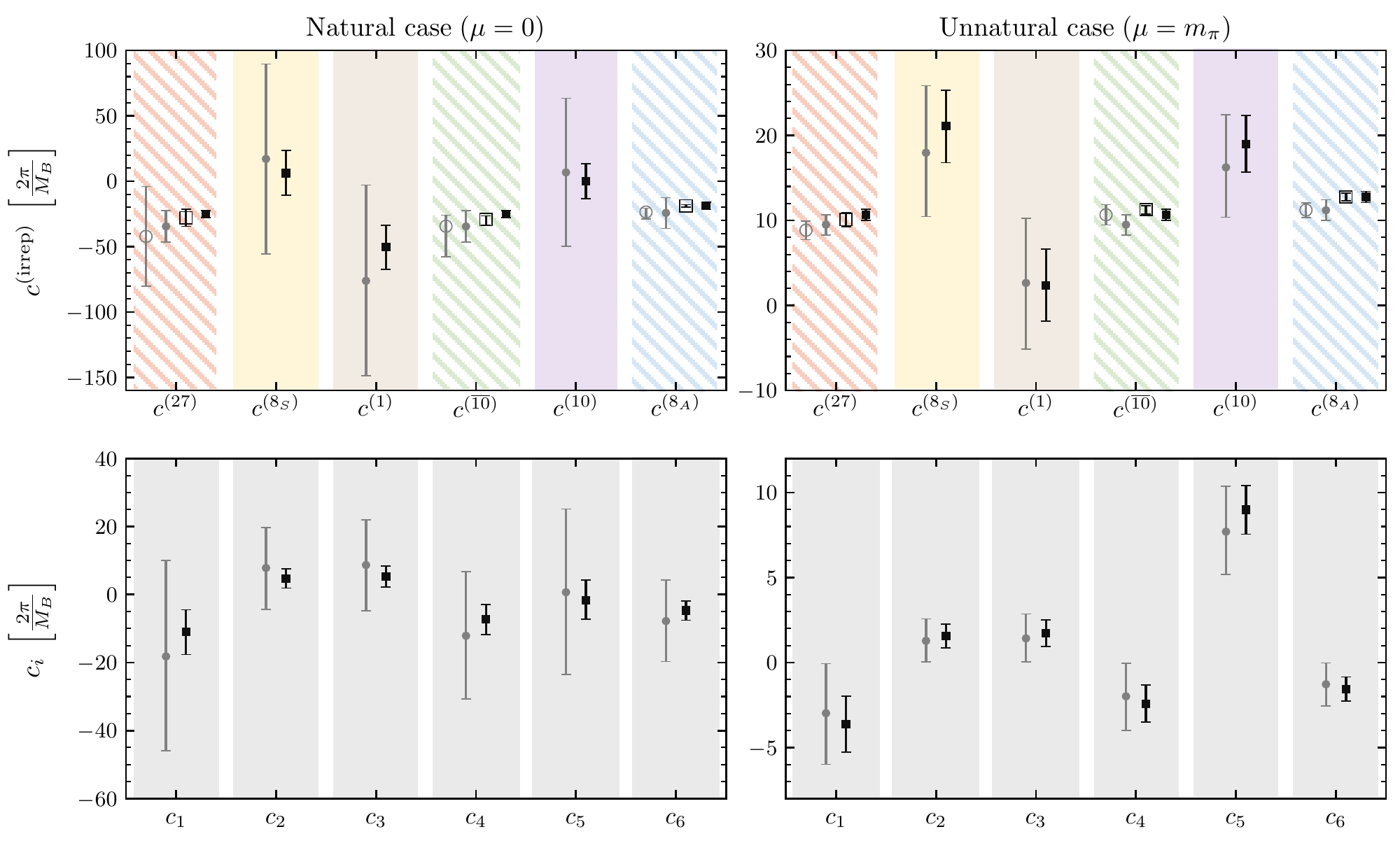}
\caption{The predicted (filled markers) LO $SU(3)$ coefficients $c^{(\text{irrep})}$ (upper panels), as well as Savage-Wise coefficients $c_i$ (lower panels) reconstructed from the $SU(6)$ relations are compared with the directly-extracted LECs (empty markers) under the assumption of natural (left panels) and unnatural (right panels) interactions, in units of $[\frac{2\pi}{M_{B}}]$, where $M_B$ is the centroid of the octet-baryon masses. The gray-circle symbols denote quantities that have been extracted using the scattering parameters obtained from the ERE fit (method I), while black-square symbols denote those that are obtained from scattering lengths constrained by binding momenta (method II). The hashed background in the upper panels denotes coefficients whose values were used to constrain $a$ and $b$, and hence are not predictions.}
\label{fig:abSU3coeff}
\end{figure}

Given the extracted values of $a$ and $b/3$, several checks can be performed, and several predictions can be made. The simplest check is to compute all of the LO $SU(3)$ LECs, $c^{(\text{irrep})}$, using the relations in Eqs.~\eqref{eq:SU6eq}.
The results are shown in the first rows of Table~\ref{tab:abSU3coeff} and the upper panels of Fig.~\ref{fig:abSU3coeff}. Columns with hashed backgrounds are the coefficients whose values were used as an input to make predictions for other coefficients, presented in panels with solid colored backgrounds.
These input coefficients ($c^{(27)}$, $c^{(\overline{10})}$, and $c^{(8_a)}$) can be reevaluated using the average values of $a$ and $b/3$, which therefore gives back consistent values but with different uncertainties (for $c^{(27)}$, the average of the values given in Table~\ref{tab:LECtab2} is computed).
The large uncertainties in the $c^{(8_s)}$, $c^{(1)}$, and $c^{(10)}$ coefficients are due to the fact that $b/3$, with a larger uncertainty than $a$, is numerically more important in these cases; see Eqs.~\eqref{eq:SU6eq}.
Additionally, the Savage-Wise coefficients $c_i$ can be computed by inverting the relations in Eqs.~\eqref{eq:SU3eq}, and the resulting values are presented in the last rows of Table~\ref{tab:abSU3coeff} and the lower panels of Fig.~\ref{fig:abSU3coeff}.
Due to large uncertainties in the natural case, no conclusions can be made regarding the relative size of the coefficients. In the unnatural case and at the chosen value of the renormalization scale, the $c_5$ coefficient has a larger value than the rest of the coefficients. 
The relative importance of $c_5$ is a remnant of an accidental approximate $SU(16)$ symmetry of $s$-wave two-baryon interactions that is more pronounced in the $SU(3)$-symmetric study with $m_{\pi}\sim 806$ MeV in Ref.~\cite{Wagman:2017tmp}.
It will be interesting to explore whether the remnant of this symmetry remains visible in studies closer to the physical quark masses.
\begin{table}[t!]
\caption{Predicted $SU(3)$ LECs, $c^{(\text{irrep})}$, as well as the Savage-Wise coefficients, $c_i$, obtained from the Kaplan-Savage $SU(6)$ coefficients $a$ and $b$ using the relations in Eqs.~\eqref{eq:SU6eq} and~\eqref{eq:SU3eq}. Quantities are expressed in units of $[\frac{2\pi}{M_{B}}]$, where $M_B$ is the centroid of the octet-baryon masses.}
\label{tab:abSU3coeff}
\begin{ruledtabular}
\begin{tabular}{cccccccc}
$\mu$	&	Method	&	$c^{(27)}$	&	$c^{(8_s)}$	&	$c^{(1)}$	&	$c^{(\overline{10})}$	&	$c^{(10)}$	&	$c^{(8_a)}$	\\ \hline
\multirow{2}{*}{$0$}	&	I	&	$-35_{(-12)}^{(+12)}$	&	$17_{(-73)}^{(+73)}$	&	$-76_{(-73)}^{(+73)}$	&	$-35_{(-12)}^{(+12)}$	&	$7_{(-57)}^{(+57)}$	&	$-24_{(-12)}^{(+12)}$	\\
	&	II	&	$-25_{(-3)}^{(+3)}$	&	$6_{(-17)}^{(+17)}$	&	$-50_{(-17)}^{(+17)}$	&	$-25_{(-3)}^{(+3)}$	&	$0_{(-13)}^{(+14)}$	&	$-19_{(-3)}^{(+3)}$	\\
\multirow{2}{*}{$m_{\pi}$}	&	I	&	$9.5_{(-1.2)}^{(+1.2)}$	&	$18.0_{(-7.5)}^{(+7.9)}$	&	$2.7_{(-7.8)}^{(+7.6)}$	&	$9.5_{(-1.2)}^{(+1.2)}$	&	$16.2_{(-5.9)}^{(+6.2)}$	&	$11.2_{(-1.2)}^{(+1.2)}$	\\
	&	II	&	$10.7_{(-0.7)}^{(+0.6)}$	&	$21.1_{(-4.3)}^{(+4.2)}$	&	$2.4_{(-4.2)}^{(+4.3)}$	&	$10.7_{(-0.7)}^{(+0.6)}$	&	$19.0_{(-3.3)}^{(+3.3)}$	&	$12.8_{(-0.7)}^{(+0.6)}$	\\ \hline\hline
$\mu$	&	Method	&	$c_1$	&	$c_2$	&	$c_3$	&	$c_4$	&	$c_5$	&	$c_6$	\\ \hline
\multirow{2}{*}{$0$}	&	I	&	$-18_{(-28)}^{(+28)}$	&	$8_{(-12)}^{(+12)}$	&	$9_{(-13)}^{(+13)}$	&	$-12_{(-19)}^{(+19)}$	&	$1_{(-24)}^{(+24)}$	&	$-8_{(-12)}^{(+12)}$	\\
	&	II	&	$-11_{(-7)}^{(+6)}$	&	$5_{(-3)}^{(+3)}$	&	$5_{(-3)}^{(+3)}$	&	$-7_{(-4)}^{(+4)}$	&	$-2_{(-6)}^{(+6)}$	&	$-5_{(-3)}^{(+3)}$	\\
\multirow{2}{*}{$m_{\pi}$}	&	I	&	$-3.0_{(-3.0)}^{(+2.9)}$	&	$1.3_{(-1.3)}^{(+1.3)}$	&	$1.4_{(-1.4)}^{(+1.4)}$	&	$-2.0_{(-2.0)}^{(+1.9)}$	&	$7.7_{(-2.5)}^{(+2.7)}$	&	$-1.3_{(-1.3)}^{(+1.3)}$	\\
	&	II	&	$-3.6_{(-1.6)}^{(+1.7)}$	&	$1.6_{(-0.7)}^{(+0.7)}$	&	$1.7_{(-0.8)}^{(+0.8)}$	&	$-2.4_{(-1.1)}^{(+1.1)}$	&	$9.0_{(-1.4)}^{(+1.4)}$	&	$-1.6_{(-0.7)}^{(+0.7)}$	\\
\end{tabular}
\end{ruledtabular}
\end{table}
\begin{table}[b!]
\caption{Predicted inverse scattering lengths, $a^{-1}$, for the systems where an ERE fit was not possible, using the $SU(6)$ LECs $a$ and $b$. Quantities are expressed in lattice units.}
\label{tab:predainv}
\begin{tabular}{c @{\qquad \quad} c @{\qquad \quad} r @{\qquad \quad} r}
\toprule
$\mu$ & Method & \multicolumn{1}{c @{\qquad \quad}}{$a^{-1}_{\Sigma N \;(\3s1)}$} & \multicolumn{1}{c}{$a^{-1}_{\Xi \Xi \;(\3s1)}$} \\ \hline
\multirow{2}{*}{$0$} & I & $-0.02_{(-07)}^{(+11)}$ & $-0.02_{(-06)}^{(+10)}$ \\
& II & $0.06_{(-44)}^{(+33)}$ & $0.05_{(-40)}^{(+30)}$ \\\hline
\multirow{2}{*}{$m_{\pi}$} & I & $0.14_{(-07)}^{(+04)}$ & $0.15_{(-06)}^{(+03)}$ \\
& II & $0.16_{(-02)}^{(+02)}$ & $0.17_{(-01)}^{(+02)}$ \\ \botrule
\end{tabular}
\end{table}
The values of the $SU(6)$ coefficients $a$ and $b$ allow predictions to be made for the scattering lengths of the systems that could not be constrained in this study by an ERE fit, namely the $\Sigma N \; (\3s1)$ and $\Xi \Xi \; (\3s1)$ channels.
Using the $c^{(\text{irrep})}$ coefficients computed previously, the relations in Eq.~\eqref{eq:scattparam1} can be inverted to obtain $a^{-1}$, assuming that the values of $\bm{c}^{\chi}_i$ are negligible compared with those of $c^{(\text{irrep})}$ (an observation that is only confirmed for given linear combinations of these LECs but is assumed to hold in general given the hints of an approximate $SU(3)$ symmetry in this study).
This exercise leads to consistent results for the inverse scattering length for systems for which the ERE allowed a direct extraction of this parameter, while it provides predictions for the channels shown in Table~\ref{tab:predainv}.
For the case of natural interactions, the scattering lengths are not constrained well, although they are consistent within uncertainties with those in the unnatural case, demonstrating the renormalization-scale independence of the scattering length. For the unnatural case, both methods are consistent and give rise to inverse scattering lengths that are positive and larger than those obtained for the rest of the systems studied in this work.
This is in agreement with the parameters found when fitting the results for $k^*\cot\delta$ in these channels beyond the $t$-channel cut, see Table~\ref{tab:scattPar-beyond}.
\section{Conclusions \label{sec:con}}
\noindent
Nuclear and hypernuclear interactions are key inputs into investigations of the properties of matter, and their knowledge continues to be limited in systems with multiple neutrons or when hyperons are present.
In recent years, LQCD has reached the stage where controlled first-principles studies of nuclei are feasible, and may soon constrain nuclear and hypernuclear few-body interactions in nature. The present work demonstrates such a capability in the case of two-baryon interactions, albeit at an unphysically large value of the quark masses corresponding to a pion mass of $\sim 450$ MeV.
It illustrates how Euclidean two-point correlation functions of systems with the quantum numbers of two baryons computed with LQCD can be used to constrain a wealth of quantities, from scattering phase shifts to low-energy scattering parameters and binding energies, to EFTs of forces, or precisely the LECs describing the interactions of two baryons.
This same approach can be expected to be followed in upcoming computations with the physical quark masses, and its output, in form of both finite-volume energy spectra and constrained EFT interactions, can serve as input into quantum many-body studies of larger isotopes, at both unphysical and physical values of quark masses; see e.g., Refs.~\cite{Barnea:2013uqa,Contessi:2017rww,Bansal:2017pwn} for previous studies in the nuclear sector.
By supplementing the missing experimental input for scattering and spectra of two-baryon systems, such LQCD analyses can constrain phenomenological models and EFTs of hypernuclear forces. 

In summary, the present paper includes a computation of the lowest-lying spectra of several two-octet baryon systems with strangeness ranging from $0$ to $-4$.
These results have been computed in three different volumes, using a single lattice spacing, and with unphysical values of the light-quark masses. The finite-volume nature of the energies provides a means to constrain the elastic scattering amplitudes in these systems through the use of L\"uscher's formalism.
Assuming small discretization artifacts given the improved LQCD action that is employed, our results reveal interesting features about the nature of two-baryon forces with larger-than-physical values of the quark masses.
In particular, the determination of scattering parameters of two-baryon systems at low energies has enabled constraints on the LO and NLO interactions of a pionless EFT, for both the $SU(3)$ flavor-symmetric and the symmetry-breaking interactions.
While the two-baryon channels studied in this work only allowed two sets of leading $SU(3)$ symmetry-breaking LECs to be constrained, and those values are seen to be consistent with zero, the present study is the first such analysis to access these interactions, extending the previous EFT matching presented in Ref.~\cite{Wagman:2017tmp} at an $SU(3)$-symmetric point with $m_\pi = m_K \sim 806$ MeV.
Given the limited knowledge of flavor-symmetry-breaking effects in the two-baryon sector in nature, this demonstrates the potential of  LQCD to improve the situation. 
Finally, the observation of an approximate $SU(3)$ symmetry in the two-baryon systems of this work led to an investigation of the large-$N_c$ predictions of Ref.~\cite{Kaplan:1995yg}, through matching the LQCD results for scattering amplitudes to the EFT.
In particular, the $s$-wave interactions at LO are found to exhibit an $SU(6)$ spin-flavor symmetry at this pion mass, as also observed in Ref.~\cite{Wagman:2017tmp} at a larger value of the pion mass. Both of the two independent spin-flavor-symmetric interactions at LO are found to contribute to the amplitude. Nonetheless, the extracted values of the coefficients of the LO $SU(3)$-symmetric EFT suggest a remnant of an approximate accidental $SU(16)$ symmetry observed in the $SU(3)$ flavor-symmetric study at $m_\pi \sim 806$~MeV~\cite{Wagman:2017tmp}.
It will be interesting to examine these symmetry considerations in the hypernuclear forces at the physical values of the quark masses, particularly given the conjectured connections between the nature of forces in nuclear physics and the quantum entanglement in the underlying systems~\cite{Beane:2018oxh}.
While no attempt is made in the current work to constrain forces within the EFTs at the physical point, a naive extrapolation is performed using the results of this work and those at $m_\pi \sim 806$ MeV, with simple extrapolation functions, to make predictions for the binding energies of several two-baryon channels.
The results for ground-state energies of two-nucleon systems are found to be compatible with the experimental values. Furthermore, stronger evidence for the existence of bound states in the $\Xi \Xi \; (\1s0)$ and $\Xi N \; (\3s1)$ channels is observed compared with other two-baryon systems.
Such predictions are in agreement with current phenomenological models and EFT predictions. However, conclusive results can only be reached by performing LQCD studies of multi-baryon systems at or near the physical values of the quark masses and upon taking the continuum limit using multiple values of lattice spacing, a program that will be pursued in the upcoming years.

%%%%%%%%%%%%%%%%%%%%%%%%%

\begin{acknowledgments}
We thank Joan Soto and Isaac Vida\~{n}a for enlightening discussions on the effective field theory formalism and on the physics of neutron stars and the hyperon puzzle, respectively. We also thank them, as well as Andr{\'e} Walker-Loud, for comments on the first version of this manuscript.

M.I.\ is supported by the Universitat de Barcelona through the scholarship APIF. M.I.\ and A.P.\ acknowledge support from the Spanish Ministerio de Econom\'{\i}a y Competitividad (MINECO) under Project No.\ MDM-2014-0369 of ICCUB (Unidad de Excelencia “Mar\'{\i}a de Maeztu”), from the European FEDER funds under Contract No.\ \mbox{FIS2017-87534-P} and by the EU STRONG-2020 project under the program \mbox{H2020-INFRAIA-2018-1}, Grant Agreement No.\ 824093. M.I.\ acknowledges the University of Maryland and the Massachusetts Institute of Technology for hospitality and partial support during preliminary stages of this work.
S.R.B.\ is supported in part by the U.S.\ Department of Energy, Office of Science, Office of Nuclear Physics under Contract No.\ \mbox{DE-FG02-97ER-41014}.
Z.D.\ is supported by Alfred P.\ Sloan fellowship and by Maryland Center for Fundamental Physics at the University of Maryland, College Park.
W.D.\ and P.E.S.\ acknowledge support from the U.S.\ DOE Grant No.\ \mbox{DE-SC0011090}. W.D.\ is also supported within the framework of the TMD Topical Collaboration of the U.S.\ DOE Office of Nuclear Physics, and by the SciDAC4 Award No.\ \mbox{DE-SC0018121}. P.E.S.\ is additionally supported by the National Science Foundation under CAREER Award No.\ 1841699 and under EAGER Grant No.\ 2035015, by the U.S. DOE Early Career Award No.\ \mbox{DE-SC0021006}, by a NEC research award, and by the Carl G.\ and Shirley Sontheimer Research Fund.
K.O.\ and F.W.\ are supported by U.S.\ DOE Grant No.\ \mbox{DE-FG02-04ER41302} and by Jefferson Science Associates, LLC under U.S.\ DOE Contract No.\ \mbox{DE-AC05-06OR23177}.
F.W.\ is additionally supported by the USQCD Scientific Discovery through the Advanced Computing (SciDAC) project funded by U.S.\ Department of Energy, Office of Science, Offices of Advanced Scientific Computing Research, Nuclear Physics and High Energy Physics.
M.J.S.\ is supported by the Institute for Nuclear Theory with DOE Grant No.\ \mbox{DE-FG02-00ER41132}.
This manuscript has been authored by Fermi Research Alliance, LLC under Contract No.\ \mbox{DE-AC02-07CH11359} with the U.S.\ Department of Energy, Office of Science, Office of High Energy Physics.

The results presented in this manuscript were obtained using ensembles of isotropic-clover gauge-field configurations produced several years ago with resources obtained by researchers at the College of William and Mary and the Thomas Jefferson National Accelerator Facility and by the NPLQCD collaboration. Computations were performed using a College of William and Mary led XSEDE and NERSC allocation, and NPLQCD PRACE allocations on Curie and MareNostrum, on LLNL machines, on the HYAK computational infrastructure at the UW, and through ALCC allocations. 
Calculations of propagators and their contractions were performed using computational resources provided by the Extreme Science and Engineering Discovery Environment, which is supported by National Science Foundation Grant No.\ OCI-1053575, by NERSC (supported by U.S.\ Department of Energy Grant No.\ \mbox{DE-AC02-05CH11231}), and by the USQCD collaboration. This research used resources of the Oak Ridge Leadership Computing Facility at the Oak Ridge National Laboratory, which is supported by the Office of Science of the U.S.\ Department of Energy under Contract No.\ \mbox{DE-AC05-00OR22725}. The authors thankfully acknowledge the computer resources at MareNostrum and the technical support provided by Barcelona Supercomputing Center (\mbox{RES-FI-2019-2-0032} and \mbox{RES-FI-2019-3-0024}). Parts of the calculations used the Chroma~\cite{Edwards:2004sx} and QUDA~\cite{Clark:2009wm,Babich:2010mu} software suites. We thank Andr{\'e} Walker-Loud and Thomas Luu for contributions during initial stages of production prior to 2014.

After this manuscript was completed, an additional preprint \cite{Horz:2020zvv} appeared regarding two-nucleon scattering, using a set of scattering interpolating operators similar to Ref.~\cite{Francis:2018qch}.

\end{acknowledgments}

%%%%%%%%%%%%%%%%%%%%%%%
\bibliography{BB450_bib}

%%%%%%%%%%%%%%%%%%%
\appendix

\section{On the validity of the extraction of the lowest-lying energies and the corresponding scattering amplitudes \label{sec:appen-checks}}
\noindent
In Refs.~\cite{Iritani:2017rlk,Iritani:2018talk}, several criteria were presented to validate studies of two-baryon systems that rely on the extraction of finite-volume energies from Euclidean LQCD correlation functions for use in L\"uscher's formalism. The results of the present work are examined and validated with regard to these criteria. Similar investigations were performed in Refs.~\cite{Wagman:2017tmp,Beane:2017edf} for the study at $m_{\pi} \sim806$ MeV in Ref.~\cite{Wagman:2017tmp}.
\begin{figure}[t!]
\includegraphics[width=\textwidth]{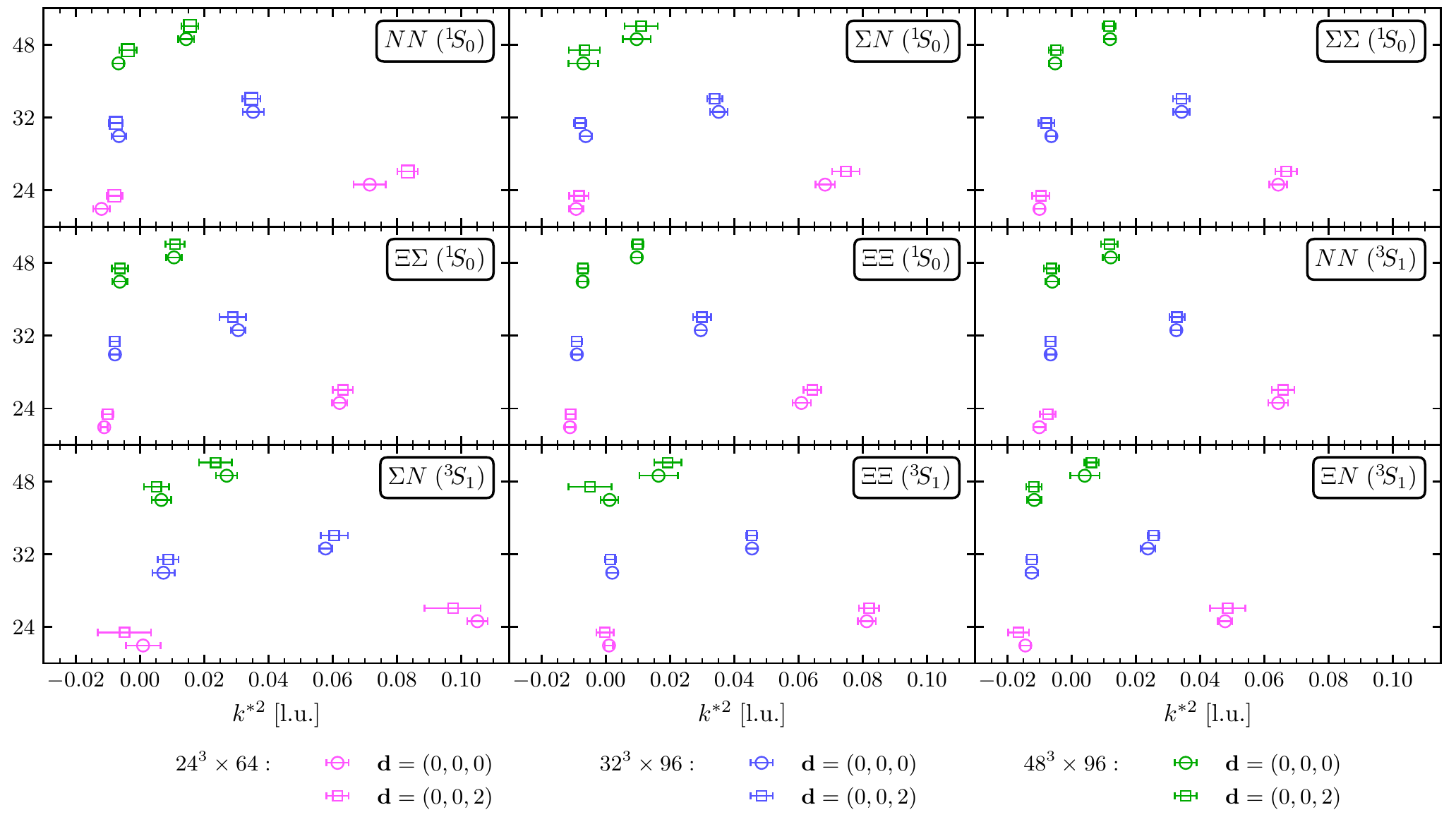}
\caption{The values of $k^{*2}$ for all systems analyzed in this work. Quantities are expressed in lattice units.}
\label{fig:appenA_k2}
\end{figure}
\begin{itemize}
\item[-] \textit{Interpolating-operator independence}: The two different source-sink operator structures, denoted SP and SS and described in Sec.~\ref{subsec:comp}, yield the same energies for both the ground and the first excited states obtained in this work. This consistency can be verified by examining the late-time behavior of the effective-energy and effective-energy-shift functions constructed from the SS and SP correlation functions in Figs.~\ref{fig:B1_EMP}-\ref{fig:XN3s1_EMP}. Moreover, the c.m.\ momenta $k^{*2}$ obtained from the correlation functions with $\bm{d}=(0,0,0)$ and $\bm{d}=(0,0,2)$ must be consistent, up to negligible relativistic and small $\mathcal{O}\left((m_1^2-m_2^2)/E^{*2}\right)$ corrections~\cite{Davoudi:2011md}, a feature that is observed in the results presented here, as shown in Fig.~\ref{fig:appenA_k2}. The largest difference is seen in the $NN~(\1s0)$ channel for the $n=2$ level on the ensemble with $L=24$, for which the c.m.\ momenta in the unboosted and boosted cases exhibit a $\sim 2 \sigma$ difference. 
\begin{figure}[t!]
\includegraphics[width=\textwidth]{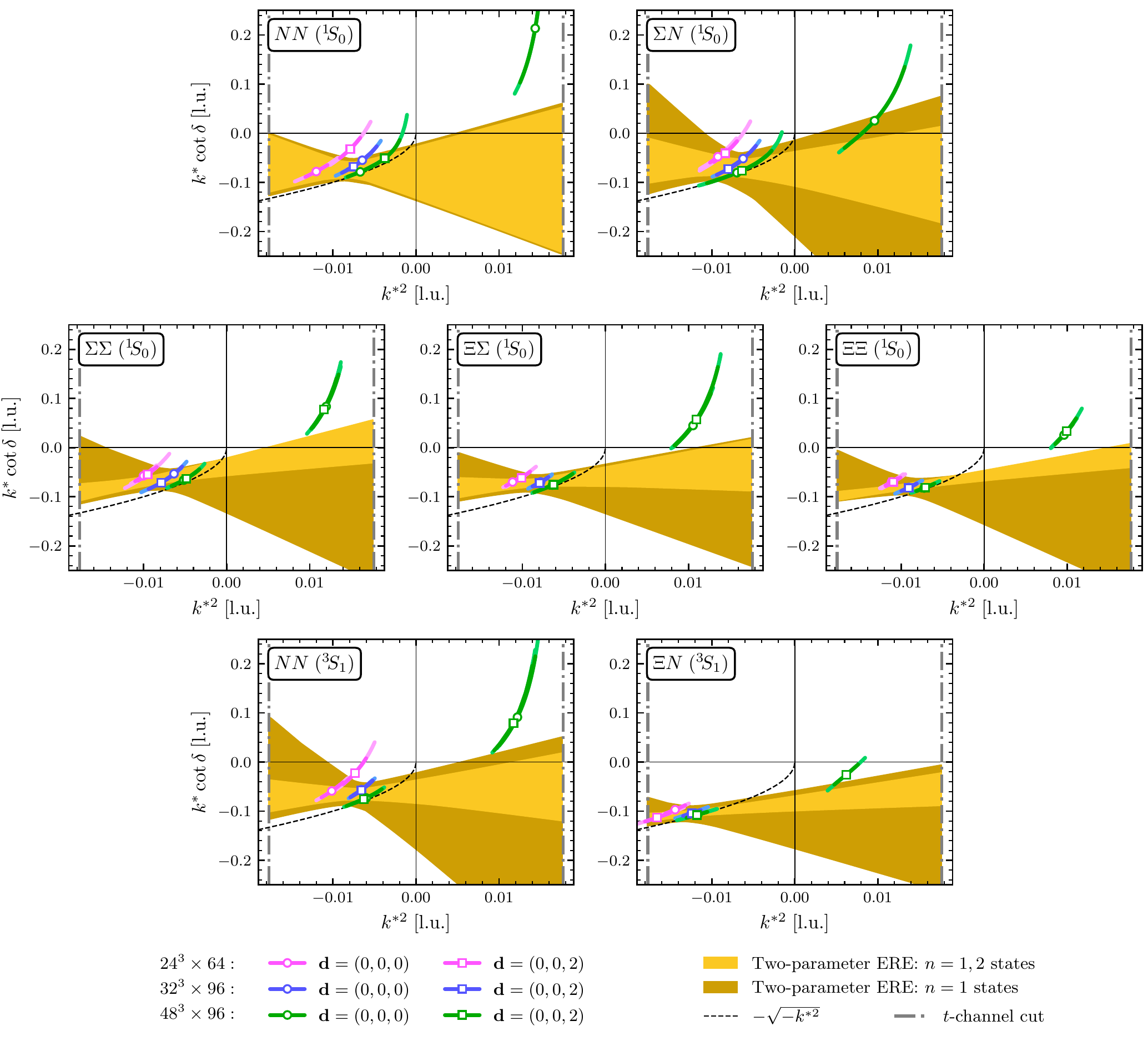}
\caption{$k^{*2}\cot\delta$ values as a function of the c.m.\ momenta $k^{*2}$, together with bands representing the two-parameter ERE using all the energy levels (ground state $n=1$ and excited states $n=2$) in lighter yellow, or using just the ground state in darker yellow. Quantities are expressed in lattice units.}
\label{fig:appenA_n12}
\end{figure}
\item[-] \textit{Consistency between ERE parameters for $k^{*2}<0$ and $k^{*2}>0$}: In the two-baryon channels studied in this work, there are not sufficient data points for $k^*\cot\delta$ below the $t$-channel cut to extract precise scattering parameters, as pointed out in Sec.~\ref{subsec:fitalg}. Nonetheless, for the cases for which two sets of data at positive and negative values of $k^{*2}$ are available, the ERE fits obtained by fitting to all $k^{*2}$ versus only fitting to $k^{*2}<0$ values are fully consistent with each other, as is shown in Fig.~\ref{fig:appenA_n12}.

\item[-] \textit{Non-singular scattering parameters}: None of the scattering parameters extracted show singular behavior, as can be seen from the values in Table~\ref{tab:scattPar}.

\item[-] \textit{Requirement on the residue for the scattering amplitude at the bound-state pole}: In order to support a physical bound state, the slope of the ERE as a function of $k^{*2}$ must be smaller than the slope of the $-\sqrt{-k^{*2}}$ at the bound-state pole. The two slopes and associated uncertainty bands are depicted in Fig.~\ref{fig:appenA_slope} for all two-baryon channels and the two-parameter EREs obtained, demonstrating that the needed inequality is satisfied. The values of binding momenta used in this analysis are taken from Table~\ref{tab:binding_momenta} (the $\bm{d}=\{(0,0,0),(0,0,2)\}$ column).
\begin{figure}[t!]
\includegraphics[width=\textwidth]{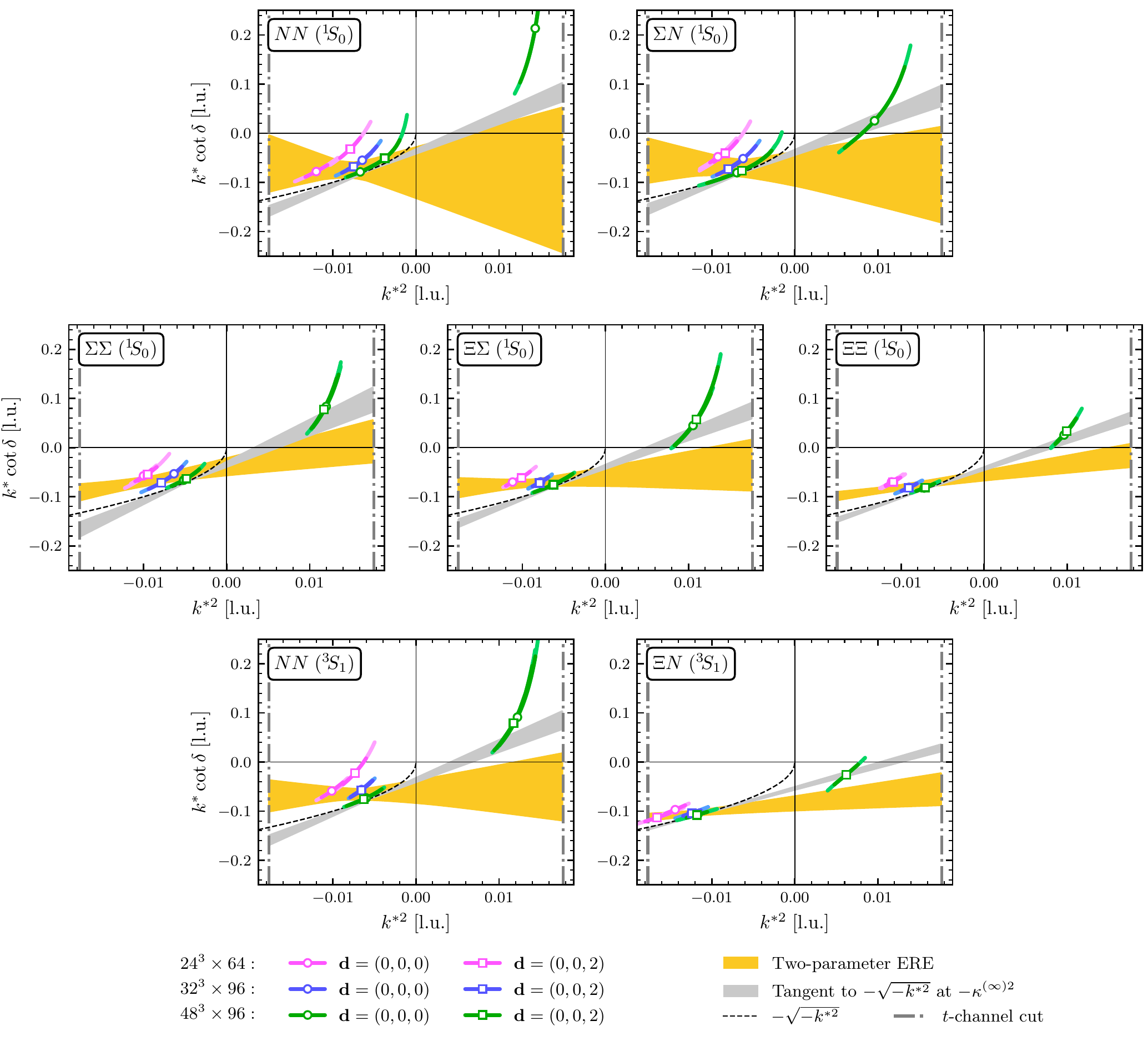}
\caption{Comparison between the two-parameter ERE and the slope of $-\sqrt{-k^{*2}}$ at $k^{*2}=-\kappa^{(\infty)2}$, where $\kappa^{(\infty)}$ is taken from the $\bm{d}=\{(0,0,0),(0,0,2)\}$ column of Table~\ref{tab:binding_momenta}. Quantities are expressed in lattice units.}
\label{fig:appenA_slope}
\end{figure}
\item[-] \textit{The absence of more than one bound state with an ERE parametrization of amplitudes}: None of the systems analyzed exhibit more than one bound state; i.e., the ERE does not cross the $-\sqrt{-k^{*2}}$ curve more than once. Therefore, applying the ERE parametrization of the $s$-wave scattering amplitude in all channels appears to be justified.
\item[-] \textit{Constrained range for ERE parameters in the presence of a bound state}: If the system presents a bound state, the ratio $r/a$ must be smaller than 1/2 for the two-parameter ERE to cross the $-\sqrt{-k^{*2}}$ function once from below, which is the condition for a physical bound state. Moreover, the ERE must cross the $\mathcal{Z}$-functions corresponding to different volumes to satisfy L\"uscher's quantization condition, introducing more constraints on scattering parameters. With the use of the two-dimensional $\chi^2$ in this work to fit the $k^*\cot\delta$ values, the confidence region of the ERE parameters does not cross these prohibited areas, as was demonstrated in Fig.~\ref{fig:ere-parameters}.
\end{itemize}

\section{Comparison with previous LQCD results and those obtained from low-energy theorems \label{sec:appen-vs2015}}
\noindent
A subset of the correlation functions used in this work has already been analyzed in Ref.~\cite{Orginos:2015aya}, where the $NN~(\1s0)$ and $NN~(\3s1)$ channels were studied. In the following, we present the outcome of a careful comparison of the results obtained using both analyses, along with a comparison of the updated scattering parameters from this work and those obtained from low-energy theorems in Ref.~\cite{Baru:2015ira}.

%%%%%%%%%%%%%%%%%%%%%%%%%%%%%%%%%%%%%%%%%%%
\subsection{Differences in the fitting strategy}
\begin{figure}[t!]
\includegraphics[width=0.8\textwidth]{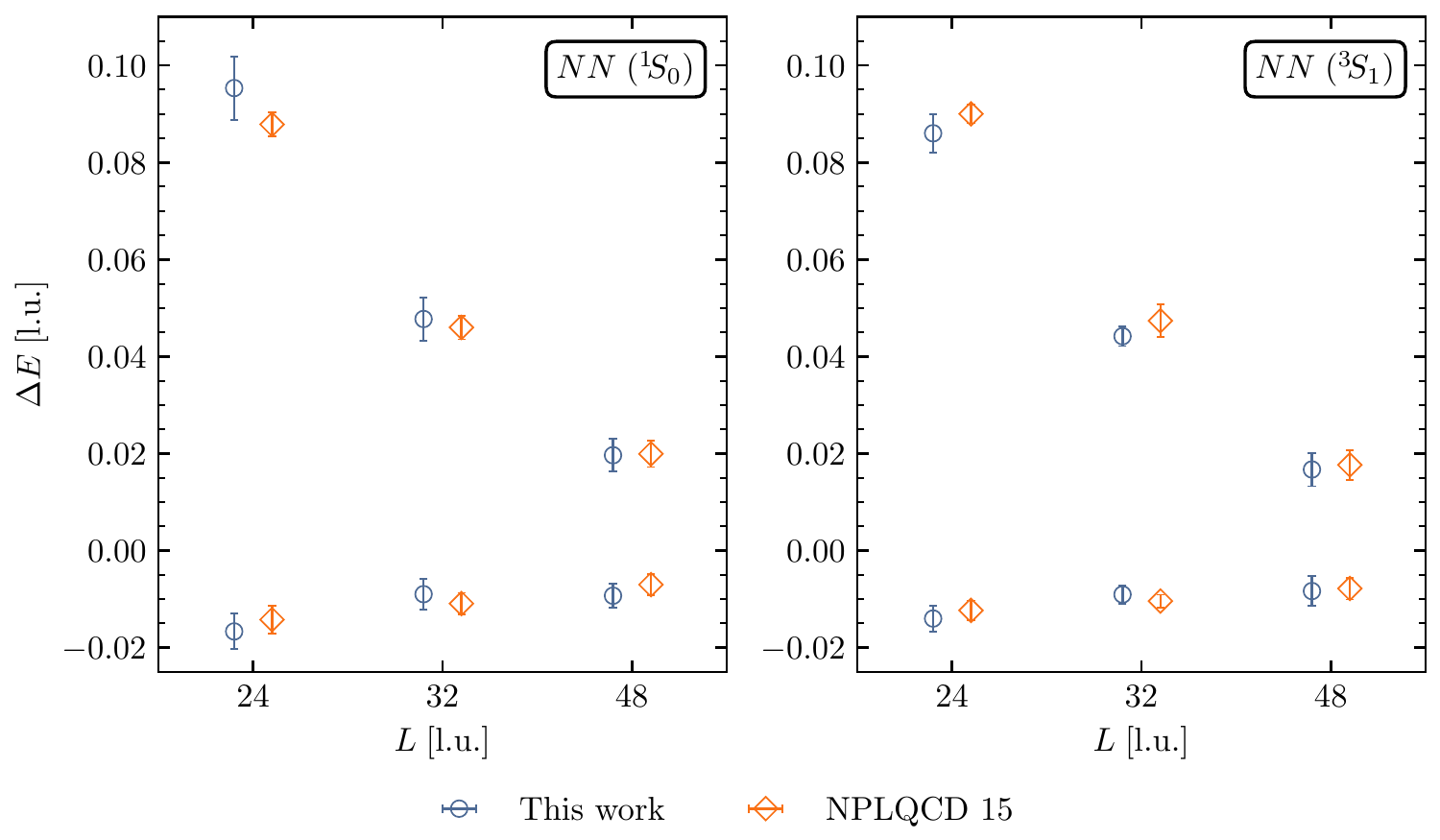}
\caption{ Comparison of the ground-state and first excited-state energies obtained in this work (blue circles) and from Ref.~\cite{Orginos:2015aya} (orange diamonds), labeled as NPLQCD 15. The figure shows results with statistical and systematic uncertainties combined in quadrature. Quantities are expressed in lattice units.}
\label{fig:20152020_energiescompar}
\end{figure}
The ground-state and first excited-state energies obtained in this work and those from Ref.~\cite{Orginos:2015aya} are shown in Fig.~\ref{fig:20152020_energiescompar}.
While all numbers are in agreement within uncertainties, it is clear that, in general, the analysis performed in Ref.~\cite{Orginos:2015aya} led to smaller uncertainties (one exception is the $NN\; (\3s1)$ first excited state with $L=32$).
That analysis consisted of the following: (1) taking linear combinations of the SP and SS correlation functions (except for the $L=48$ ensemble, where only SP correlation functions were computed); (2) the use of the Hodges-Lehmann (HL) robust estimator under bootstrap resampling to estimate the ensemble-averaged correlation functions; and (3) fitting constants to the effective-(mass) energy functions built from the combinations mentioned above.
In the present analysis, multi-exponential fits are performed to both SP and SS correlation functions in a correlated way (when available), using the mean under bootstrap resampling. 

\begin{figure}[b!]
\includegraphics[width=1.0\textwidth]{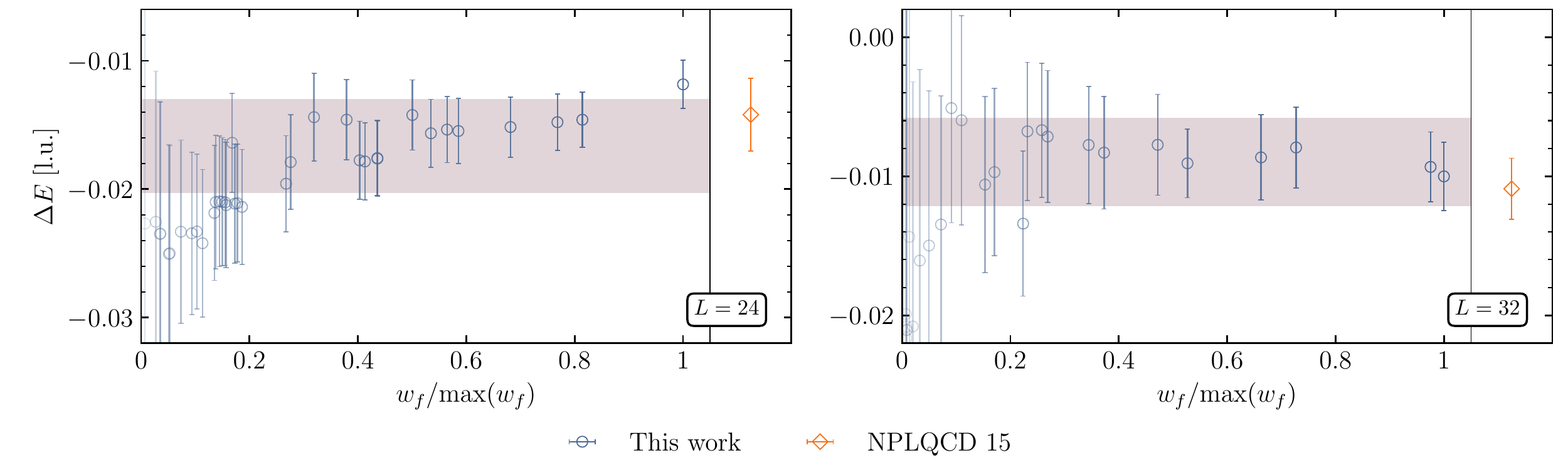}
\caption{Ground-state energies for the $NN\; (\1s0)$ system computed on ensembles with $L=24$ (left) and $L=32$ (right), sorted by their weight. The weight of each individual fit is indicated by the level of transparency of each point (darker points have larger weight). The band shows the final result, obtained by combining the individual points with the corresponding weight according to Eq.~\eqref{eq:weights_fitting}, with statistical and systematic uncertainties combined in quadrature. To facilitate the comparison, the orange point in the right panel of each figure shows the result of Ref.~\cite{Orginos:2015aya}, labeled as NPLQCD 15.}
\label{fig:nn1s0_example}
\end{figure}
Taking a closer look at how the statistical and systematic uncertainties are computed, it is worth examining the individual fits from all accepted time windows.
These are shown in Fig.~\ref{fig:nn1s0_example} for the $NN\; (\1s0)$ $L=24$ and $L=32$ ground states, sorted by their weight, $w_f$, as defined in Eq.~\eqref{eq:weights}. As can be seen, there are cases for which the size of the uncertainty is similar to or smaller than that presented in Ref.~\cite{Orginos:2015aya}.
However, the final combined uncertainty, represented by the band in Fig.~\ref{fig:nn1s0_example}, is larger. This can be understood as using a more conservative procedure for quantifying the systematic uncertainty, as well as a more thorough one: not only are variations of the fitting range considered, but also variations in the fitting form, including forms with multiple exponentials, see Sec.~\ref{subsec:fitalg}.
\begin{figure}[t!]
\centering
\renewcommand{\arraystretch}{1.5}
\begin{tabular}{rrccccl}
& & \multicolumn{2}{c}{One input (SS)} & \multicolumn{2}{c}{Multiple inputs (SS and SP)} & \\
& & Effective energy & Corr. & Effective energy & Corr. & \multirow{3}{*}[-20ex]{\includegraphics[width=0.6cm]{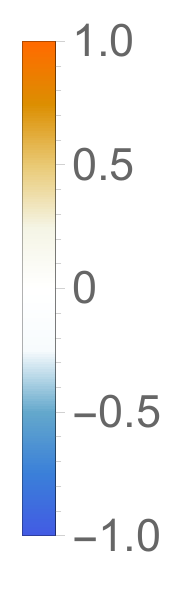}}\\
\multirow{2}{5ex}{\rotatebox{90}{$L=24$}} & \rotatebox{90}{Mean} & \includegraphics[width=3.2cm]{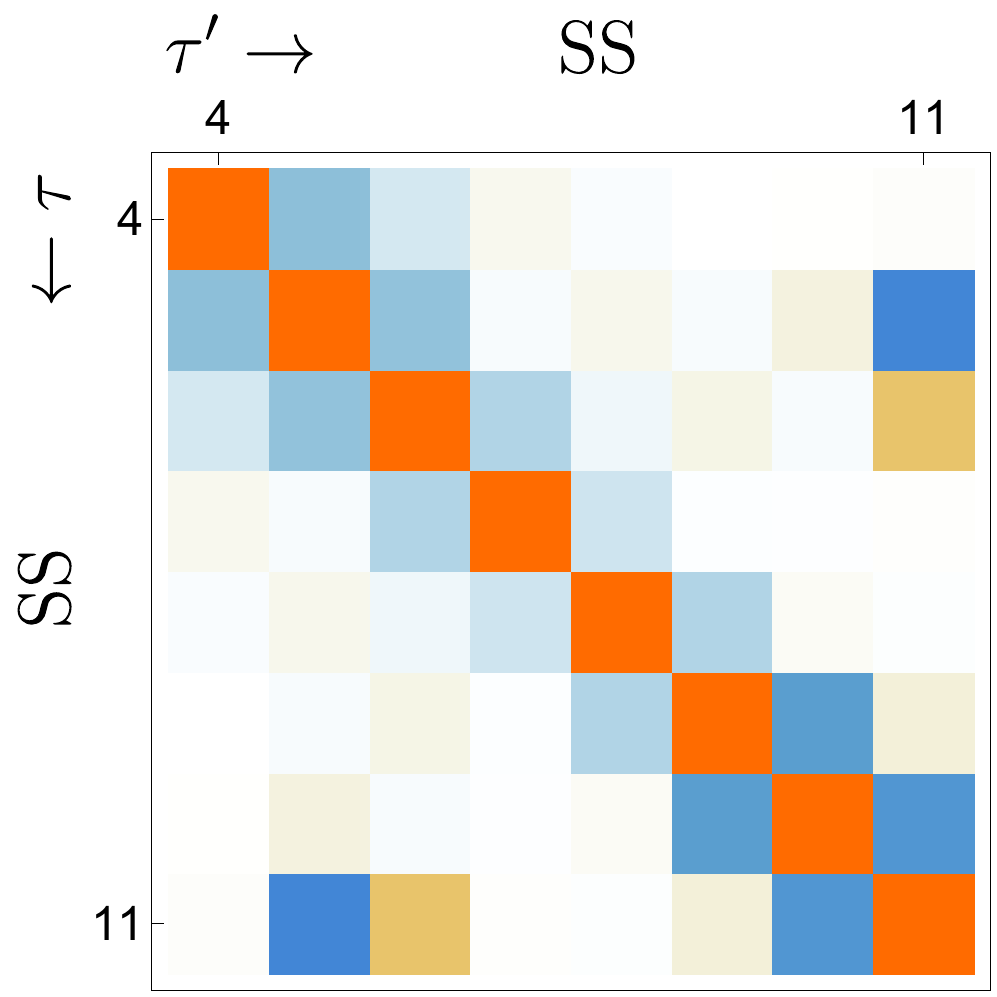} & \includegraphics[width=3.2cm]{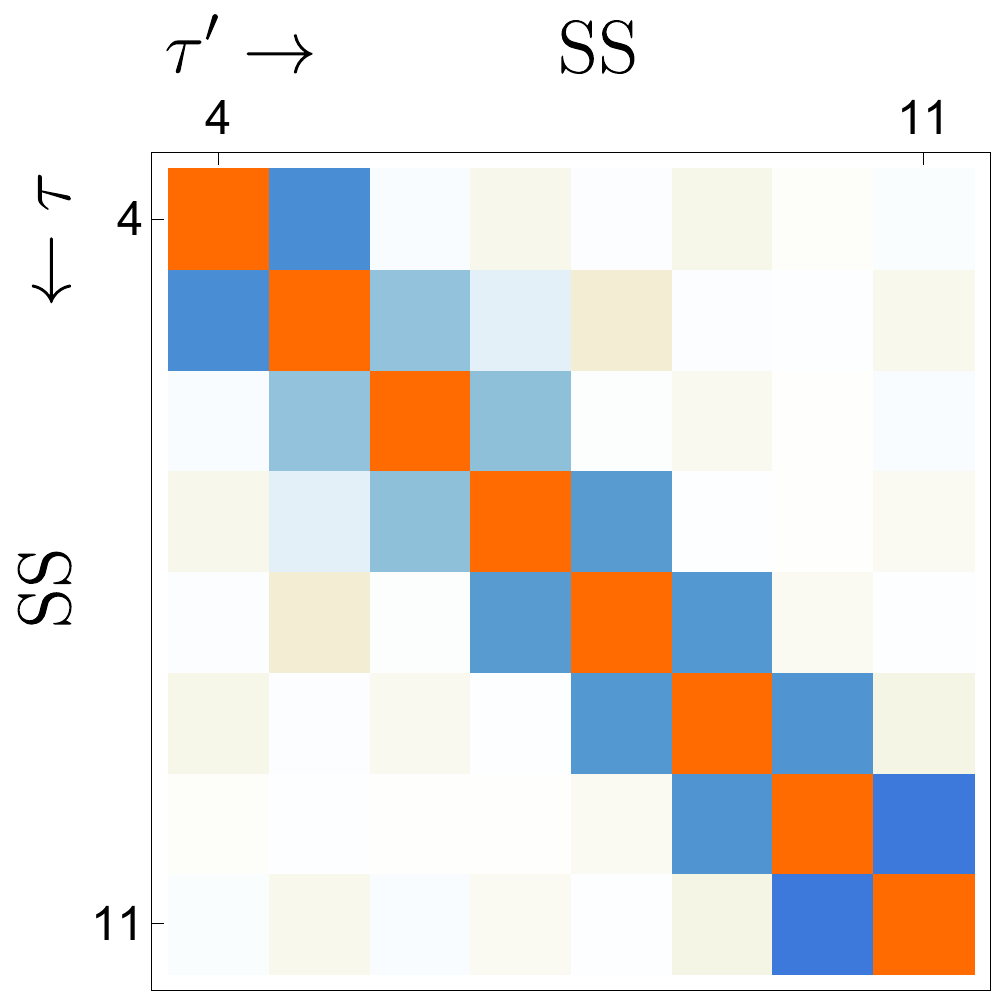} & \includegraphics[width=3.2cm]{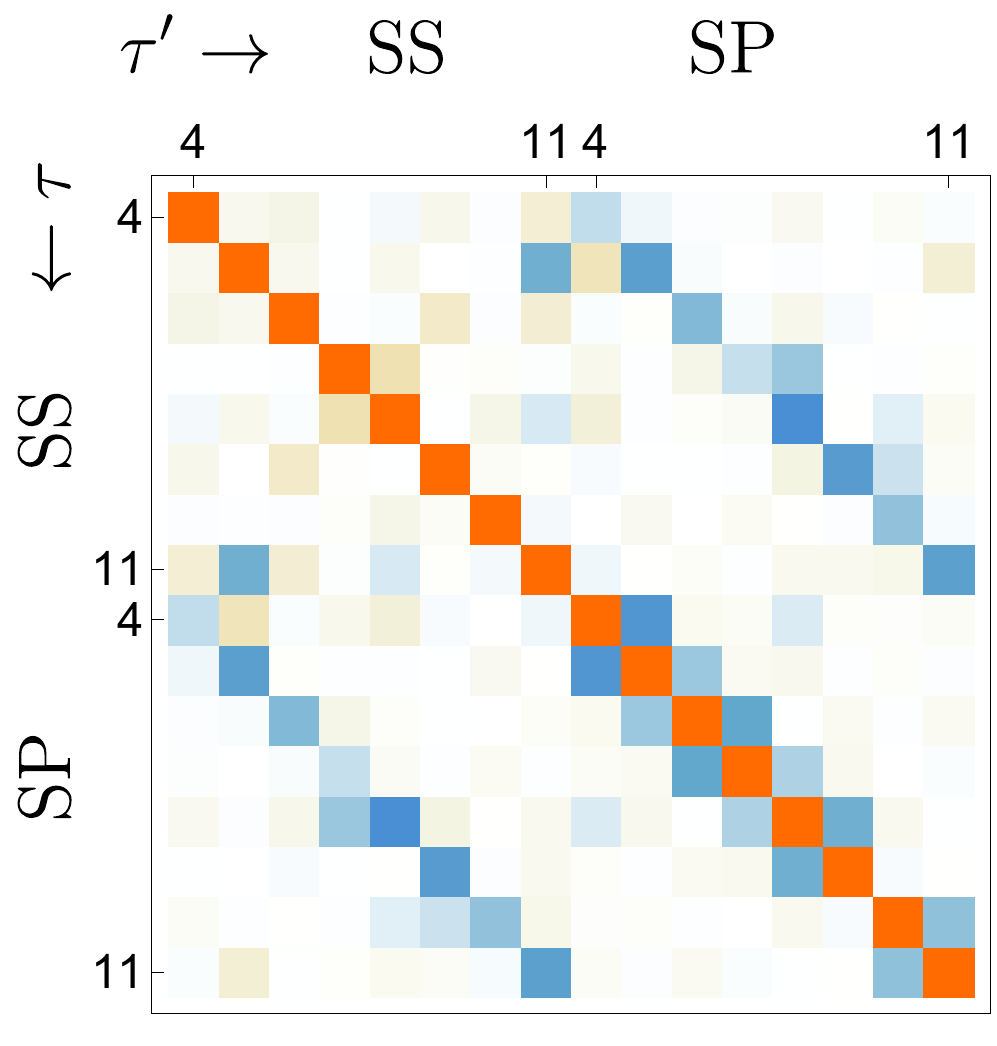} & \includegraphics[width=3.2cm]{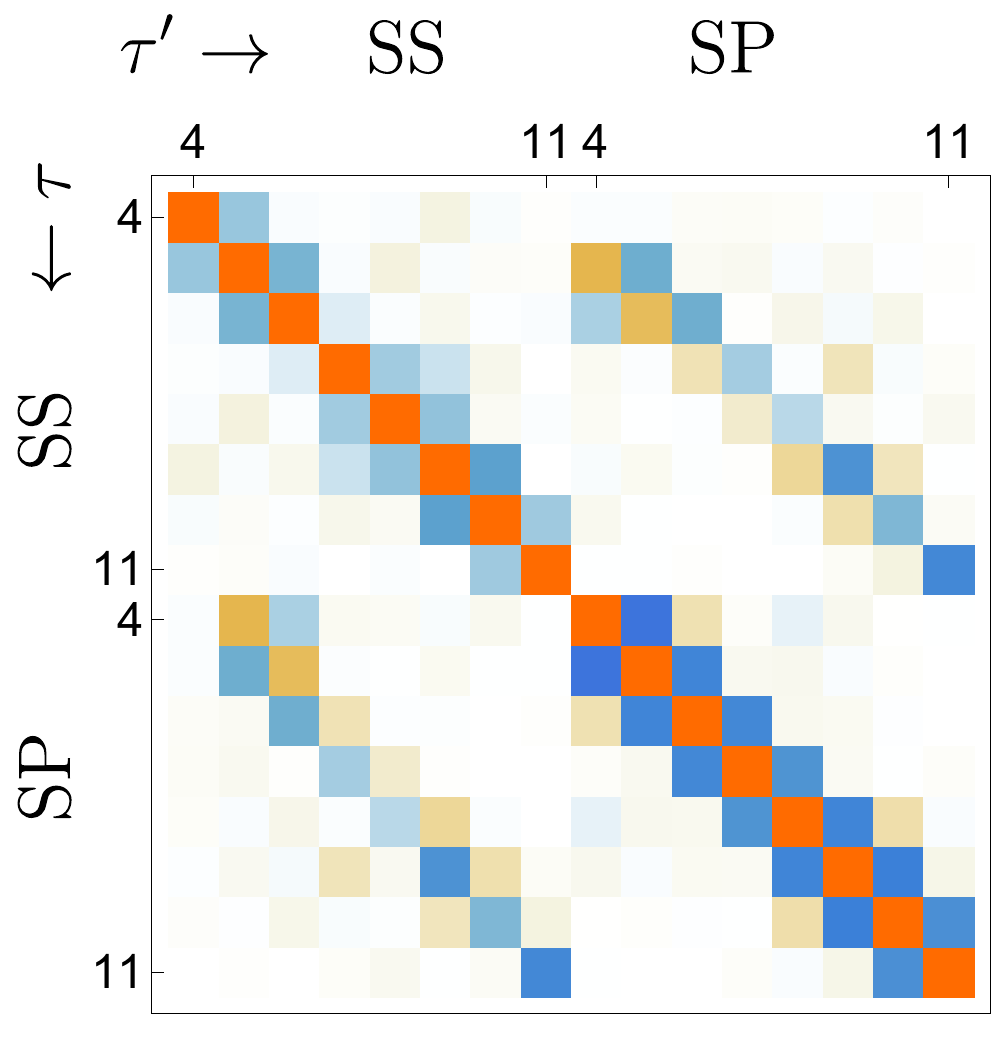} & \\ 
& \rotatebox{90}{HL estimator} & \includegraphics[width=3.2cm]{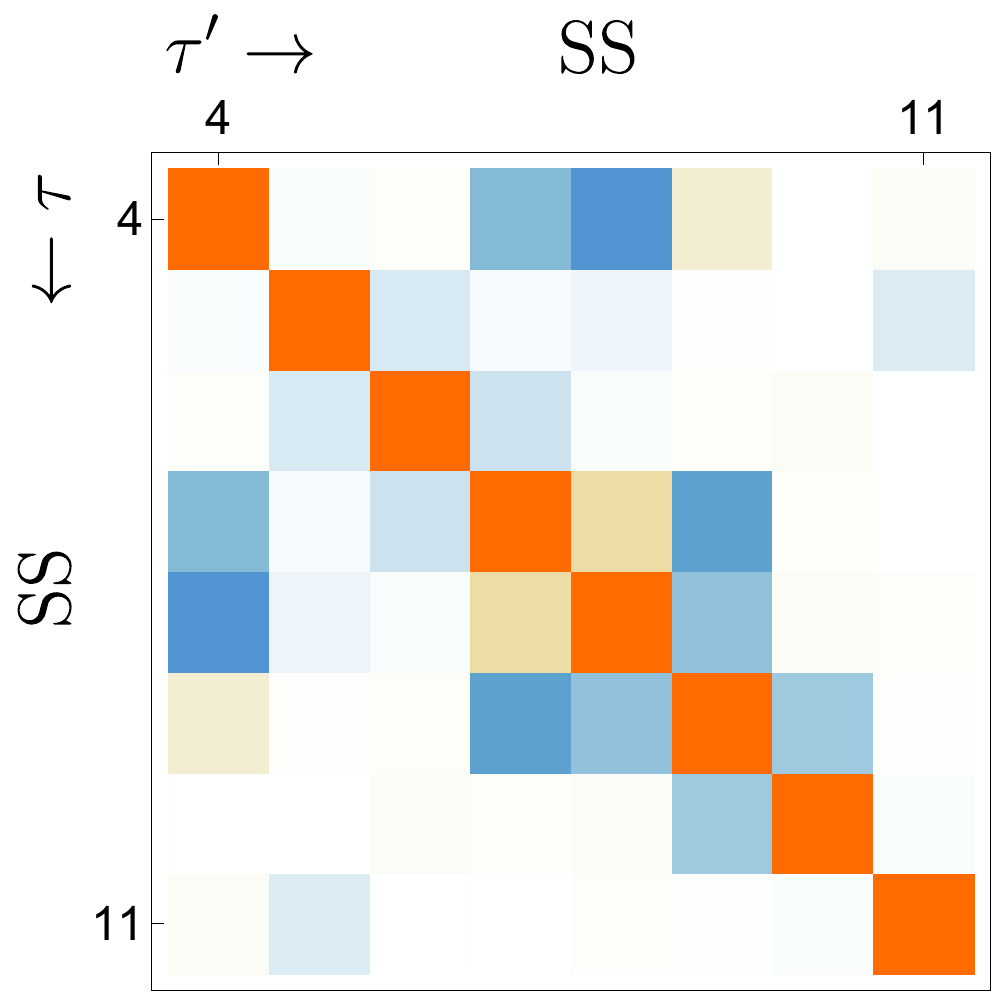} & \includegraphics[width=3.2cm]{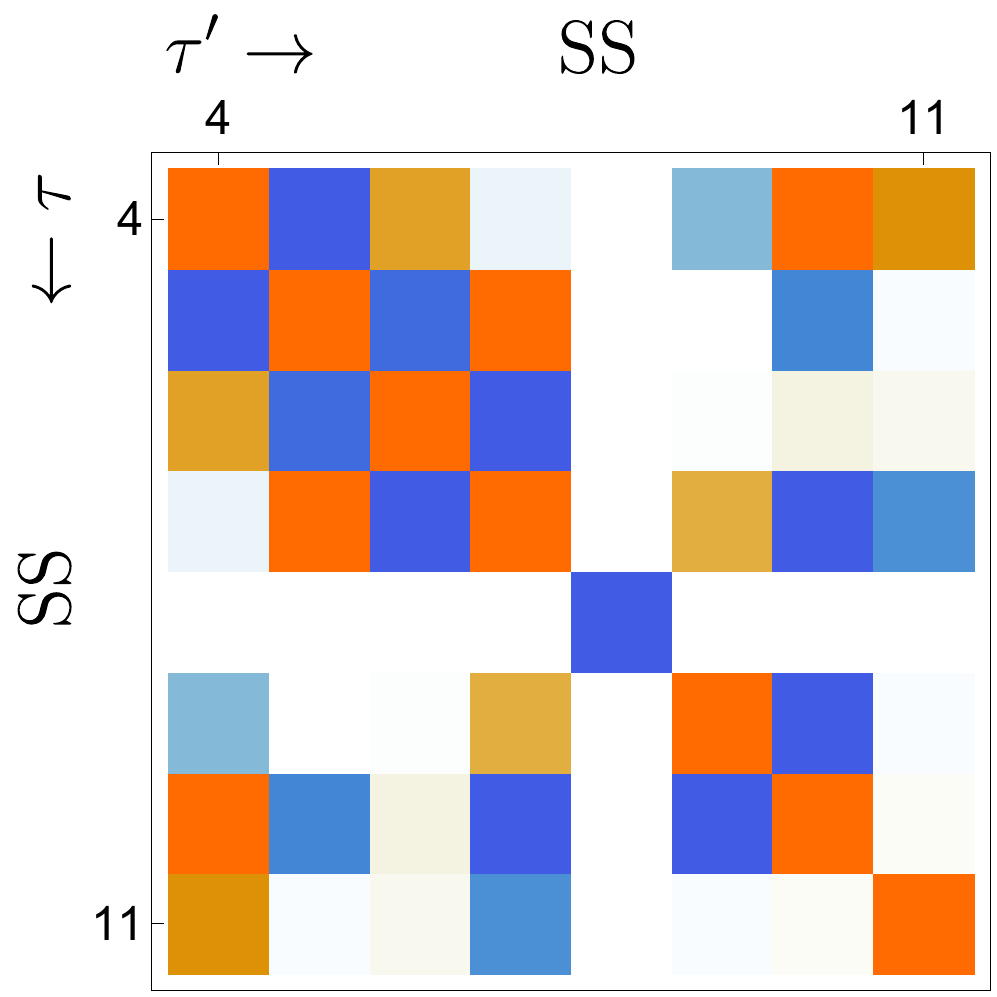} & \includegraphics[width=3.2cm]{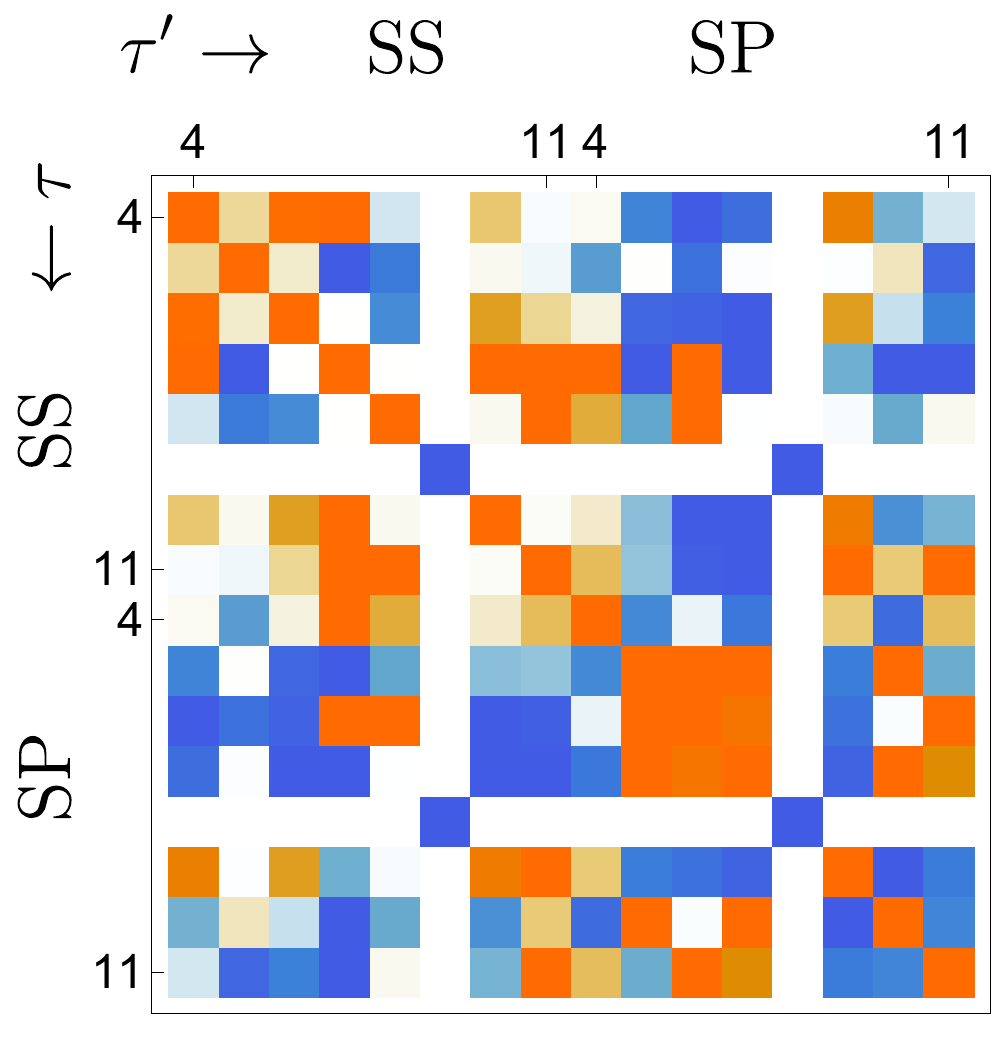} & \includegraphics[width=3.2cm]{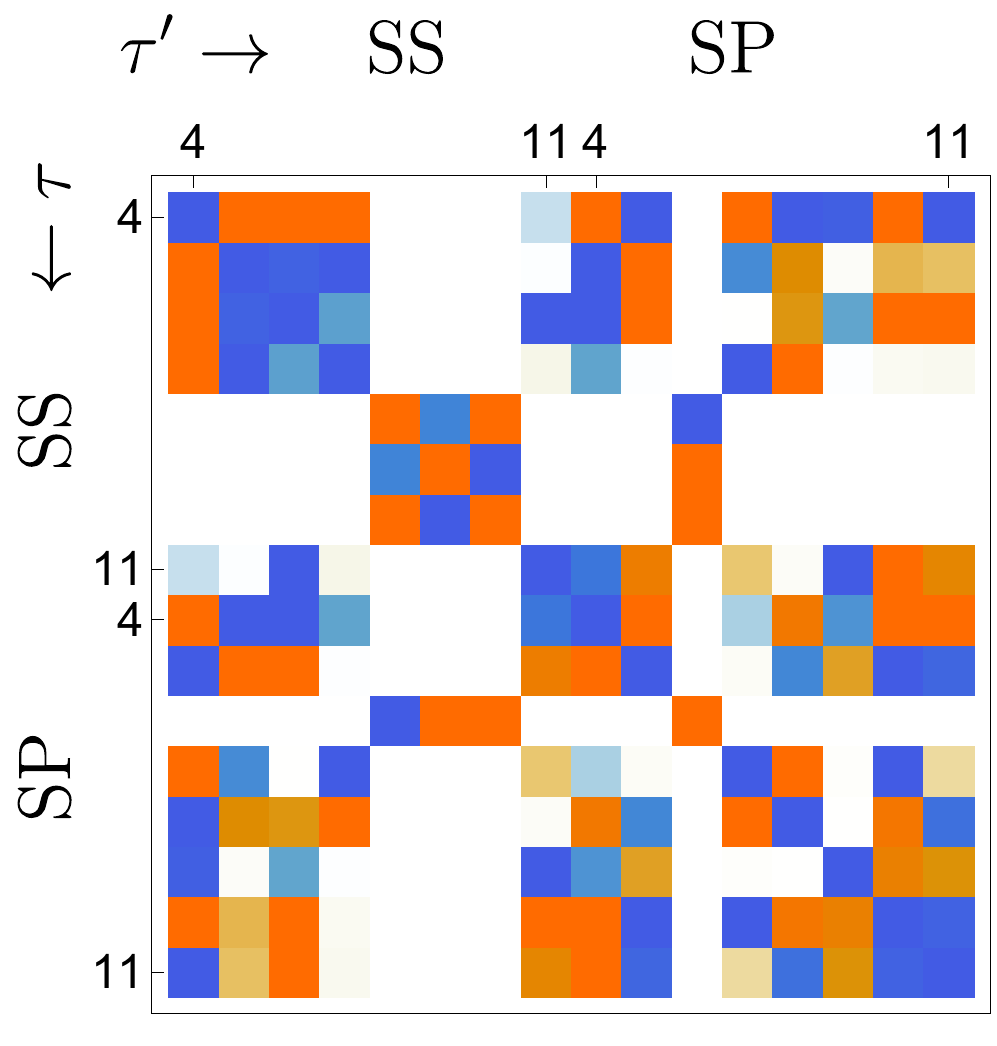} & \\
\multirow{2}{5ex}{\rotatebox{90}{$L=32$}} & \rotatebox{90}{Mean} & \includegraphics[width=3.2cm]{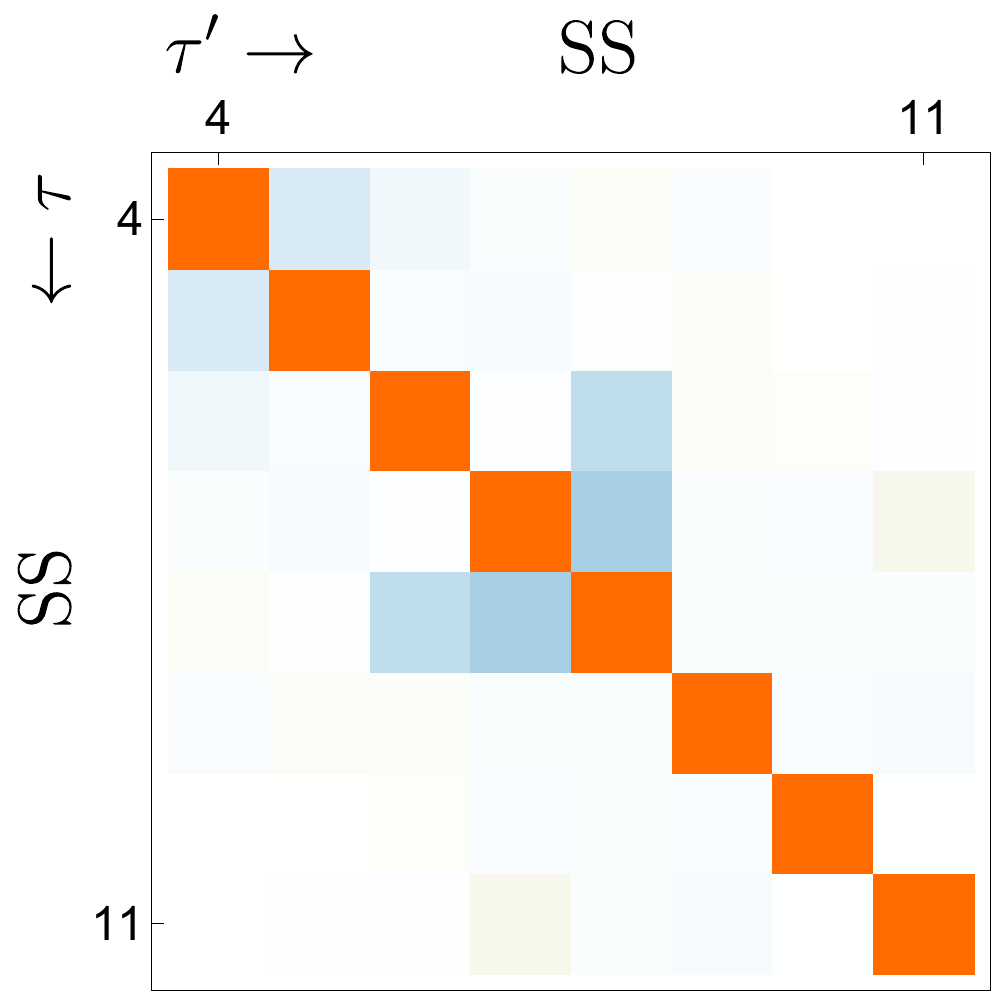} & \includegraphics[width=3.2cm]{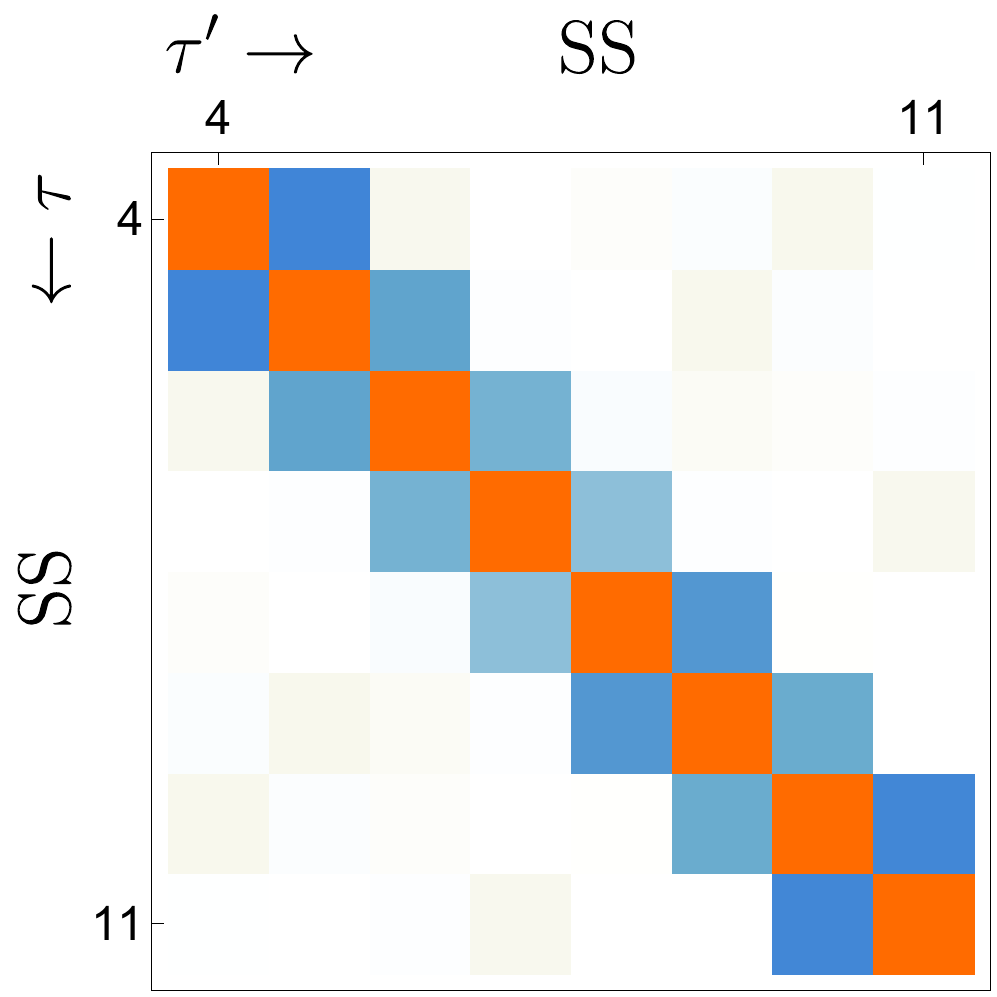} & \includegraphics[width=3.2cm]{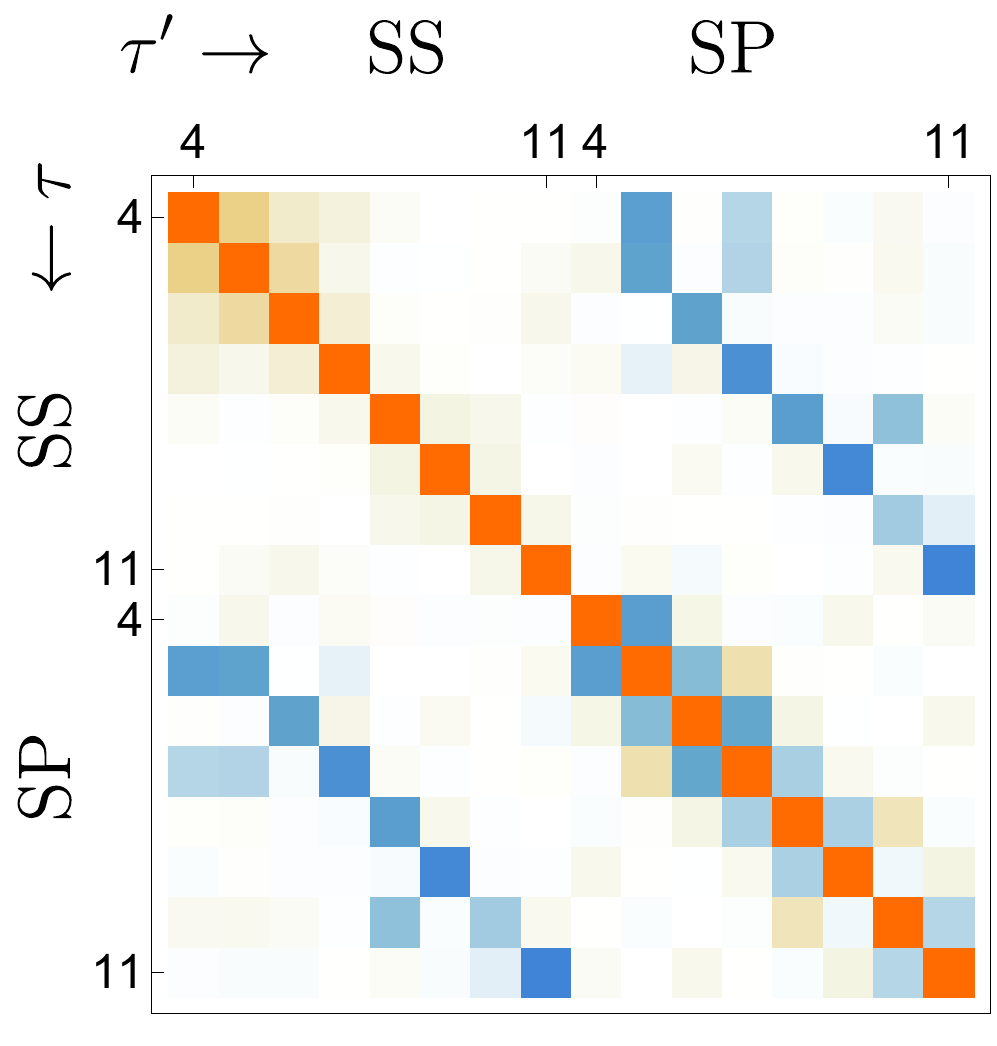} & \includegraphics[width=3.2cm]{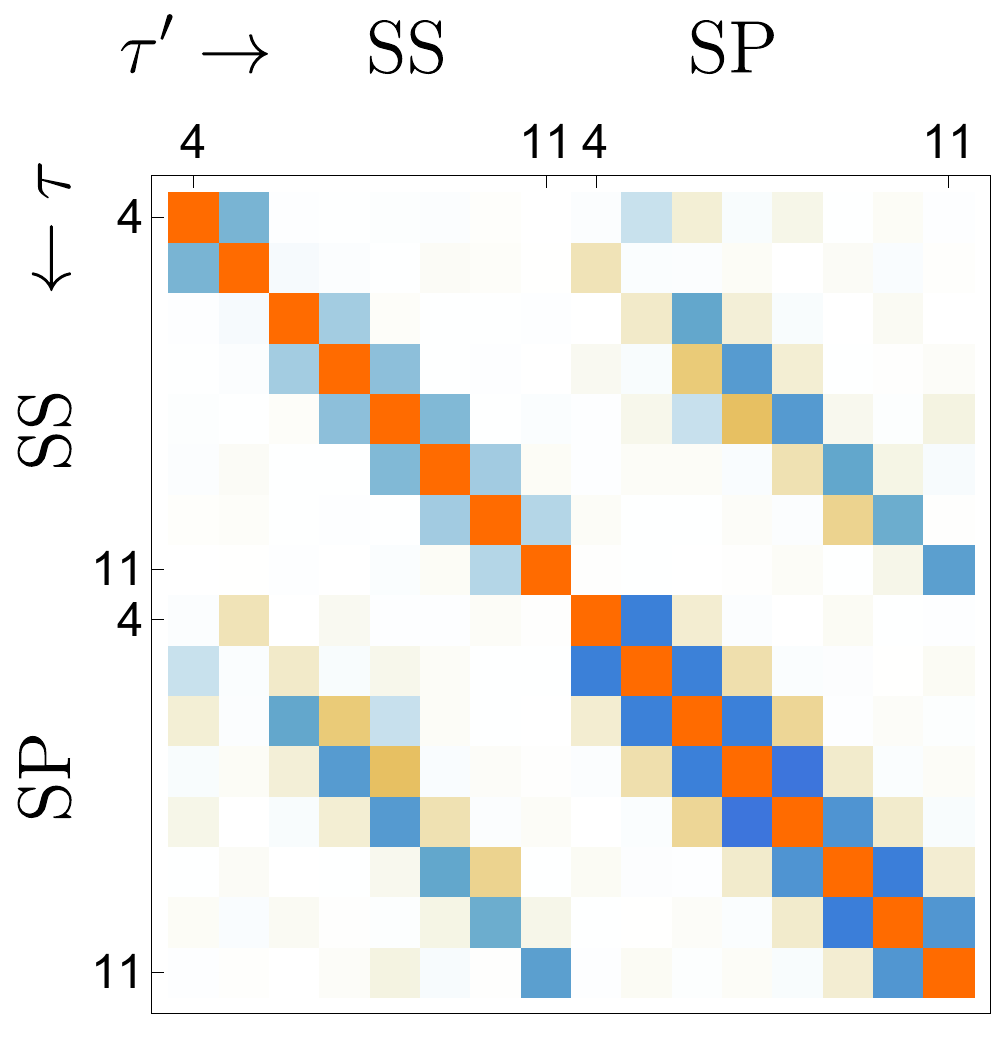} & \multirow{2}{*}[3ex]{\includegraphics[width=0.6cm]{figures/appendix-B/covmatrices/legend.pdf}}\\ 
& \rotatebox{90}{HL estimator} & \includegraphics[width=3.2cm]{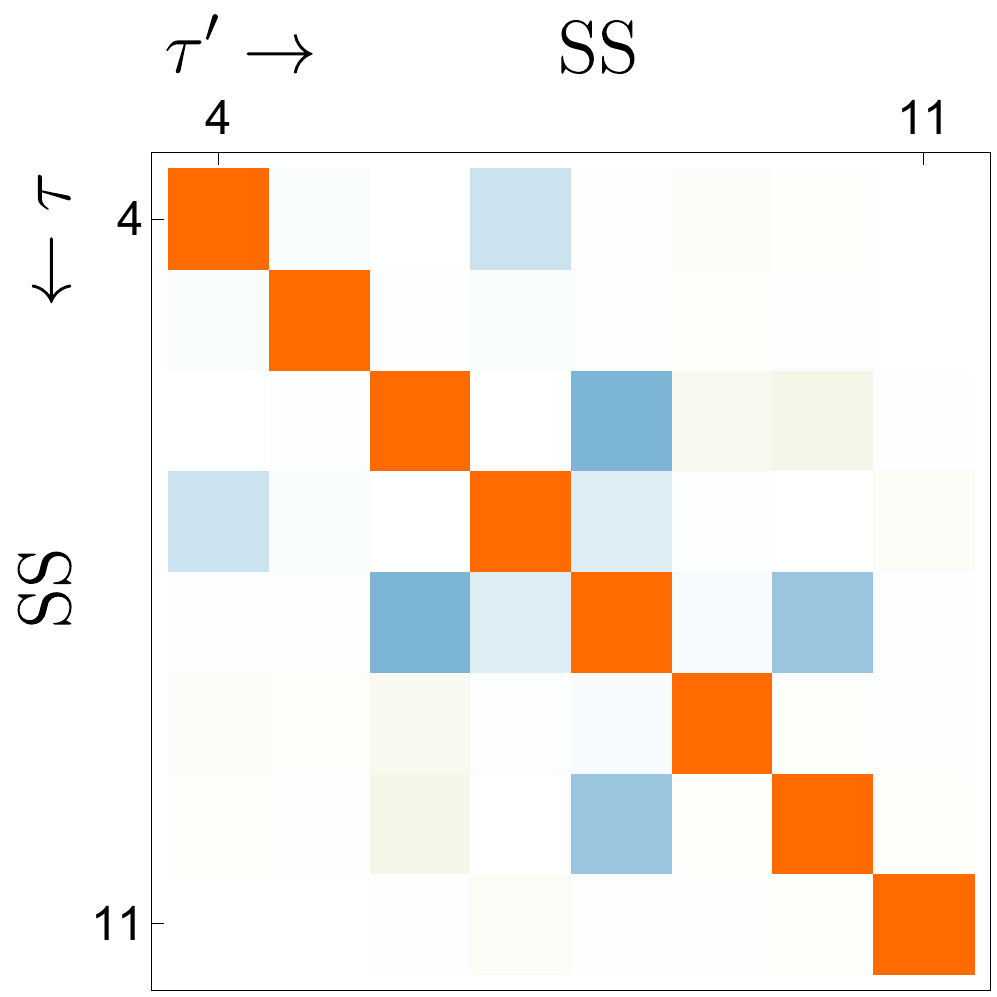} & \includegraphics[width=3.2cm]{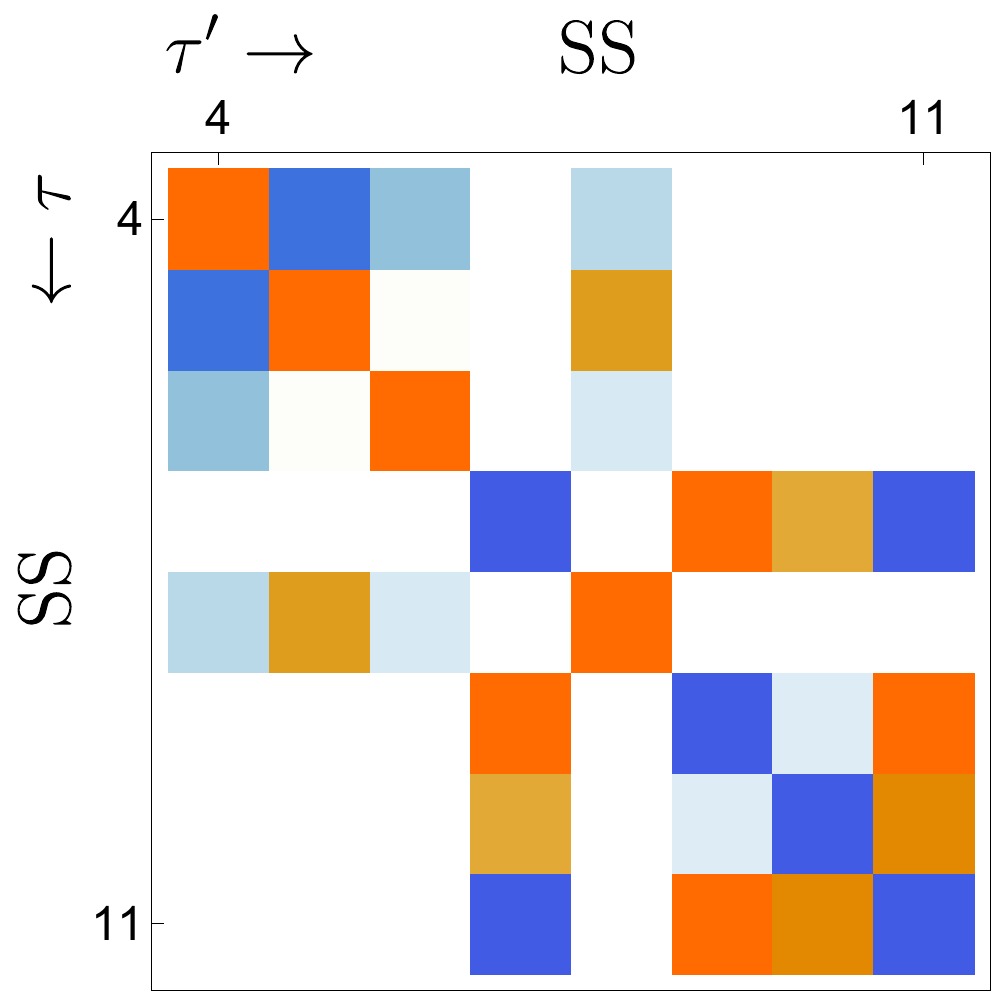} & \includegraphics[width=3.2cm]{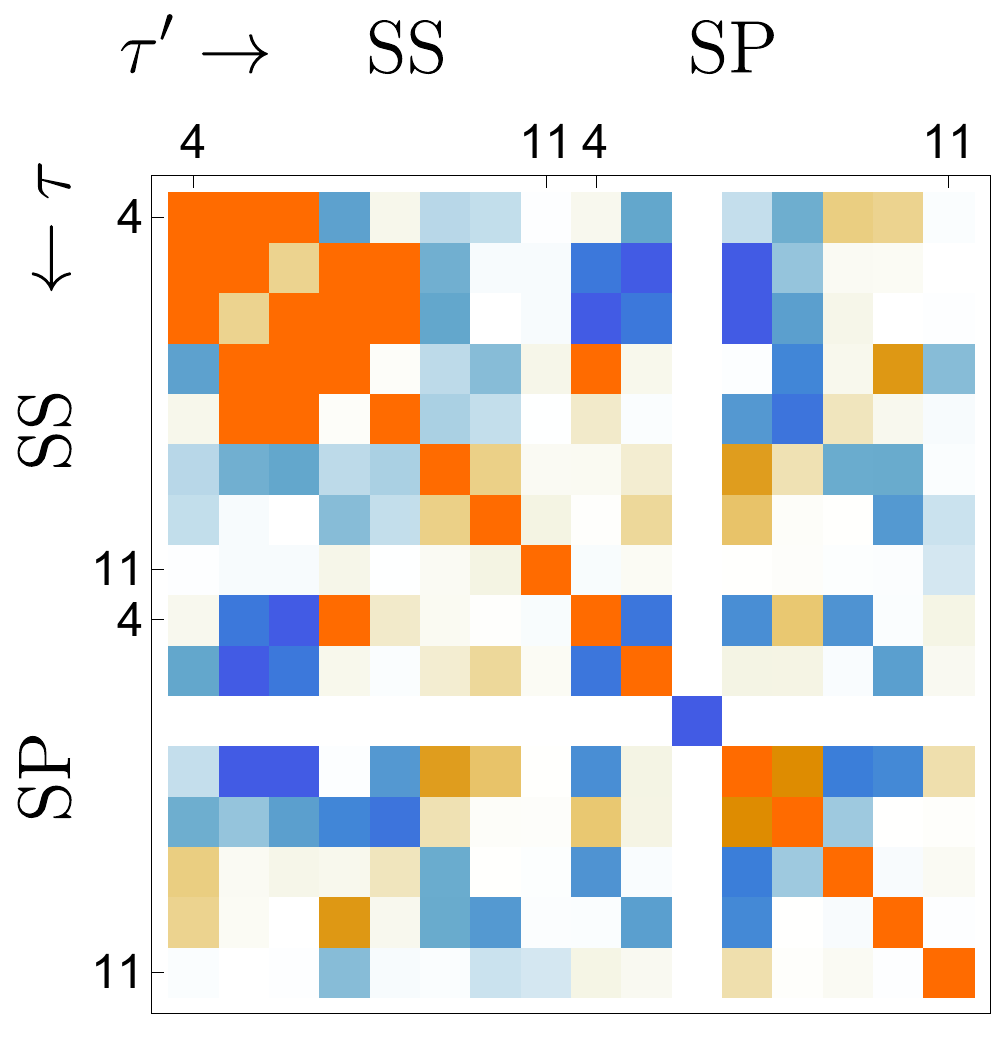} & \includegraphics[width=3.2cm]{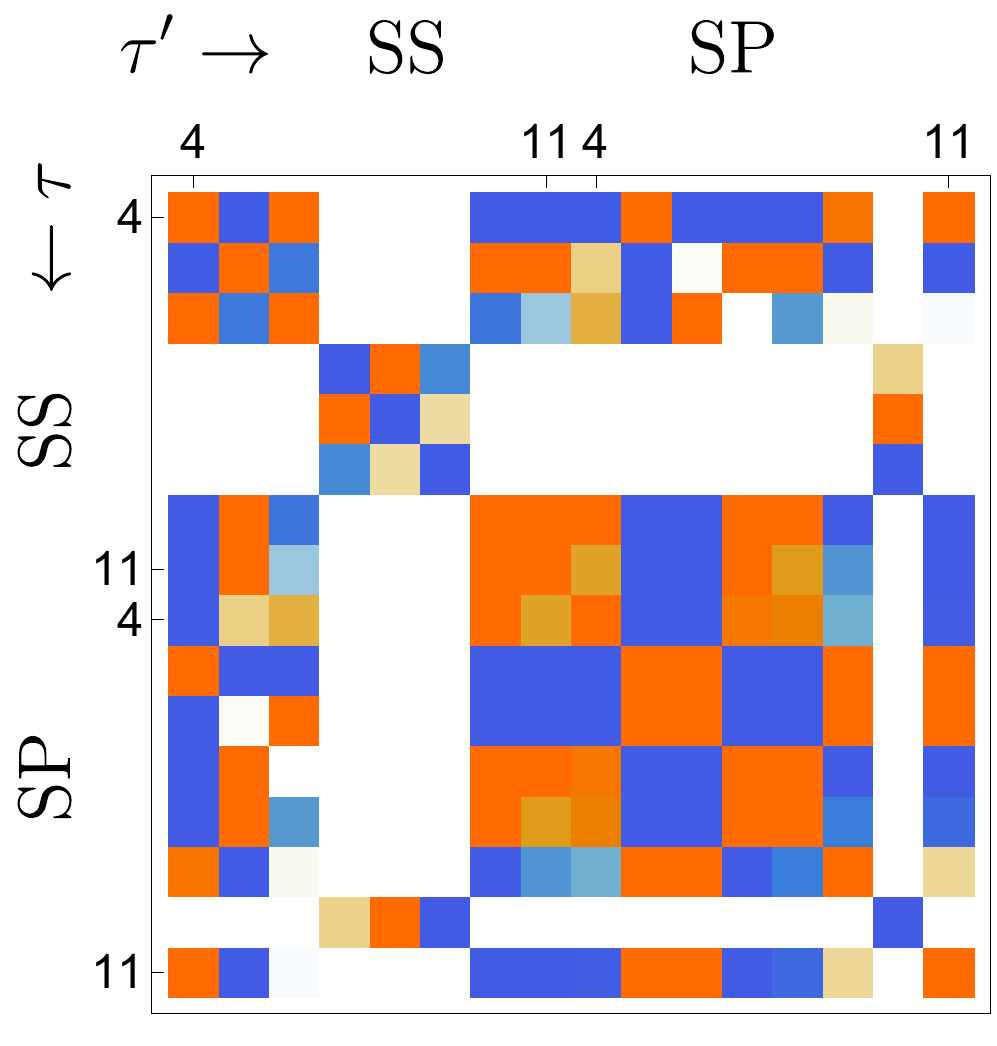} &
\end{tabular}
\caption{Normalized inverse covariance matrices computed for the $NN\, (\1s0)$ ground state with $L=24$ (top) and $L=32$ (bottom) for $\tau\in [4,11]$ l.u.\ using the mean and HL estimators applied to the effective-energy function and correlation function. }
\label{fig:covmatrices}
\end{figure}
Next, the implications of using the HL estimator (instead of the mean) on the individual SP and SS correlation functions are analyzed. When correlations are fully taken into account, the covariance matrix associated with the HL estimator is computed with the Median Absolute Deviation (MAD):
\begin{equation}
\mathcal{C}(\tau,\tau')=\text{Median}\left[(\tilde{C}(\tau)-\text{Median}[\tilde{C}(\tau)])(\tilde{C}(\tau')-\text{Median}[\tilde{C}(\tau')])\right],
\end{equation}
where $\tilde{C}(\tau)$ is the bootstrap ensemble computed with the HL estimator of the original correlation function $C(\tau)$. However, in some cases the resulting covariance matrix is found not to be positive semi-definite, and it only becomes well behaved when a single type of correlation function is used (or a linear combination of several) in the form of an effective-(mass) energy function.
To illustrate this, Fig.~\ref{fig:covmatrices} shows the normalized inverse covariance matrix, $\mathcal{C}^{-1}(\tau,\tau')/\sqrt{\mathcal{C}^{-1}(\tau,\tau)\mathcal{C}^{-1}(\tau',\tau')}$, for the $NN\, (\1s0)$ ground state with $L=24$ and $L=32$ for all possible choices, i.e., HL estimator versus mean and correlation function versus the effective-energy function.

\begin{figure}[t!]
\includegraphics[width=0.9\textwidth]{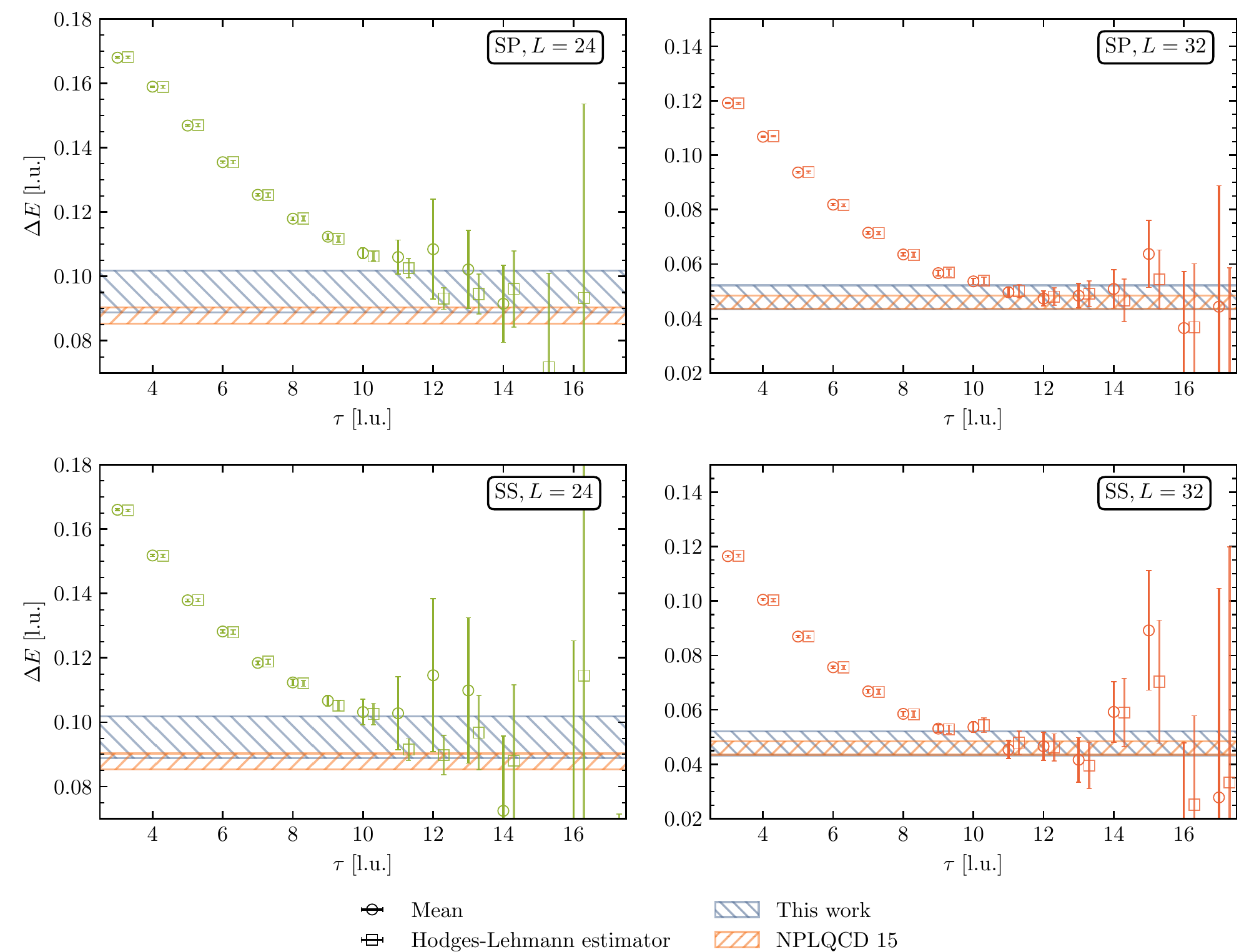}
\caption{Comparison of the effective energy-shift plots of the SP and SS correlation functions for the $NN\; (\1s0)$ $L=24$ (left panel) and $L=32$ (right panel) first excited states computed using the mean (dark green/red circles) and the HL estimator (light green/red squares, shifted horizontally for clarity). The bands show the results of this work and of Ref.~\cite{Orginos:2015aya}, labeled as NPLQCD 15.}
\label{fig:nn1s0EMP_24vs32}
\end{figure}

Therefore, in order to incorporate the HL estimator into the fitting strategy used here, only the fully uncorrelated covariance matrix can be used, and this leads to results which are compatible with the ones presented here using the mean.
In Fig.~\ref{fig:nn1s0EMP_24vs32}, the effective-energy functions computed with the mean and HL are compared for the $NN\; (\1s0)$ first excited states, showing agreement within uncertainties.

\begin{figure}[tb]
\includegraphics[width=0.93\textwidth]{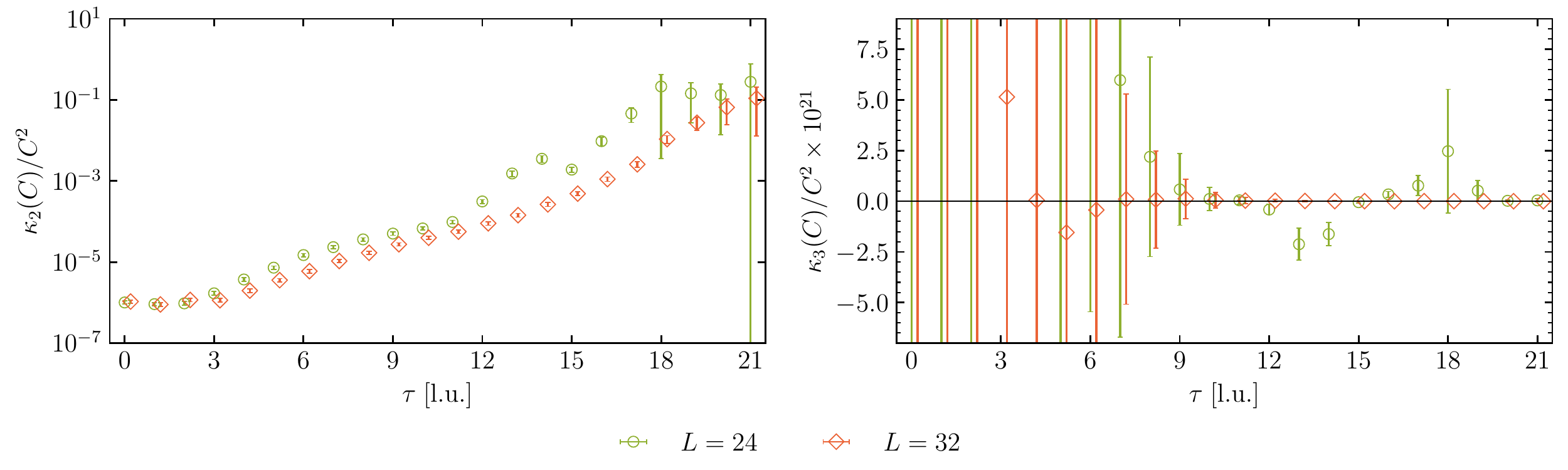}
\caption{The bootstrap estimates of the variance $\kappa_2(C(\tau))/C^2(\tau)$ and skewness $\kappa_3(C(\tau))/C^2(\tau)$ for the SS correlation functions corresponding to the $NN\; (\1s0)$ first excited state with $L=24$ (green circles) and $L=32$ (red diamonds). The $L=32$ points have been shifted slightly along the $\tau$ axis for clarity.}
\label{fig:var_skew}
\end{figure}

To understand the ill-behaved behavior of some of the HL correlation functions, it is important to recall that baryonic correlation functions exhibit distributions that are largely non-Gaussian with heavy tails, and the mean becomes Gaussian only in the limit of large statistics.
However, at late times, the signal-to-noise degradation worsens, and outliers occur more frequently in the distribution. For the $L=32$ and $L=48$ cases, the point at which the HL estimator gives different results compared with the usual estimator (mean and standard deviation), which would indicate a deviation from Gaussian behavior, occurs at a much later time compared with the maximum time included in the fits using the automated fitter of this work.
For the $L=24$ case, the data are more noisy than on the other two ensembles, showing non-Gaussianity at earlier times. To illustrate the different behavior between the $L=24$ and $L=32$ ensembles, the second and third cumulants of $C(\tau)$, defined as
\begin{equation}
\kappa_n(C(\tau))=\left\langle C(\tau)^n \right\rangle -\sum_{m=1}^{n-1} \binom{n-1}{m-1} \kappa_m(C(\tau)) \left\langle C(\tau)^{n-m} \right\rangle,
\end{equation}
with $n\in\{2,3\}$, respectively, are shown in Fig.~\ref{fig:var_skew} for the two ensembles in the case of the $NN~(\1s0)$ first excited state.
Looking at the second cumulant (variance), $\kappa_2$, it is clear that $L=24$ is more noisy than $L=32$, and looking at the third cumulant (skewness), $\kappa_3$, it is clear that $L=24$ deviates from zero, an indication of the non-Gaussian behavior. The use of robust estimators is, therefore, questionable in this case. This is the main reason for abandoning the use of the HL estimator in the analysis of correlation functions in the present study.
%

%%%%%%%%%%%%%%%%%%%%%%%%%%%%%%%%%%%%%%%%%%%
\subsection{Differences in the scattering parameters}
\begin{figure}[t!]
\includegraphics[width=0.86\textwidth]{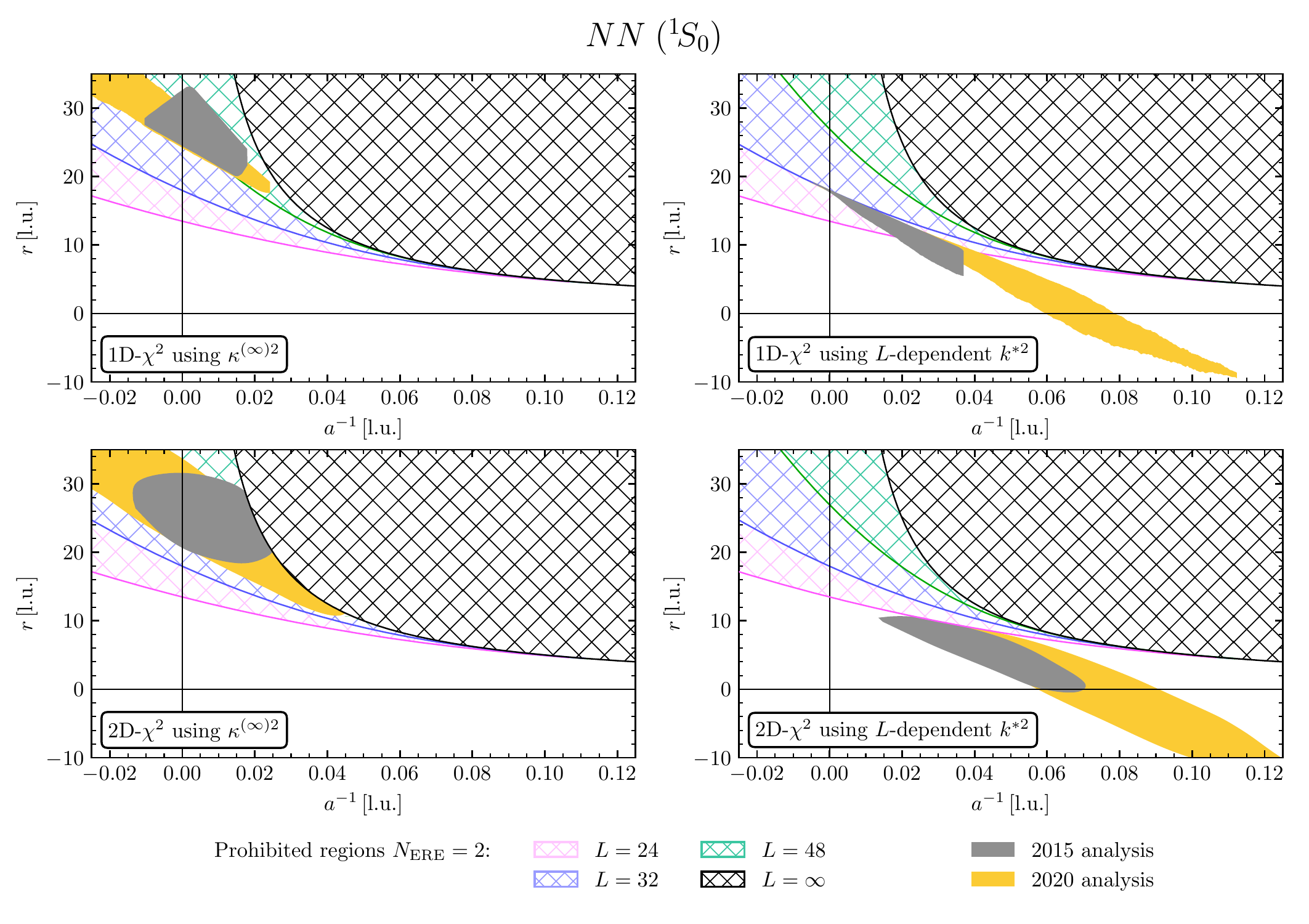}
\includegraphics[width=0.86\textwidth]{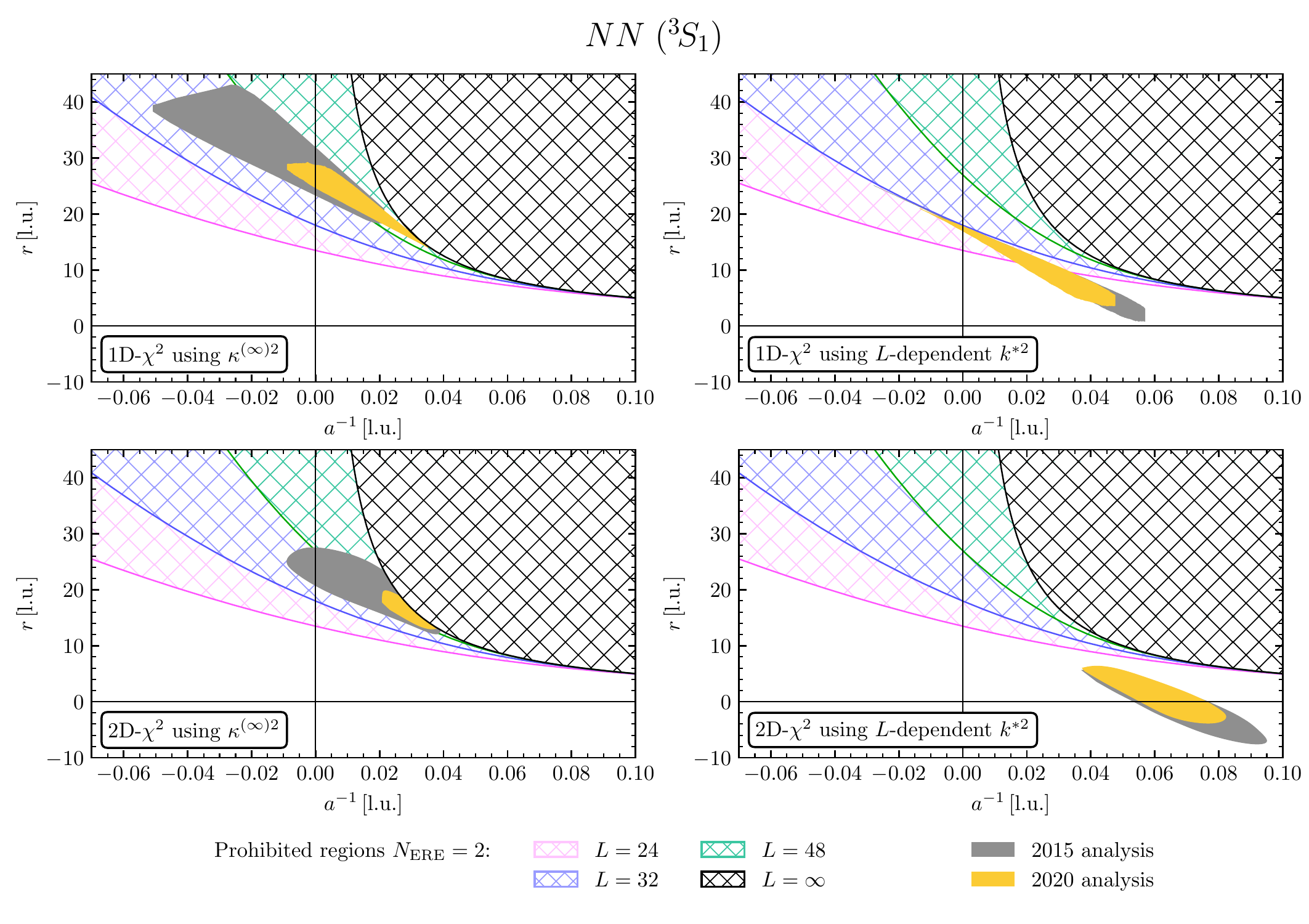}
\caption{Comparison of the 68\% confidence region of the scattering parameters obtained using the energy levels extracted in this work (yellow area) and from Ref.~\cite{Orginos:2015aya} (gray area) with four different analyses. The regions include both statistical and systematic uncertainties combined in quadrature. The prohibited regions where the two-parameter ERE does not cross the $\mathcal{Z}$-function are the crossed areas. Quantities are expressed in lattice units.}
\label{fig:20152020_EREcompar_tests}
\end{figure}

\begin{figure}[t!]
\includegraphics[width=0.9\textwidth]{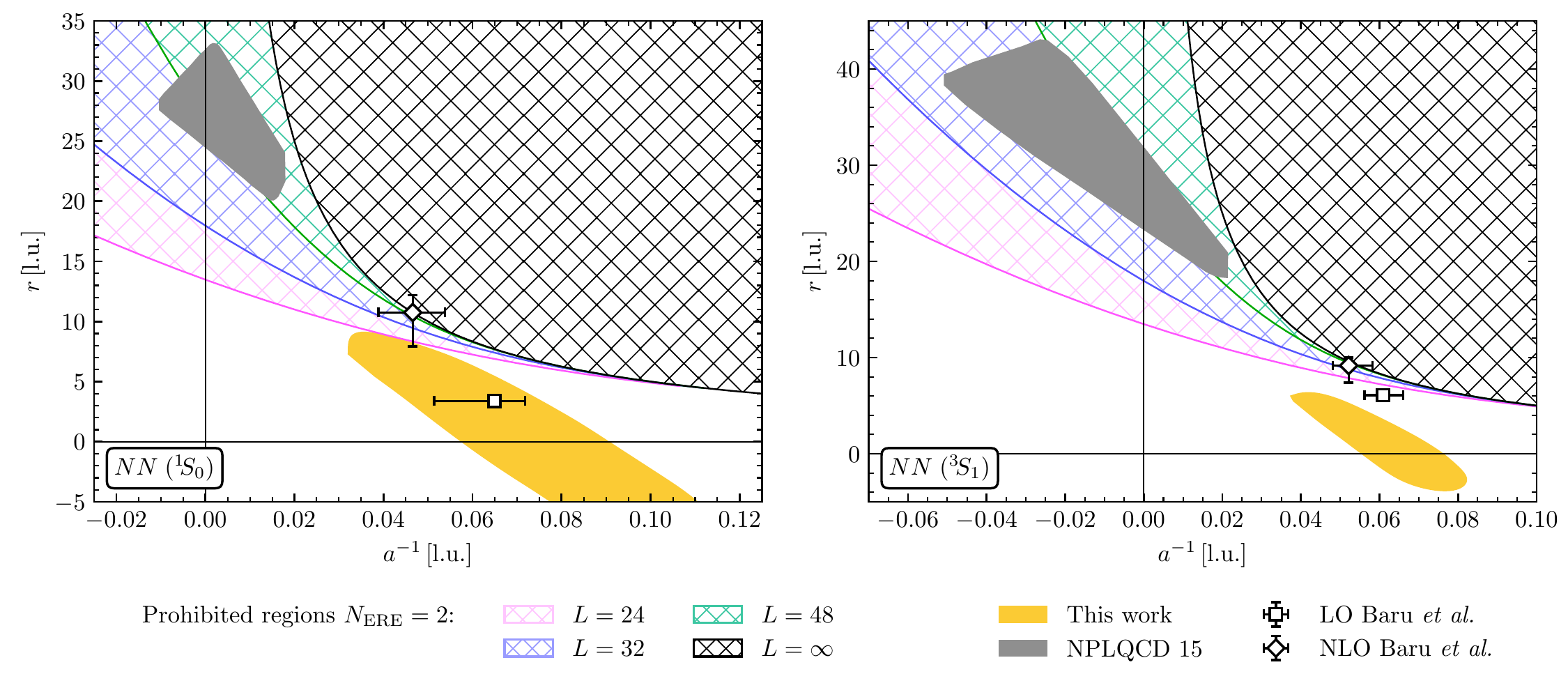}
\caption{Comparison of the 68\% confidence region of the scattering parameters obtained in this work (yellow area), from Ref.~\cite{Orginos:2015aya} (gray area, labeled as NPLQCD 15), and predictions of low-energy theorems from Ref.~\cite{Baru:2016evv} (LO and NLO results). The regions include both statistical and systematic uncertainties combined in quadrature. The prohibited regions where the two-parameter ERE does not cross the $\mathcal{Z}$-functions at given volumes or in the infinite-volume limit are denoted as hashed areas. Quantities are expressed in lattice units.}
\label{fig:20152020Baru_EREcompar}
\end{figure}

The 68\% confidence region of the scattering parameters from a two-parameter ERE extracted in this work and in Ref.~\cite{Orginos:2015aya} are shown in Fig.~\ref{fig:20152020Baru_EREcompar}.
It can be seen that the values of the parameters obtained in the two analyses do not fully agree at the $1\sigma$ level, although the uncertainties are rather large.
There are two significant differences between the two analyses: (1) the use of the new definition for the $\chi^2$ function (2D-$\chi^2$) in the present work, as opposed to the usual $\chi^2$ function (1D-$\chi^2$) used in Ref.~\cite{Orginos:2015aya}, and (2) the use of the $L$-dependent ground-state $k^{*2}$ values in the fits to ERE in the present work, instead of using only the infinite-volume extrapolated value, $\kappa^{(\infty)}$, used in Ref.~\cite{Orginos:2015aya}. To see the effects of each, a comprehensive analysis has been performed, the results of which are shown in Fig.~\ref{fig:20152020_EREcompar_tests}. Here, four different possibilities, corresponding to the types of the $\chi^2$ function (1D or 2D) and the use of ground-state $k^{*2}$ data ($L$-dependent or extrapolated), are tested using the lowest-lying spectra obtained in Ref.~\cite{Orginos:2015aya} and those in the present work.

From these tests, several interesting features are observed. First, the use of the 1D-$\chi^2$, either with the $L$-dependent $k^{*2}$ or the extrapolated one, is insensitive to the conditions imposed by L\"uscher's quantization condition, and as a result, the confidence regions of the scattering parameters could lie on top of the prohibited regions. This is because the distance minimized in the 1D-$\chi^2$ is the vertical one, and not the one along the $\mathcal{Z}$-function, so the ERE is not forced to cross it. Second, when the 2D-$\chi^2$ is used with the extrapolated $k^{*2}$ value, $\kappa^{(\infty)2}$, the only region that is avoided is the one corresponding to $L=\infty$ in the figures, which is expected: with the value of the pole position given by Eq.~\eqref{eq:kinf3}, the function $k^*\cot\delta|_{k^* = i \kappa^{(\infty)}}$ equals $-\sqrt{-k^{*2}}$ and the ERE crosses the $-\sqrt{-k^{*2}}$ function, imposing the $r/a<1/2$ constraint on the scattering parameters. Third, it is reassuring that the regions obtained using the two different energy inputs, from this work or from Ref.~\cite{Orginos:2015aya}, are always overlapping.

Perhaps the most significant observation is that the choice of including the points in the negative $k^{*2}$ region in the fit, i.e., the infinite-volume extrapolated value of the momenta versus the $L$-dependent values, has far more impact on the differences observed than which $\chi^2$ function is used. What the new $\chi^2$ function does is to move the scattering parameters to the allowed region by the $\mathcal{Z}$-functions. Furthermore, with the new fitting methodology, several questions raised about the validity of the ERE fits are addressed, as was presented in Appendix~\ref{sec:appen-checks}. An important one is that the updated results of this work recover the position of the bound state pole obtained via the infinite-volume extrapolation of the energies, and do not yield a second pole near threshold, which would be incompatible with the use of the ERE. As a final remark, it should be noted that the data fitted to extract these parameters are highly non-Gaussian, as can be seen from the correlation between $k^{*2}$ and $k\cot\delta$ in Fig.~\ref{fig:kcotdelta}, and exhibit large uncertainties. This can be compared with the results of Refs.~\cite{Beane:2013br,Wagman:2017tmp} at $m_{\pi}\sim 806$ MeV, where more finite-volume energy eigenvalues, with better precision, could be used in the ERE fitting. As a result, it has been verified that either the $L$-dependent or the infinite-volume extrapolated value of $k^{*2}$ in the ERE fitting gives compatible scattering parameters.

In Ref.~\cite{Baru:2016evv}, low-energy theorems~\cite{Baru:2015ira} were used to compute the scattering parameters from the binding energies of the $NN$ systems obtained in Ref.~\cite{Orginos:2015aya}, and it was pointed out that there were some tensions with the scattering parameters obtained from the LQCD data using Lüscher's method, i.e., those reported in Ref.~\cite{Orginos:2015aya}. Since the binding energies obtained in this work are in full agreement with those obtained in Ref.~\cite{Orginos:2015aya}, the results obtained in Ref.~\cite{Baru:2016evv} can be compared with the updated scattering parameters of this work. As is depicted in Fig.~\ref{fig:20152020Baru_EREcompar}, the tension has reduced considerably. For the two-parameter ERE results, the scattering length is now completely consistent with the low-energy theorem predictions, at both LO and NLO. For the effective range, since the NLO predictions of the low-energy theorems enter the prohibited region for the two-parameter ERE, the comparison may only be made with the LO results. As is seen, for both the $\1s0$ and $\3s1$ channels, the effective ranges are also in agreement (with the $\1s0$ state having a better overlap).

\section{ On leading flavor-symmetry breaking coefficients in the EFT \label{sec:appen-EFT}}
\noindent
Table 10 of Ref.~\cite{Petschauer:2013uua} lists the $SU(3)$ flavor-symmetry-breaking LECs $c^i_{\chi}$ for all of the two-(octet) baryon channels. These coefficients are a combination of different terms in the Lagrangian shown in Table 9 of the same reference (terms 29-40). The relations between the $c^i_{\chi}$ from Ref.~\cite{Petschauer:2013uua} and the ones in Eq.~\eqref{eq:LagNLO2} introduced in the present work are presented in Table~\ref{tab:coeff_petshauervsme}. Instead of the $(^{2s+1} L_J,I)$ notation, the channels are labeled as $(^{2I+1}_{2s+1})$ for brevity, as $L=0$ in all cases.

In Table~\ref{tab:full_eft_coeff}, a list of the two-baryon channels one needs to study in order to obtain independently all the LECs of this work is provided. There are 6 LO and 12 NLO symmetry-breaking coefficients that are referred to as momentum independent in this paper, as well as 6 NLO momentum-dependent coefficients, making a total of 24 parameters that need to be constrained in a more exhaustive study in the future. For the momentum-independent coefficients, the choice of the systems is not unique, as there are 37 different channels that can be used to constrain only 18 parameters (assuming $SU(2)$ flavor symmetry and no electromagnetic interaction). For the momentum-dependent coefficients, no extra channels are needed besides those used for the momentum-independent coefficients. For simplicity, only channels that do not change the baryon content are used (e.g., $\Sigma N \rightarrow \Sigma N$, denoted as $\Sigma N$ in short).
\begin{table}[h!]
\caption{Comparison between the symmetry-breaking LECs of this work and those in Ref.~\cite{Petschauer:2013uua} for the two-baryon channels for which only one $c^i_{\chi}$ appears in that reference.}
\label{tab:coeff_petshauervsme}
\centering
\renewcommand{\arraystretch}{1.6}
\begin{ruledtabular}
\begin{tabular}{ccc}
Channel $(^{2I+1}_{2s+1})$	&	Ref.~\cite{Petschauer:2013uua} & Coefficients in Eq.~\eqref{eq:LagNLO2}	\\ \hline
$NN\rightarrow NN \; (^3_1)$ & $\frac{c^1_{\chi}}{2}$ & $4(c^{\chi}_3 - c^{\chi}_4)$ \\
$\Lambda N\rightarrow \Lambda N \; (^2_1)$	&	$c^2_{\chi}$	&	$\frac{1}{3}(4c^{\chi}_1-4c^{\chi}_2+9c^{\chi}_3-9c^{\chi}_4-4c^{\chi}_5+4c^{\chi}_6-c^{\chi}_9+c^{\chi}_{10}+4c^{\chi}_{11}-4c^{\chi}_{12})$ 	\\
$\Lambda N\rightarrow \Sigma N	 \; (^2_1)$	&	$-c^3_{\chi}$	&	$c^{\chi}_3-c^{\chi}_4+2c^{\chi}_5-2c^{\chi}_6+c^{\chi}_9-c^{\chi}_{10}$	\\
$\Sigma N\rightarrow \Sigma N \; (^2_1)$	&	$c^4_{\chi}$	&	$-c^{\chi}_3+c^{\chi}_4-3c^{\chi}_9+3c^{\chi}_{10}$	\\
$\Lambda \Lambda\rightarrow \Lambda \Lambda \; (^1_1)$	&	$\frac{c^5_{\chi}}{2}$	& $\frac{8}{9}(2c^{\chi}_1-2c^{\chi}_2+2c^{\chi}_3-2c^{\chi}_4-4c^{\chi}_5+4c^{\chi}_6-2c^{\chi}_7+2c^{\chi}_8-2c^{\chi}_9+2c^{\chi}_{10}+3c^{\chi}_{11}-3c^{\chi}_{12})$	\\
$\Xi N\rightarrow \Xi N \; (^3_1)$	&	$c^6_{\chi}$	&	$2(-2c^{\chi}_5+2c^{\chi}_6+c^{\chi}_{11}-c^{\chi}_{12})$	\\
$NN\rightarrow NN \; (^1_3)$	&	$\frac{c^7_{\chi}}{2}$	&	$4(c^{\chi}_3+ c^{\chi}_4)$	\\
$\Lambda N\rightarrow \Lambda N \; (^2_3)$	&	$c^8_{\chi}$	&	$\frac{1}{3}(4c^{\chi}_1+4c^{\chi}_2+7c^{\chi}_3+7c^{\chi}_4+12c^{\chi}_5+12c^{\chi}_6+9c^{\chi}_9+9c^{\chi}_{10}+4c^{\chi}_{11}+4c^{\chi}_{12})$	\\
$\Lambda N\rightarrow \Sigma N \; (^2_3)$	&	$-c^9_{\chi}$	&	$-c^{\chi}_3-c^{\chi}_4+2c^{\chi}_5+2c^{\chi}_6+3c^{\chi}_9+3c^{\chi}_{10}$ \\
$\Sigma N\rightarrow \Sigma N \; (^2_3)$	&	$c^{10}_{\chi}$	&	$c^{\chi}_3+c^{\chi}_4+3c^{\chi}_9+3c^{\chi}_{10}$ 	\\ 
$\Xi N\rightarrow \Xi N \; (^1_3)$	&	$c^{11}_{\chi}$	&	 $2(2c^{\chi}_5+2c^{\chi}_6+2c^{\chi}_7+2c^{\chi}_8+2c^{\chi}_9+2c^{\chi}_{10}+c^{\chi}_{11}+c^{\chi}_{12})$	\\ 
$\Xi N\rightarrow \Xi N \; (^3_3)$	&	$c^{12}_{\chi}$	&	$2(2c^{\chi}_5+2c^{\chi}_6+c^{\chi}_{11}+c^{\chi}_{12})$
\end{tabular}
\end{ruledtabular}
\renewcommand{\arraystretch}{1}
\end{table}
\begin{table}[!htbp]
\renewcommand{\arraystretch}{1.6}
\caption{Combinations of two-baryon channels necessary to constrain independently all of the LO+NLO EFT LECs introduced in Sec.~\ref{subsec:lecs}.}
\label{tab:full_eft_coeff}
\begin{ruledtabular}
\begin{tabular}{c >{\footnotesize}c c >{\footnotesize}c}
Coefficient & \multicolumn{3}{c}{Channels $(^{2I+1}_{2s+1})$}\\
\hline
$c^{(27)}$ & \multicolumn{3}{ >{\footnotesize}c}{$2 \Xi \Sigma (^4_1)- \Xi\Xi (^3_1)$} \\
$c^{(8_s)}$ & \multicolumn{3}{ >{\footnotesize}c}{$\frac{15}{4} \Lambda \Lambda (^1_1)+\frac{35}{36} \Sigma\Sigma (^1_1)-5\Xi\Lambda (^2_1)-\frac{5}{3}\Xi N (^1_1)+\frac{5}{9}\Xi \Sigma (^2_1)+\frac{1}{3}\Xi \Sigma (^4_1) +\frac{37}{18}\Xi\Xi (^3_1)$} \\
$c^{(1)}$ & \multicolumn{3}{ >{\footnotesize}c}{$-6 \Lambda \Lambda (^1_1)+\frac{10}{9} \Sigma\Sigma (^1_1)+8\Xi\Lambda (^2_1) +\frac{8}{3}\Xi N (^1_1)-\frac{8}{9}\Xi \Sigma (^2_1)-\frac{2}{3}\Xi \Sigma (^4_1) -\frac{29}{9}\Xi\Xi (^3_1)$} \\
$c^{(\overline{10})}$ & \multicolumn{3}{ >{\footnotesize}c}{$\frac{1}{3}NN(^1_3)+\frac{8}{9}\Sigma N (^2_3)+\frac{2}{9}\Sigma N (^4_3)+\frac{4}{3}\Xi\Lambda (^2_3)-\frac{2}{3}\Xi N (^1_3)-\frac{2}{3}\Xi N (^3_3)-\frac{4}{9}\Xi\Sigma (^2_3)+\frac{4}{9}\Xi\Sigma (^4_3)-\frac{4}{9}\Xi\Xi (^1_3)$}\\
$c^{(10)}$ & \multicolumn{3}{ >{\footnotesize}c}{$\frac{1}{3}NN(^1_3)-\frac{4}{9}\Sigma N (^2_3)+\frac{8}{9}\Sigma N (^4_3)-\frac{2}{3}\Xi\Lambda (^2_3)+\frac{1}{3}\Xi N (^1_3)+\frac{1}{3}\Xi N (^3_3)+\frac{2}{9}\Xi\Sigma (^2_3)-\frac{2}{9}\Xi\Sigma (^4_3)+\frac{2}{9}\Xi\Xi (^1_3)$} \\
$c^{(8_a)}$ & \multicolumn{3}{ >{\footnotesize}c}{$-\frac{11}{12}NN(^1_3)+\frac{17}{9}\Sigma N (^2_3)-\frac{7}{9}\Sigma N (^4_3)-\frac{1}{6}\Xi\Lambda (^2_3)-\frac{17}{12}\Xi N (^1_3)+\frac{19}{12}\Xi N (^3_3)+\frac{37}{18}\Xi\Sigma (^2_3)-\frac{5}{9}\Xi\Sigma (^4_3)-\frac{25}{36}\Xi\Xi (^1_3)$} \\[1.5ex]
$c^{\chi}_1$ & \multicolumn{3}{ >{\footnotesize}c}{$\begin{aligned}
&\tfrac{9}{16}\Lambda\Lambda (^1_1)-\tfrac{1}{48}NN(^1_3) -\tfrac{5}{36}\Sigma N (^2_3)-\tfrac{1}{2}\Sigma N (^4_1)+\tfrac{1}{36}\Sigma N (^4_3)+\tfrac{7}{48}\Sigma\Sigma (^1_1)-\tfrac{5}{24}\Xi \Lambda (^2_3)-\tfrac{1}{4}\Xi N (^1_1)\\
&+\tfrac{5}{48}\Xi N (^1_3)-\tfrac{1}{4}\Xi N (^3_1)+\tfrac{5}{48}\Xi N (^3_3)+\tfrac{5}{72}\Xi\Sigma (^2_3)+\tfrac{5}{6}\Xi \Sigma (^4_1)+\tfrac{1}{18}\Xi \Sigma (^4_3)+\tfrac{1}{144}\Xi \Xi (^1_3)-\tfrac{13}{24}\Xi \Xi (^3_1)
\end{aligned}$} \\[3.5ex]
$c^{\chi}_2$ & \multicolumn{3}{ >{\footnotesize}c}{$\begin{aligned}
&-\tfrac{9}{16}\Lambda\Lambda (^1_1)-\tfrac{1}{48}NN(^1_3) -\tfrac{5}{36}\Sigma N (^2_3)+\tfrac{1}{2}\Sigma N (^4_1)+\tfrac{1}{36}\Sigma N (^4_3)-\tfrac{7}{48}\Sigma\Sigma (^1_1)-\tfrac{5}{24}\Xi \Lambda (^2_3)+\tfrac{1}{4}\Xi N (^1_1)\\
&+\tfrac{5}{48}\Xi N (^1_3)+\tfrac{1}{4}\Xi N (^3_1)+\tfrac{5}{48}\Xi N (^3_3)+\tfrac{5}{72}\Xi\Sigma (^2_3)-\tfrac{5}{6}\Xi \Sigma (^4_1)+\tfrac{1}{18}\Xi \Sigma (^4_3)+\tfrac{1}{144}\Xi \Xi (^1_3)+\tfrac{13}{24}\Xi \Xi (^3_1)
\end{aligned}$} \\[3.5ex]
$c^{\chi}_3$ & \multicolumn{3}{ >{\footnotesize}c}{$\begin{aligned}
&\tfrac{1}{12}NN(^1_3)-\tfrac{1}{9}\Sigma N (^2_3)+\tfrac{1}{4}\Sigma N (^4_1)-\tfrac{1}{36}\Sigma N (^4_3)-\tfrac{1}{6}\Xi\Lambda (^2_3)+\tfrac{1}{12}\Xi N (^1_3)\\
&+\tfrac{1}{12}\Xi N (^3_3)+\tfrac{1}{18}\Xi\Sigma (^2_3)-\tfrac{1}{2}\Xi\Sigma (^4_1)-\tfrac{1}{18}\Xi\Sigma (^4_3)+\tfrac{1}{18}\Xi\Xi (^1_3)+\tfrac{1}{4}\Xi\Xi (^3_1)
\end{aligned}$} \\[3.5ex]
$c^{\chi}_4$ & \multicolumn{3}{ >{\footnotesize}c}{$\begin{aligned}
&\tfrac{1}{12}NN(^1_3)-\tfrac{1}{9}\Sigma N (^2_3)-\tfrac{1}{4}\Sigma N (^4_1)-\tfrac{1}{36}\Sigma N (^4_3)-\tfrac{1}{6}\Xi\Lambda (^2_3)+\tfrac{1}{12}\Xi N (^1_3)\\
&+\tfrac{1}{12}\Xi N (^3_3)+\tfrac{1}{18}\Xi\Sigma (^2_3)+\tfrac{1}{2}\Xi\Sigma (^4_1)-\tfrac{1}{18}\Xi\Sigma (^4_3)+\tfrac{1}{18}\Xi\Xi (^1_3)-\tfrac{1}{4}\Xi\Xi (^3_1)
\end{aligned}$} \\[3.5ex]
$c^{\chi}_5$ & \multicolumn{3}{ >{\footnotesize}c}{$\begin{aligned}
&\tfrac{1}{24}NN(^1_3)-\tfrac{1}{18}\Sigma N (^2_3)+\tfrac{1}{4}\Sigma N (^4_1)+\tfrac{1}{36}\Sigma N (^4_3)-\tfrac{3}{8}\Xi\Lambda (^2_1)+\tfrac{1}{24}\Xi\Lambda (^2_3)+\tfrac{1}{24}\Xi N (^1_3)\\
&+\tfrac{1}{24}\Xi N (^3_3)+\tfrac{1}{24}\Xi\Sigma (^2_1)-\tfrac{7}{72}\Xi\Sigma (^2_3)-\tfrac{5}{12}\Xi\Sigma (^4_1)-\tfrac{1}{36}\Xi\Sigma (^4_3)-\tfrac{1}{72}\Xi\Xi (^1_3)+\tfrac{1}{2}\Xi\Xi (^3_1)
\end{aligned}$} \\[3.5ex]
$c^{\chi}_6$ & \multicolumn{3}{ >{\footnotesize}c}{$\begin{aligned}
&\tfrac{1}{24}NN(^1_3)-\tfrac{1}{18}\Sigma N (^2_3)-\tfrac{1}{4}\Sigma N (^4_1)+\tfrac{1}{36}\Sigma N (^4_3)+\tfrac{3}{8}\Xi\Lambda (^2_1)+\tfrac{1}{24}\Xi\Lambda (^2_3)+\tfrac{1}{24}\Xi N (^1_3)\\
&+\tfrac{1}{24}\Xi N (^3_3)-\tfrac{1}{24}\Xi\Sigma (^2_1)-\tfrac{7}{72}\Xi\Sigma (^2_3)+\tfrac{5}{12}\Xi\Sigma (^4_1)-\tfrac{1}{36}\Xi\Sigma (^4_3)-\tfrac{1}{72}\Xi\Xi (^1_3)-\tfrac{1}{2}\Xi\Xi (^3_1)
\end{aligned}$} \\[3.5ex]
$c^{\chi}_7$ & \multicolumn{3}{ >{\footnotesize}c}{$\begin{aligned}
\tfrac{1}{12}&NN(^1_3)-\tfrac{1}{9}\Sigma N (^2_3)+\tfrac{1}{2}\Sigma N (^4_1)+\tfrac{1}{18}\Sigma N (^4_3)-\tfrac{3}{4}\Xi\Lambda (^2_1)+\tfrac{1}{12}\Xi\Lambda (^2_3)+\tfrac{1}{12}\Xi N (^1_3)+\tfrac{1}{4}\Xi N (^3_1)\\
&-\tfrac{1}{6}\Xi N (^3_3)-\tfrac{1}{12}\Xi\Sigma (^2_1)-\tfrac{1}{36}\Xi\Sigma (^2_3)-\tfrac{11}{12}\Xi\Sigma (^4_1)+\tfrac{1}{36}\Xi\Sigma (^4_3)-\tfrac{1}{36}\Xi\Xi (^1_3)+\Xi\Xi (^3_1)
\end{aligned}$} \\[3.5ex]
$c^{\chi}_8$ & \multicolumn{3}{ >{\footnotesize}c}{$\begin{aligned}
\tfrac{1}{12}&NN(^1_3)-\tfrac{1}{9}\Sigma N (^2_3)-\tfrac{1}{2}\Sigma N (^4_1)+\tfrac{1}{18}\Sigma N (^4_3)+\tfrac{3}{4}\Xi\Lambda (^2_1)+\tfrac{1}{12}\Xi\Lambda (^2_3)+\tfrac{1}{12}\Xi N (^1_3)-\tfrac{1}{4}\Xi N (^3_1)\\
&-\tfrac{1}{6}\Xi N (^3_3)+\tfrac{1}{12}\Xi\Sigma (^2_1)-\tfrac{1}{36}\Xi\Sigma (^2_3)+\tfrac{11}{12}\Xi\Sigma (^4_1)+\tfrac{1}{36}\Xi\Sigma (^4_3)-\tfrac{1}{36}\Xi\Xi (^1_3)-\Xi\Xi (^3_1)
\end{aligned}$} \\[3.5ex]
$c^{\chi}_9$ & \multicolumn{3}{ >{\footnotesize}c}{$ \begin{aligned}
-\tfrac{9}{16}&\Lambda\Lambda (^1_1)+\tfrac{1}{48}NN(^1_3)-\tfrac{1}{36}\Sigma N (^2_3)-\tfrac{1}{2}\Sigma N (^4_1)+\tfrac{1}{18}\Sigma N (^4_3)+\tfrac{1}{48}\Sigma\Sigma (^1_1)+\tfrac{3}{2}\Xi \Lambda (^2_1)-\tfrac{1}{24}\Xi \Lambda (^2_3)+\tfrac{1}{8}\Xi N (^1_1)\\
&+\tfrac{7}{48}\Xi N (^1_3)-\tfrac{1}{8}\Xi N (^3_1)-\tfrac{5}{48}\Xi N (^3_3)-\tfrac{11}{72}\Xi\Sigma (^2_3)+\tfrac{5}{6}\Xi \Sigma (^4_1)+\tfrac{1}{36}\Xi \Sigma (^4_3)+\tfrac{11}{144}\Xi \Xi (^1_3)-\tfrac{31}{24}\Xi \Xi (^3_1)
\end{aligned}$} \\[3.5ex]
$c^{\chi}_{10}$ & \multicolumn{3}{ >{\footnotesize}c}{$\begin{aligned}
\tfrac{9}{16}&\Lambda\Lambda (^1_1)+\tfrac{1}{48}NN(^1_3)-\tfrac{1}{36}\Sigma N (^2_3)+\tfrac{1}{2}\Sigma N (^4_1)+\tfrac{1}{18}\Sigma N (^4_3)-\tfrac{1}{48}\Sigma\Sigma (^1_1)-\tfrac{3}{2}\Xi \Lambda (^2_1)-\tfrac{1}{24}\Xi \Lambda (^2_3)-\tfrac{1}{8}\Xi N (^1_1)\\
&+\tfrac{7}{48}\Xi N (^1_3)+\tfrac{1}{8}\Xi N (^3_1)-\tfrac{5}{48}\Xi N (^3_3)-\tfrac{11}{72}\Xi\Sigma (^2_3)-\tfrac{5}{6}\Xi \Sigma (^4_1)+\tfrac{1}{36}\Xi \Sigma (^4_3)+\tfrac{11}{144}\Xi \Xi (^1_3)+\tfrac{31}{24}\Xi \Xi (^3_1)
\end{aligned}$} \\[3.5ex]
$c^{\chi}_{11}$ & \multicolumn{3}{ >{\footnotesize}c}{$\begin{aligned}
&-\tfrac{9}{16}\Lambda\Lambda (^1_1)-\tfrac{1}{16}NN(^1_3)-\tfrac{1}{12}\Sigma N (^2_3)+\tfrac{1}{2}\Sigma N (^4_1)-\tfrac{1}{12}\Sigma N (^4_3)-\tfrac{7}{48}\Sigma\Sigma (^1_1)-\tfrac{1}{8}\Xi \Lambda (^2_3)+\tfrac{1}{4}\Xi N (^1_1)\\
&+\tfrac{1}{16}\Xi N (^1_3)+\tfrac{1}{4}\Xi N (^3_1)+\tfrac{1}{16}\Xi N (^3_3)+\tfrac{1}{24}\Xi\Sigma (^2_3)-\tfrac{13}{12}\Xi \Sigma (^4_1)+\tfrac{1}{12}\Xi \Sigma (^4_3)+\tfrac{5}{48}\Xi \Xi (^1_3)+\tfrac{19}{24}\Xi \Xi (^3_1)
\end{aligned}$} \\[3.5ex]
$c^{\chi}_{12}$ & \multicolumn{3}{ >{\footnotesize}c}{$\begin{aligned}
&\tfrac{9}{16}\Lambda\Lambda (^1_1)-\tfrac{1}{16}NN(^1_3)-\tfrac{1}{12}\Sigma N (^2_3)-\tfrac{1}{2}\Sigma N (^4_1)-\tfrac{1}{12}\Sigma N (^4_3)+\tfrac{7}{48}\Sigma\Sigma (^1_1)-\tfrac{1}{8}\Xi \Lambda (^2_3)-\tfrac{1}{4}\Xi N (^1_1)\\
&+\tfrac{1}{16}\Xi N (^1_3)-\tfrac{1}{4}\Xi N (^3_1)+\tfrac{1}{16}\Xi N (^3_3)+\tfrac{1}{24}\Xi\Sigma (^2_3)+\tfrac{13}{12}\Xi \Sigma (^4_1)+\tfrac{1}{12}\Xi \Sigma (^4_3)+\tfrac{5}{48}\Xi \Xi (^1_3)-\tfrac{19}{24}\Xi \Xi (^3_1)
\end{aligned}$} \\[2.5ex]
\hline
$\tilde{c}^{(27)}$ & $\phantom{\Lambda\Lambda\Lambda\Lambda (^1_1)} \Xi\Xi (^3_1) \phantom{\Lambda\Lambda\Lambda\Lambda (^1_1)}$ & $\tilde{c}^{(\overline{10})}$ & $NN(^1_3)$\\
$\tilde{c}^{(8_s)}$ & $\phantom{\Lambda\Lambda\Lambda\Lambda (^1_1)} \tfrac{1}{3}\Lambda\Lambda (^1_1)+2\Sigma\Sigma (^1_1)-\tfrac{5}{3}\Xi N (^1_1) \phantom{\Lambda\Lambda\Lambda\Lambda (^1_1)}$ & $\tilde{c}^{(10)}$ & $\Xi \Xi (^1_3)$\\
$\tilde{c}^{(1)}$ & $\phantom{\Lambda\Lambda\Lambda\Lambda (^1_1)} -\tfrac{7}{6}\Lambda\Lambda (^1_1)-\tfrac{1}{2}\Sigma\Sigma (^1_1)+\tfrac{8}{3}\Xi N (^1_1) \phantom{\Lambda\Lambda\Lambda\Lambda (^1_1)}$ & $\tilde{c}^{(8_a)}$ & $\Xi N (^1_3)$
\end{tabular}
\end{ruledtabular}
\renewcommand{\arraystretch}{1}
\end{table}
\section{Supplementary figures and tables \label{sec:appen-figtab}}
\begin{figure}[hbt!]
\includegraphics[width=\textwidth]{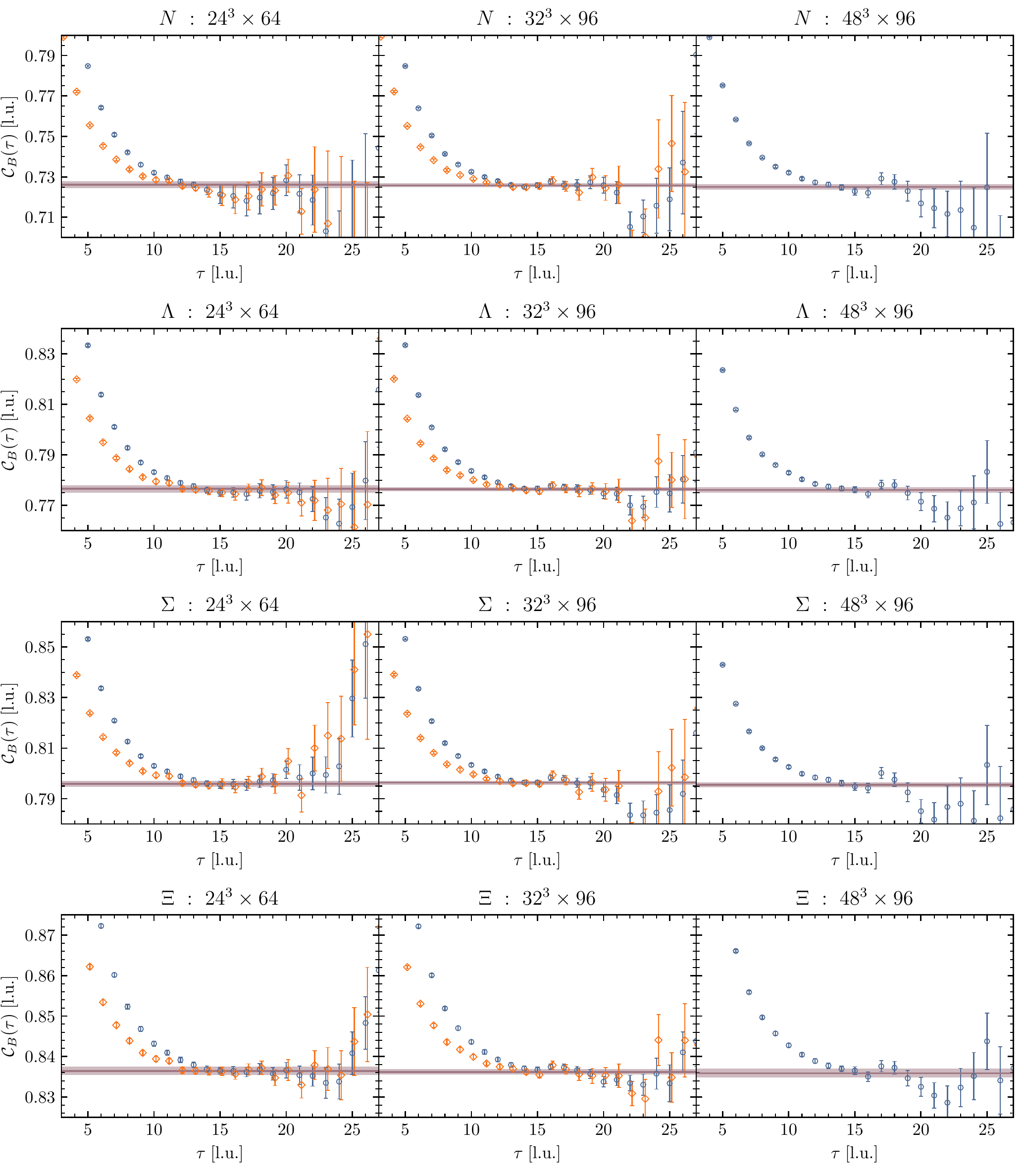}
\caption{Single-baryon EMPs for the SP (blue squares) and SS (orange diamonds) source-sink combinations. The SS points have been slightly shifted along the horizontal axis for clarity.}
\label{fig:B1_EMP}
\end{figure}

\begin{figure}[hbt!]
\includegraphics[width=0.95\textwidth]{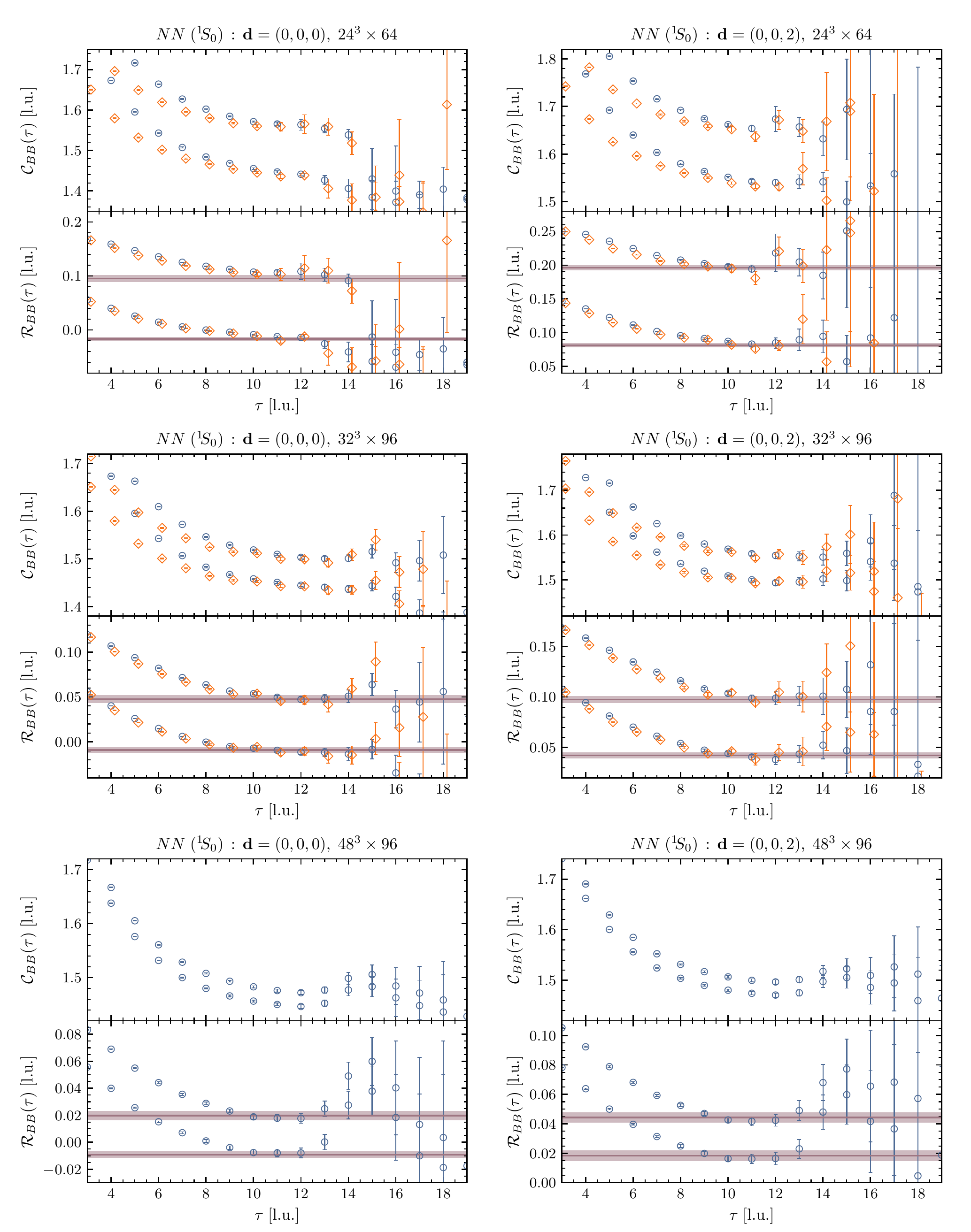}
\caption{The effective energy plots (upper panel of each segment) and the effective energy-shift plots (lower panel of each segment) for the $NN\; (\1s0)$ system at rest (left panels) and with boost $\bm{d}=(0,0,2)$ (right panels) for the SP (blue circles) and SS (orange diamonds) source-sink combinations.}
\label{fig:NN1s0_EMP}
\end{figure}

\begin{figure}[hbt!]
\includegraphics[width=0.95\textwidth]{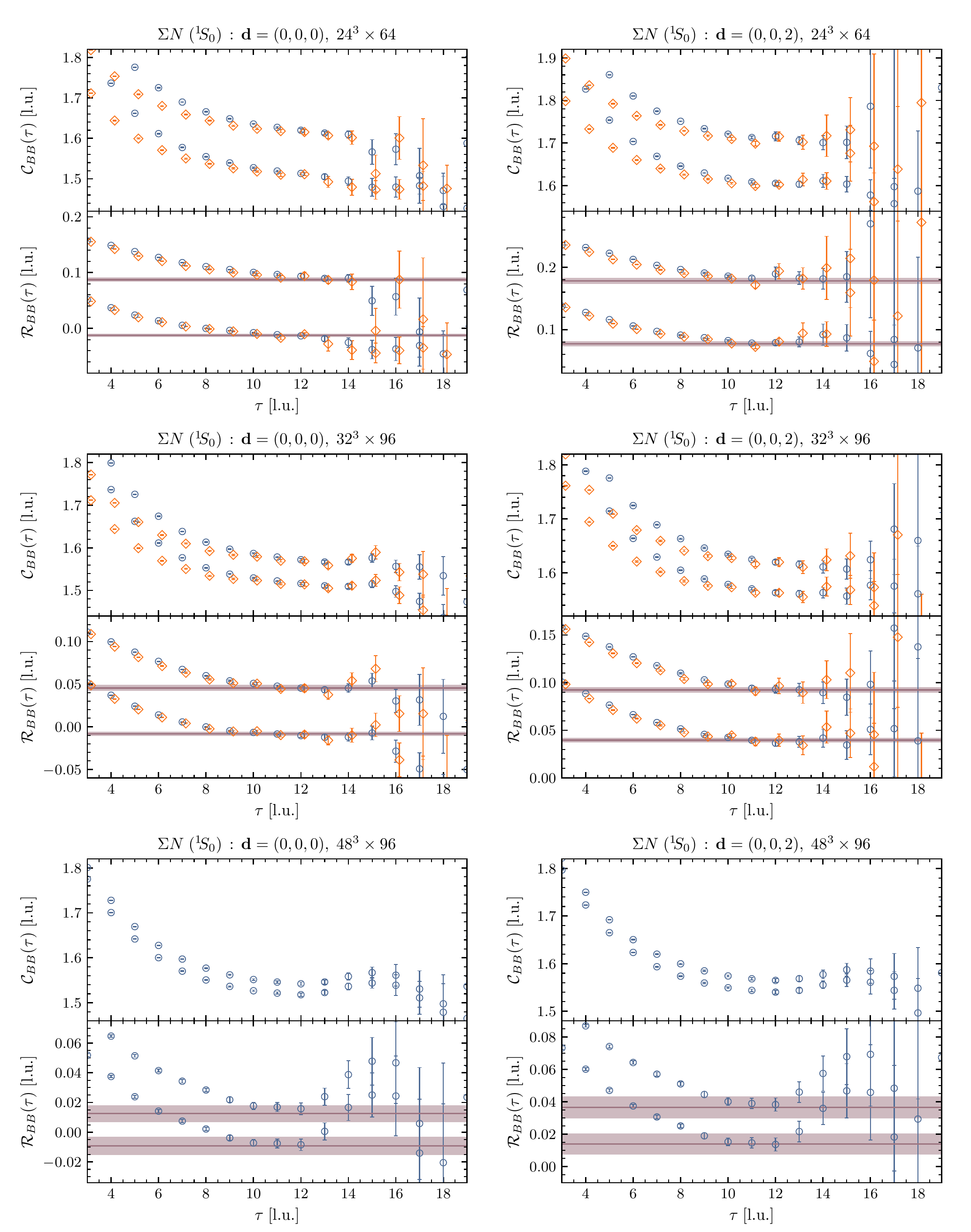}
\caption{The effective energy plots (upper panel of each segment) and the effective energy-shift plots (lower panel of each segment) for the $\Sigma N(^1S_0)$ system at rest (left panels) and with boost $\bm{d}=(0,0,2)$ (right panels) for the SP (blue circles) and SS (orange diamonds) source-sink combinations.}
\label{fig:SN1s0_EMP}
\end{figure}

\begin{figure}[hbt!]
\includegraphics[width=0.95\textwidth]{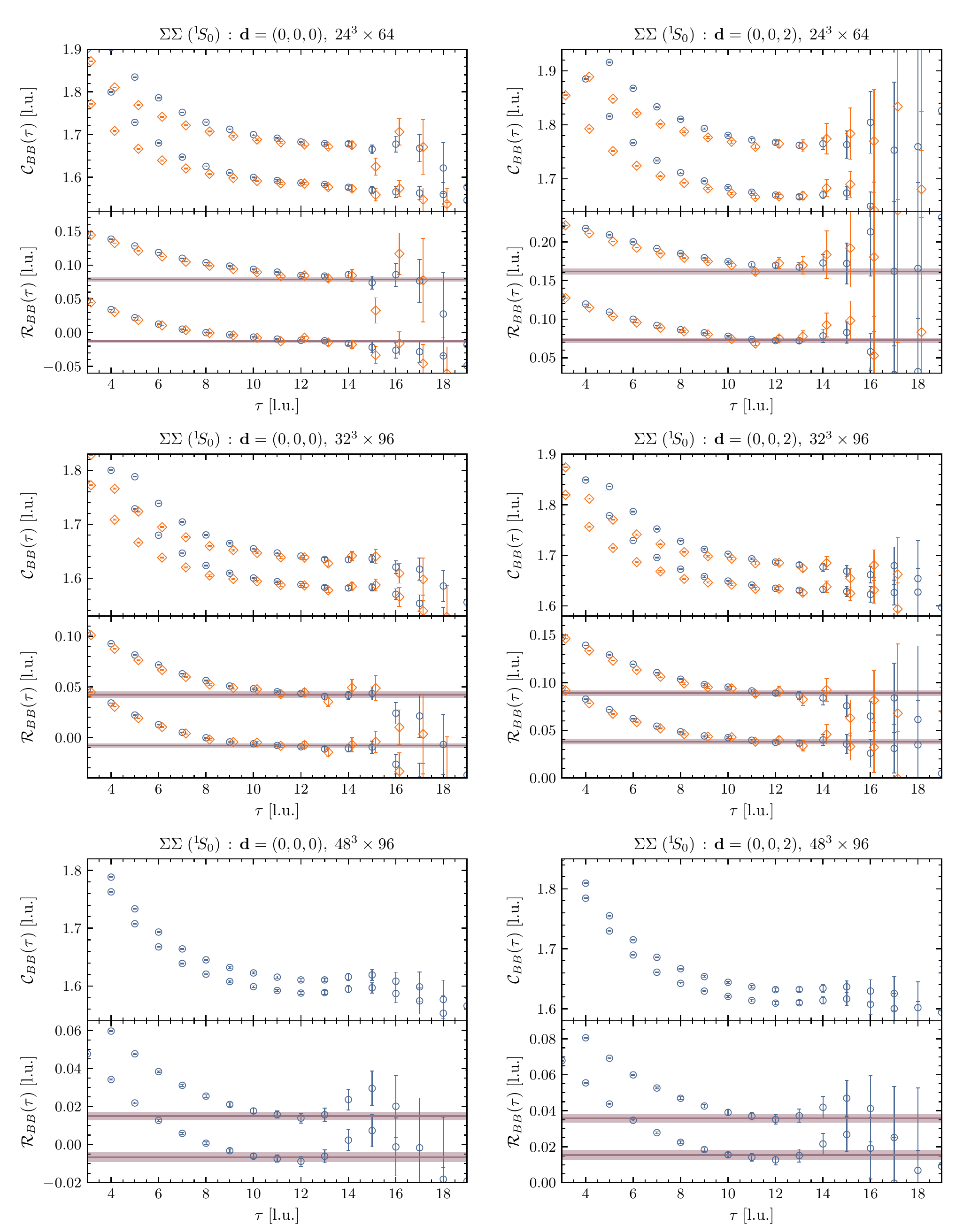}
\caption{The effective energy plots (upper panel of each segment) and the effective energy-shift plots (lower panel of each segment) for the $\Sigma \Sigma (^1S_0)$ system at rest (left panels) and with boost $\bm{d}=(0,0,2)$ (right panels) for the SP (blue circles) and SS (orange diamonds) source-sink combinations.}
\label{fig:SS1s0_EMP}
\end{figure}

\begin{figure}[hbt!]
\includegraphics[width=0.95\textwidth]{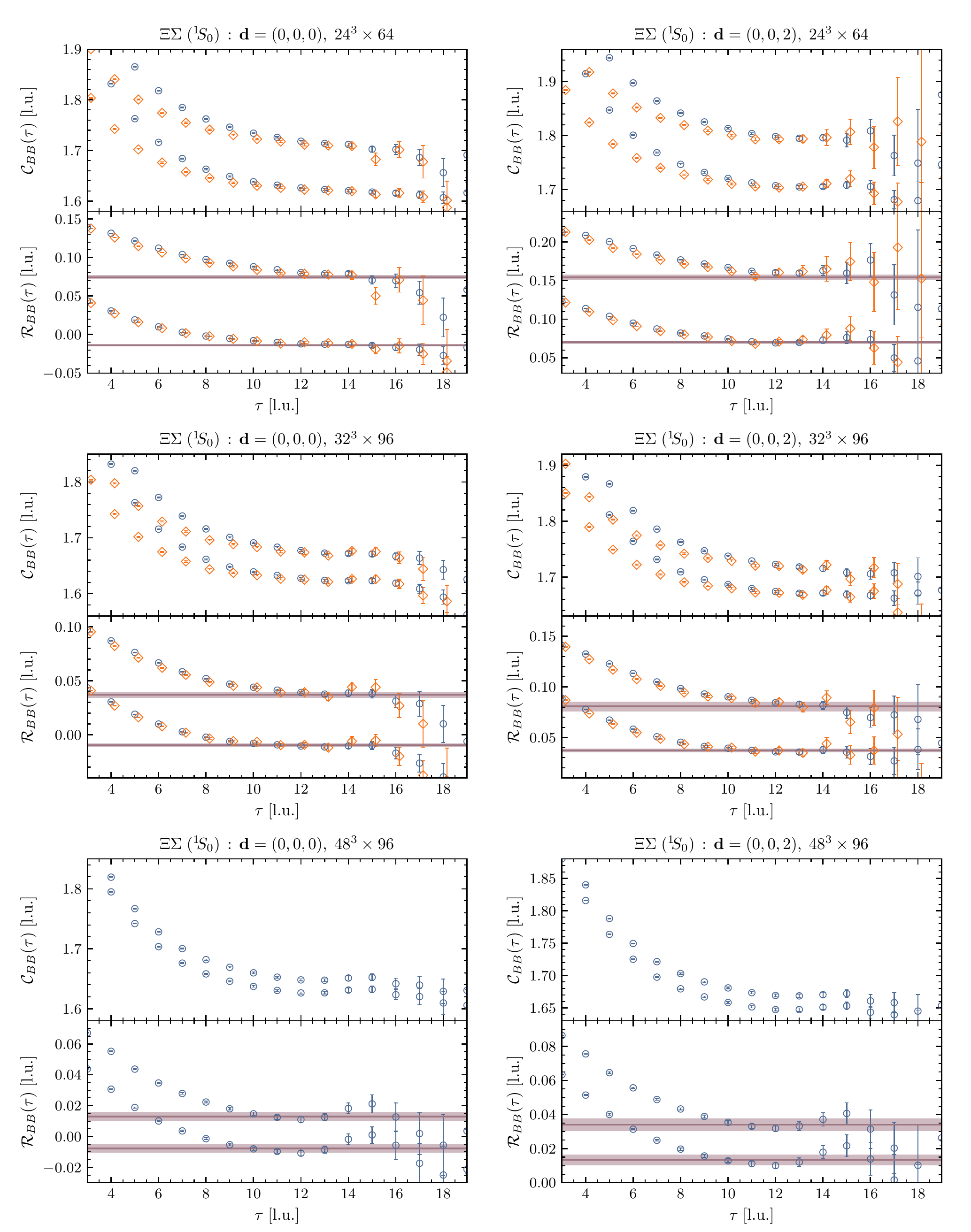}
\caption{The effective energy plots (upper panel of each segment) and the effective energy-shift plots (lower panel of each segment) for the $\Xi \Sigma (^1S_0)$ system at rest (left panels) and with boost $\bm{d}=(0,0,2)$ (right panels) for the SP (blue circles) and SS (orange diamonds) source-sink combinations.}
\label{fig:XS1s0_EMP}
\end{figure}

\begin{figure}[hbt]
\includegraphics[width=0.95\textwidth]{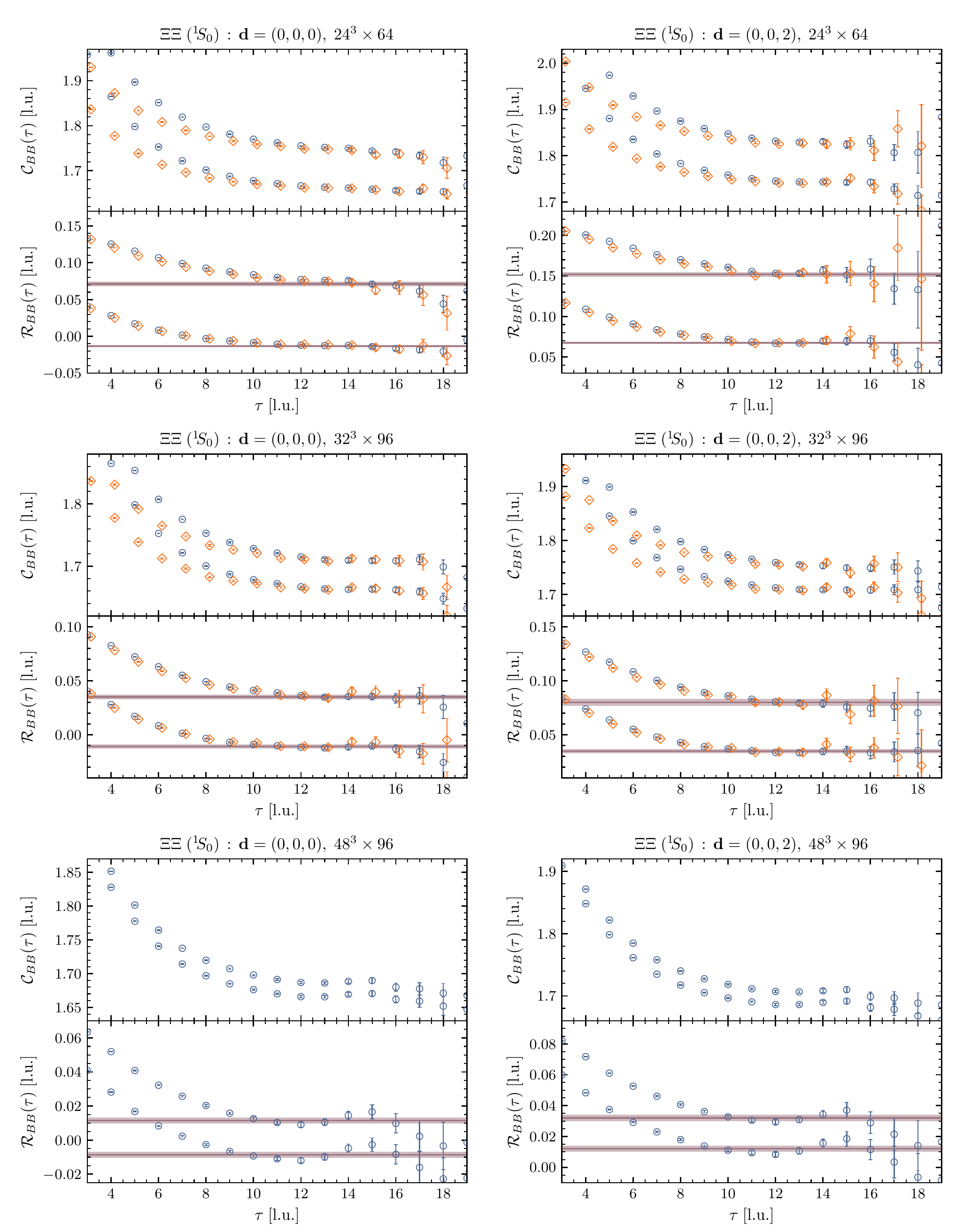}
\caption{The effective energy plots (upper panel of each segment) and the effective energy-shift plots (lower panel of each segment) for the $\Xi \Xi (^1S_0)$ system at rest (left panels) and with boost $\bm{d}=(0,0,2)$ (right panels) for the SP (blue circles) and SS (orange diamonds) source-sink combinations.}
\label{fig:XX1s0_EMP}
\end{figure}

\begin{figure}[hbt!]
\includegraphics[width=0.95\textwidth]{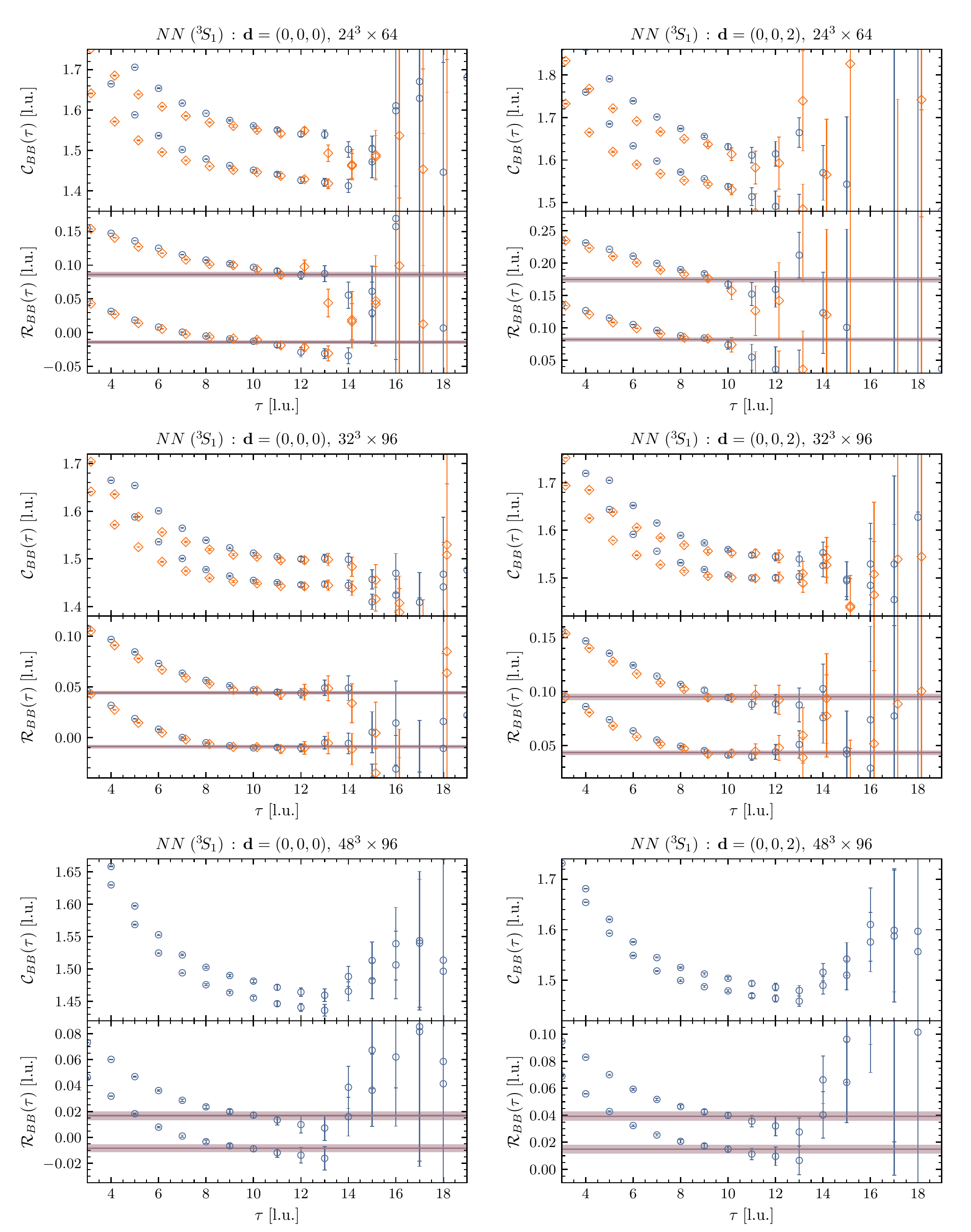}
\caption{The effective energy plots (upper panel of each segment) and the effective energy-shift plots (lower panel of each segment) for the $NN(^3S_1)$ system at rest (left panels) and with boost $\bm{d}=(0,0,2)$ (right panels) for the SP (blue circles) and SS (orange diamonds) source-sink combinations.}
\label{fig:NN3s1_EMP}
\end{figure}

\begin{figure}[hbt!]
\includegraphics[width=0.95\textwidth]{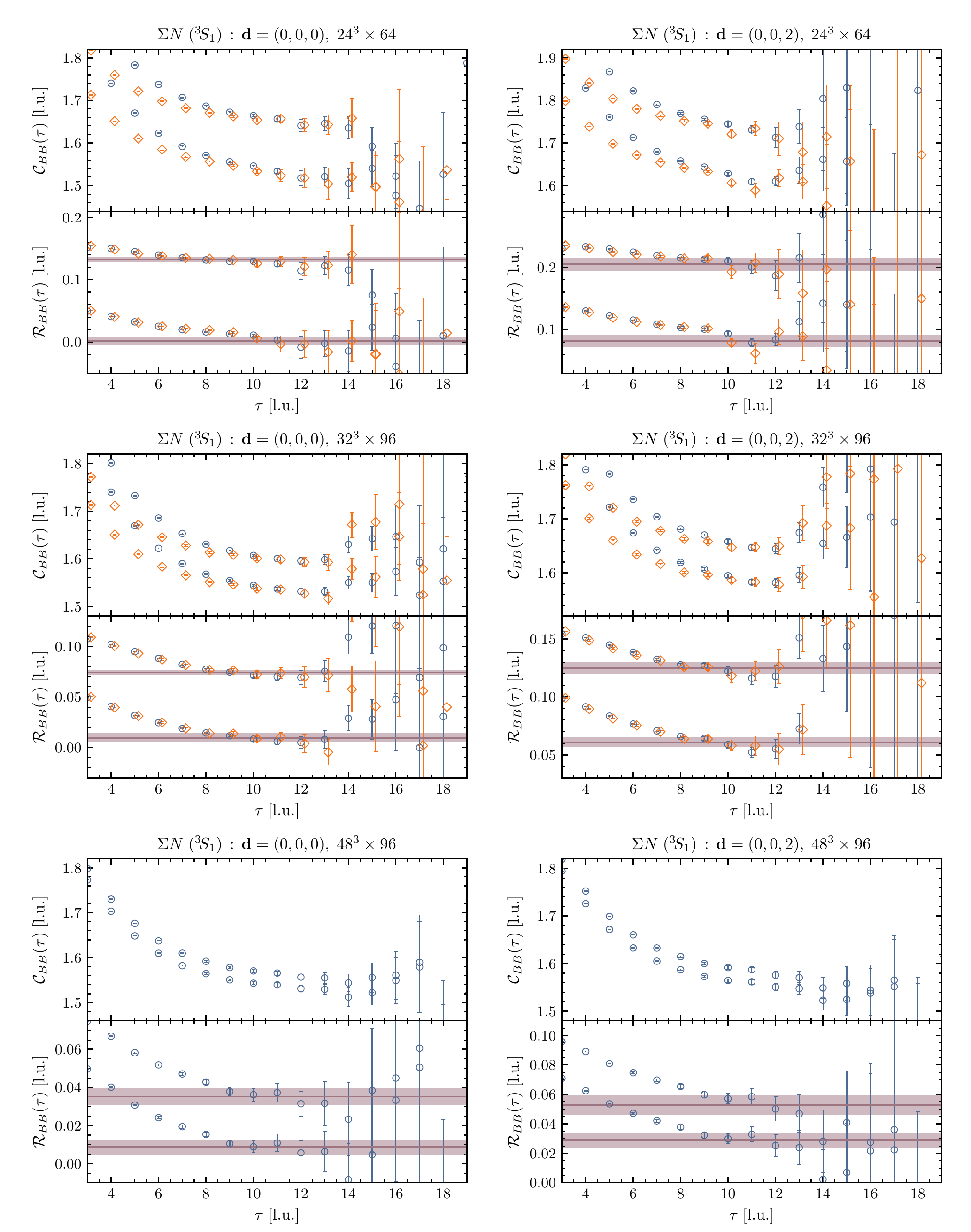}
\caption{The effective energy plots (upper panel of each segment) and the effective energy-shift plots (lower panel of each segment) for the $\Sigma N(^3S_1)$ system at rest (left panels) and with boost $\bm{d}=(0,0,2)$ (right panels) for the SP (blue circles) and SS (orange diamonds) source-sink combinations.}
\label{fig:SN3s1_EMP}
\end{figure}

\begin{figure}[hbt!]
\includegraphics[width=0.95\textwidth]{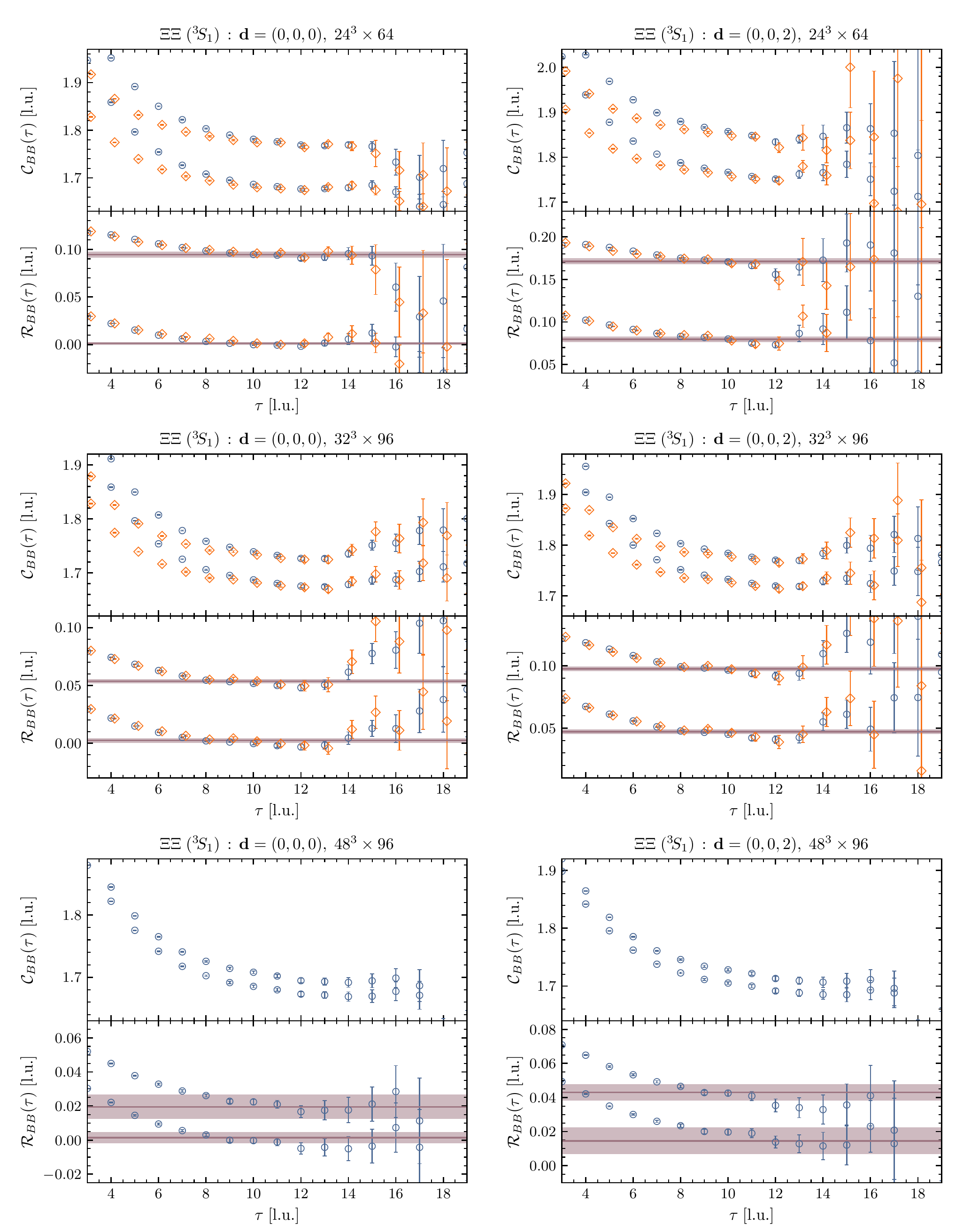}
\caption{The effective energy plots (upper panel of each segment) and the effective energy-shift plots (lower panel of each segment) for the $\Xi \Xi (^3S_1)$ system at rest (left panels) and with boost $\bm{d}=(0,0,2)$ (right panels) for the SP (blue circles) and SS (orange diamonds) source-sink combinations.}
\label{fig:XX3s1_EMP}
\end{figure}

\begin{figure}[hbt!]
\includegraphics[width=0.95\textwidth]{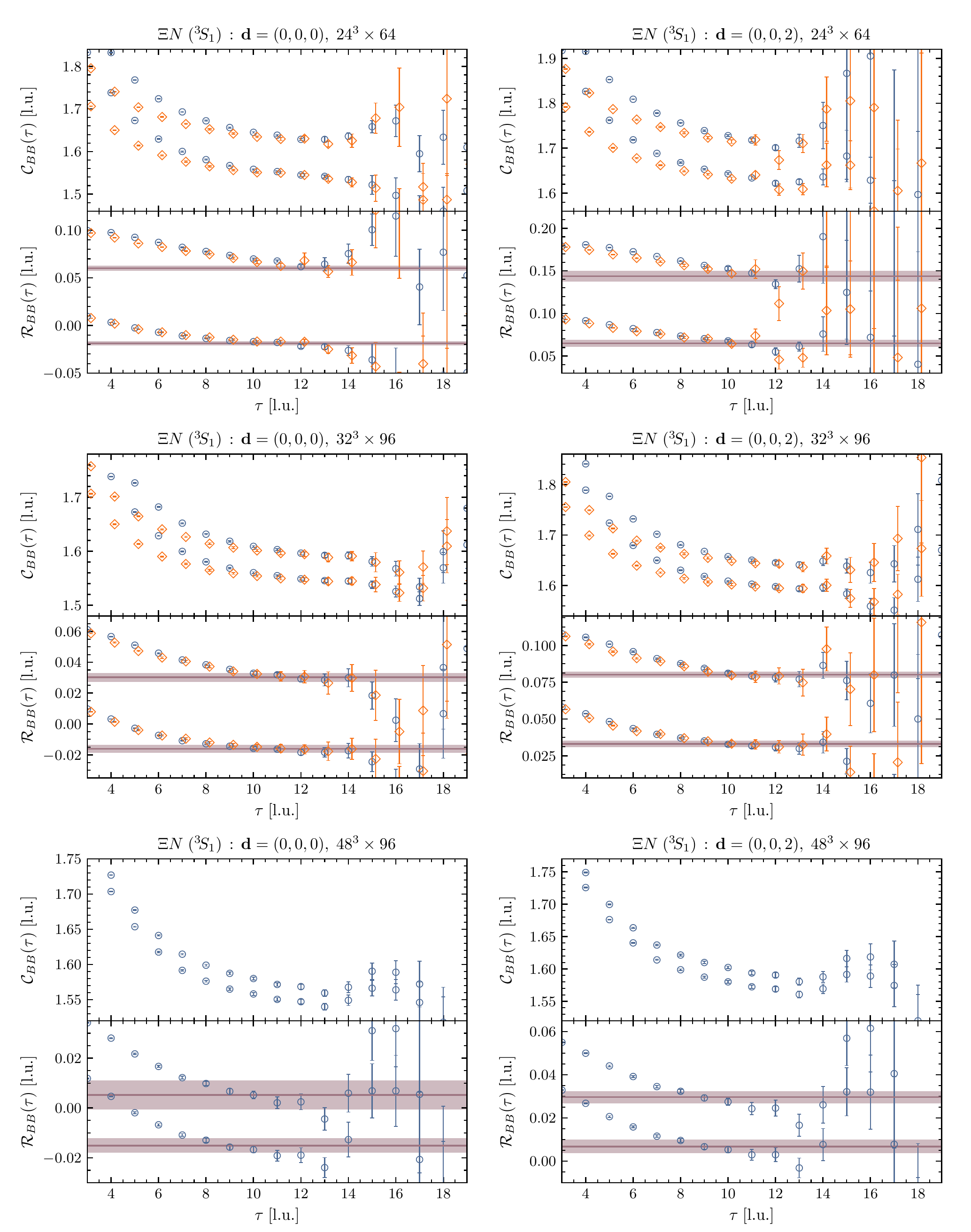}
\caption{The effective energy plots (upper panel of each segment) and the effective energy-shift plots (lower panel of each segment) for the $\Xi N(^3S_1)$ system at rest (left panels) and with boost $\bm{d}=(0,0,2)$ (right panels) for the SP (blue circles) and SS (orange diamonds) source-sink combinations.}
\label{fig:XN3s1_EMP}
\end{figure}

\noindent
This appendix contains all the figures omitted from the main body of the paper for ease of presentation. These include the effective-mass plots of the single baryons in Fig.~\ref{fig:B1_EMP}, and the effective energy and effective energy-shift plots for the two-baryon systems in Figs.~\ref{fig:NN1s0_EMP}-\ref{fig:XN3s1_EMP}. In Fig.~\ref{fig:B1_EMP}, the thin horizontal line and the horizontal band surrounding it represent, respectively, the central value of the baryon mass at each volume, and the associated statistical and systematic uncertainties combined in quadrature, obtained with the fitting procedure described in Sec.~\ref{subsec:fitalg}. Similarly, in Figs.~\ref{fig:NN1s0_EMP}-\ref{fig:XN3s1_EMP} the line and the band represent, respectively, the central value of the two-baryon energy shifts compared to non-interacting baryons at rest (bottom panels) for each volume, and the associated statistical and systematic uncertainties combined in quadrature.

The appendix also contains the numerical results that were omitted from the main body. These include the energy shifts, $\Delta E$, of the two-baryon systems, the c.m.\ momenta, $k^{*2}$, and the value of $k^*\cot\delta$ for all the systems in Tables~\ref{tab:eshift_ini}-\ref{tab:eshift_fin}. In these tables, the values in the first and second parentheses correspond to statistical and systematic uncertainties, respectively, while those in the upper and lower parentheses are, respectively, the right and left uncertainties when the error bars are asymmetric, as is generally the case for the $k^*\cot\delta$ values. When there is a dash sign in the tables, it indicates that the quantity $k\cot\delta$ diverges due to the singularities in the $\mathcal{Z}^{\bm{d}}_{00}$ function.

All quantities in the plots and tables are expressed in lattice units.

\newpage
\begin{table}[hbt!]
\caption{The values of the energy shift $\Delta E$, the c.m.\ momentum $k^{*2}$, and $k^*\cot\delta$ for the $NN \; (\1s0)$ channel.}
\label{tab:eshift_ini}
\begin{ruledtabular}
\begin{tabular}{cccrrr}
Ensemble & Boost vector & State & \multicolumn{1}{c}{$\Delta E$ [l.u.]} & \multicolumn{1}{c}{$k^{*2}$ [l.u.]} & \multicolumn{1}{c}{$k^*\cot\delta$ [l.u.]} \\
\hline
\multirow{4}{*}{$24^3\times 64$}	&	\multirow{2}{*}{$(0,0,0)$}	&	$n=1$	&	$-0.0166(19)(31)$	&	$-0.0120(13)(22)$	&	$-0.078_{(-11)(-17)}^{(+13)(+25)}$	\\
	&		&	$n=2$	&	$ 0.0953(23)(61)$	&	$ 0.0715(17)(47)$	&	\multicolumn{1}{c}{-}	\\
	&	\multirow{2}{*}{$(0,0,2)$}	&	$n=1$	&	$ 0.0812(16)(28)$	&	$-0.0079(12)(21)$	&	$-0.033_{(-18)(-27)}^{(+23)(+51)}$	\\
	&		&	$n=2$	&	$ 0.1960(16)(35)$	&	$ 0.0833(13)(29)$	&	$-0.233_{(-34)(-93)}^{(+32)(+65)}$	\\ \hline
\multirow{4}{*}{$32^3\times 96$}	&	\multirow{2}{*}{$(0,0,0)$}	&	$n=1$	&	$-0.0090(25)(20)$	&	$-0.0065(18)(14)$	&	$-0.056_{(-19)(-15)}^{(+29)(+27)}$	\\
	&		&	$n=2$	&	$ 0.0477(37)(24)$	&	$ 0.0352(28)(17)$	&	$0.7_{(-0.3)(-0.1)}^{(+3.4)(+32.0)}$	\\
	&	\multirow{2}{*}{$(0,0,2)$}	&	$n=1$	&	$ 0.0422(20)(21)$	&	$-0.0075(15)(16)$	&	$-0.068_{(-13)(-13)}^{(+18)(+21)}$	\\
	&		&	$n=2$	&	$ 0.0976(22)(27)$	&	$ 0.0347(18)(21)$	&	\multicolumn{1}{c}{-}	\\ \hline
\multirow{4}{*}{$48^3\times 96$}	&	\multirow{2}{*}{$(0,0,0)$}	&	$n=1$	&	$-0.0093(22)(11)$	&	$-0.0067(16)(08)$	&	$-0.079_{(-10)(-05)}^{(+13)(+06)}$	\\
	&		&	$n=2$	&	$ 0.0197(25)(23)$	&	$ 0.0143(18)(16)$	&	$0.2_{(-0.1)(-0.1)}^{(+0.5)(+1.8)}$	\\
	&	\multirow{2}{*}{$(0,0,2)$}	&	$n=1$	&	$ 0.0183(25)(26)$	&	$-0.0038(18)(20)$	&	$-0.051_{(-19)(-17)}^{(+38)(+80)}$	\\
	&		&	$n=2$	&	$ 0.0444(25)(24)$	&	$ 0.0156(19)(19)$	&	\multicolumn{1}{c}{-}	\\
\end{tabular}
\end{ruledtabular}
\end{table}
\begin{table}[hbt!]
\caption{The values of the energy shift $\Delta E$, the c.m.\ momentum $k^{*2}$, and $k^*\cot\delta$ for the $\Sigma N \; (\1s0)$ channel.}
\begin{ruledtabular}
\begin{tabular}{cccrrr}
Ensemble & Boost vector & State & \multicolumn{1}{c}{$\Delta E$ [l.u.]} & \multicolumn{1}{c}{$k^{*2}$ [l.u.]} & \multicolumn{1}{c}{$k^*\cot\delta$ [l.u.]} \\
\hline
\multirow{4}{*}{$24^3\times 64$}	&	\multirow{2}{*}{$(0,0,0)$}	&	$n=1$	&	$-0.0122(13)(26)$	&	$-0.0093(10)(20)$	&	$-0.048_{(-12)(-22)}^{(+14)(+35)}$	\\
	&		&	$n=2$	&	$ 0.0873(19)(32)$	&	$ 0.0682(15)(26)$	&	\multicolumn{1}{c}{-}	\\
	&	\multirow{2}{*}{$(0,0,2)$}	&	$n=1$	&	$ 0.0771(17)(35)$	&	$-0.0083(14)(27)$	&	$-0.040_{(-18)(-31)}^{(+24)(+60)}$	\\
	&		&	$n=2$	&	$ 0.1780(18)(48)$	&	$ 0.0747(16)(40)$	&	$-0.7_{(-0.2)(-1.7)}^{(+0.2)(+0.3)}$	\\ \hline
\multirow{4}{*}{$32^3\times 96$}	&	\multirow{2}{*}{$(0,0,0)$}	&	$n=1$	&	$-0.0082(21)(16)$	&	$-0.0063(16)(12)$	&	$-0.052_{(-18)(-13)}^{(+26)(+25)}$	\\
	&		&	$n=2$	&	$ 0.0456(31)(16)$	&	$ 0.0351(24)(13)$	&	$0.6_{(-0.3)(-0.1)}^{(+1.6)(+2.2)}$	\\
	&	\multirow{2}{*}{$(0,0,2)$}	&	$n=1$	&	$ 0.0396(18)(17)$	&	$-0.0080(14)(13)$	&	$-0.073_{(-12)(-10)}^{(+14)(+16)}$	\\
	&		&	$n=2$	&	$ 0.0924(18)(23)$	&	$ 0.0339(15)(19)$	&	\multicolumn{1}{c}{-}	\\ \hline
\multirow{4}{*}{$48^3\times 96$}	&	\multirow{2}{*}{$(0,0,0)$}	&	$n=1$	&	$-0.0092(48)(38)$	&	$-0.0070(36)(29)$	&	$-0.081_{(-21)(-16)}^{(+36)(+43)}$	\\
	&		&	$n=2$	&	$ 0.0126(48)(30)$	&	$ 0.0096(36)(23)$	&	$0.03_{(-06)(-03)}^{(+11)(+11)}$	\\
	&	\multirow{2}{*}{$(0,0,2)$}	&	$n=1$	&	$ 0.0139(50)(40)$	&	$-0.0066(38)(30)$	&	$-0.078_{(-23)(-18)}^{(+42)(+66)}$	\\
	&		&	$n=2$	&	$ 0.0366(51)(44)$	&	$ 0.0110(39)(34)$	&	\multicolumn{1}{c}{-}	\\
\end{tabular}
\end{ruledtabular}
\end{table}
\begin{table}[hbt!]
\caption{The values of the energy shift $\Delta E$, the c.m.\ momentum $k^{*2}$, and $k^*\cot\delta$ for the $\Sigma \Sigma \; (\1s0)$ channel.}
\begin{ruledtabular}
\begin{tabular}{cccrrr}
Ensemble & Boost vector & State & \multicolumn{1}{c}{$\Delta E$ [l.u.]} & \multicolumn{1}{c}{$k^{*2}$ [l.u.]} & \multicolumn{1}{c}{$k^*\cot\delta$ [l.u.]} \\
\hline
\multirow{4}{*}{$24^3\times 64$}	&	\multirow{2}{*}{$(0,0,0)$}	&	$n=1$	&	$-0.0126(13)(15)$	&	$-0.0100(10)(12)$	&	$-0.057_{(-11)(-12)}^{(+13)(+17)}$	\\
	&		&	$n=2$	&	$ 0.0788(21)(26)$	&	$ 0.0643(18)(21)$	&	$1.3_{(-0.4)(-0.3)}^{(+0.9)(+2.1)}$	\\
	&	\multirow{2}{*}{$(0,0,2)$}	&	$n=1$	&	$ 0.0725(14)(29)$	&	$-0.0095(11)(24)$	&	$-0.054_{(-13)(-25)}^{(+15)(+39)}$	\\
	&		&	$n=2$	&	$ 0.1618(34)(17)$	&	$ 0.0668(30)(15)$	&	\multicolumn{1}{c}{-}	\\ \hline
\multirow{4}{*}{$32^3\times 96$}	&	\multirow{2}{*}{$(0,0,0)$}	&	$n=1$	&	$-0.0080(12)(14)$	&	$-0.0063(10)(12)$	&	$-0.053_{(-12)(-13)}^{(+14)(+20)}$	\\
	&		&	$n=2$	&	$ 0.0424(19)(25)$	&	$ 0.0342(16)(20)$	&	$0.50_{(-15)(-15)}^{(+31)(+71)}$	\\
	&	\multirow{2}{*}{$(0,0,2)$}	&	$n=1$	&	$ 0.0381(19)(23)$	&	$-0.0079(15)(19)$	&	$-0.071_{(-13)(-14)}^{(+17)(+25)}$	\\
	&		&	$n=2$	&	$ 0.0889(22)(22)$	&	$ 0.0341(18)(19)$	&	\multicolumn{1}{c}{-}	\\ \hline
\multirow{4}{*}{$48^3\times 96$}	&	\multirow{2}{*}{$(0,0,0)$}	&	$n=1$	&	$-0.0065(19)(17)$	&	$-0.0051(15)(14)$	&	$-0.066_{(-12)(-10)}^{(+17)(+18)}$	\\
	&		&	$n=2$	&	$ 0.0150(19)(11)$	&	$ 0.0119(16)(09)$	&	$0.083_{(-42)(-23)}^{(+66)(+47)}$	\\
	&	\multirow{2}{*}{$(0,0,2)$}	&	$n=1$	&	$ 0.0154(19)(20)$	&	$-0.0049(16)(16)$	&	$-0.063_{(-13)(-13)}^{(+19)(+24)}$	\\
	&		&	$n=2$	&	$ 0.0359(19)(16)$	&	$ 0.0117(15)(14)$	&	$0.077_{(-39)(-31)}^{(+64)(+72)}$	\\
\end{tabular}
\end{ruledtabular}
\end{table}
\begin{table}[hbt!]
\caption{The values of the energy shift $\Delta E$, the c.m.\ momentum $k^{*2}$, and $k^*\cot\delta$ for the $\Xi \Sigma \; (\1s0)$ channel.}
\begin{ruledtabular}
\begin{tabular}{cccrrr}
Ensemble & Boost vector & State & \multicolumn{1}{c}{$\Delta E$ [l.u.]} & \multicolumn{1}{c}{$k^{*2}$ [l.u.]} & \multicolumn{1}{c}{$k^*\cot\delta$ [l.u.]} \\
\hline
\multirow{4}{*}{$24^3\times 64$}	&	\multirow{2}{*}{$(0,0,0)$}	&	$n=1$	&	$-0.0137(10)(09)$	&	$-0.0112(08)(08)$	&	$-0.070_{(-08)(-07)}^{(+09)(+08)}$	\\
	&		&	$n=2$	&	$ 0.0745(17)(22)$	&	$ 0.0621(14)(19)$	&	$0.81_{(-16)(-18)}^{(+25)(+44)}$	\\
	&	\multirow{2}{*}{$(0,0,2)$}	&	$n=1$	&	$ 0.0701(12)(17)$	&	$-0.0101(10)(15)$	&	$-0.062_{(-11)(-15)}^{(+13)(+19)}$	\\
	&		&	$n=2$	&	$ 0.1541(29)(21)$	&	$ 0.0631(25)(19)$	&	\multicolumn{1}{c}{-}	\\ \hline
\multirow{4}{*}{$32^3\times 96$}	&	\multirow{2}{*}{$(0,0,0)$}	&	$n=1$	&	$-0.0096(12)(13)$	&	$-0.0078(10)(10)$	&	$-0.070_{(-09)(-09)}^{(+10)(+12)}$	\\
	&		&	$n=2$	&	$ 0.0370(20)(18)$	&	$ 0.0306(17)(15)$	&	$0.233_{(-60)(-46)}^{(+83)(+98)}$	\\
	&	\multirow{2}{*}{$(0,0,2)$}	&	$n=1$	&	$ 0.0371(10)(14)$	&	$-0.0079(09)(11)$	&	$-0.071_{(-08)(-09)}^{(+09)(+13)}$	\\
	&		&	$n=2$	&	$ 0.0806(34)(34)$	&	$ 0.0288(29)(29)$	&	$0.19_{(-08)(-06)}^{(+16)(+25)}$	\\ \hline
\multirow{4}{*}{$48^3\times 96$}	&	\multirow{2}{*}{$(0,0,0)$}	&	$n=1$	&	$-0.0078(26)(12)$	&	$-0.0063(22)(09)$	&	$-0.075_{(-14)(-05)}^{(+20)(+08)}$	\\
	&		&	$n=2$	&	$ 0.0129(26)(13)$	&	$ 0.0106(22)(11)$	&	$0.045_{(-41)(-18)}^{(+65)(+41)}$	\\
	&	\multirow{2}{*}{$(0,0,2)$}	&	$n=1$	&	$ 0.0133(28)(14)$	&	$-0.0063(23)(10)$	&	$-0.075_{(-15)(-07)}^{(+22)(+11)}$	\\
	&		&	$n=2$	&	$ 0.0340(32)(18)$	&	$ 0.0109(26)(14)$	&	$0.06_{(-05)(-03)}^{(+11)(+08)}$	\\
\end{tabular}
\end{ruledtabular}
\end{table}
\begin{table}[hbt!]
\caption{The values of the energy shift $\Delta E$, the c.m.\ momentum $k^{*2}$, and $k^*\cot\delta$ for the $\Xi \Xi \; (\1s0)$ channel.}
\begin{ruledtabular}
\begin{tabular}{cccrrr}
Ensemble & Boost vector & State & \multicolumn{1}{c}{$\Delta E$ [l.u.]} & \multicolumn{1}{c}{$k^{*2}$ [l.u.]} & \multicolumn{1}{c}{$k^*\cot\delta$ [l.u.]} \\
\hline
\multirow{4}{*}{$24^3\times 64$}	&	\multirow{2}{*}{$(0,0,0)$}	&	$n=1$	&	$-0.0134(09)(15)$	&	$-0.0112(07)(12)$	&	$-0.070_{(-07)(-11)}^{(+08)(+14)}$	\\
	&		&	$n=2$	&	$ 0.0712(14)(30)$	&	$ 0.0609(12)(26)$	&	$0.66_{(-10)(-17)}^{(+13)(+43)}$	\\
	&	\multirow{2}{*}{$(0,0,2)$}	&	$n=1$	&	$ 0.0675(10)(13)$	&	$-0.0110(09)(11)$	&	$-0.070_{(-08)(-10)}^{(+09)(+13)}$	\\
	&		&	$n=2$	&	$ 0.1519(14)(27)$	&	$ 0.0643(12)(25)$	&	\multicolumn{1}{c}{-}	\\ \hline
\multirow{4}{*}{$32^3\times 96$}	&	\multirow{2}{*}{$(0,0,0)$}	&	$n=1$	&	$-0.0109(11)(14)$	&	$-0.0090(09)(12)$	&	$-0.081_{(-07)(-09)}^{(+08)(+11)}$	\\
	&		&	$n=2$	&	$ 0.0349(12)(16)$	&	$ 0.0295(10)(14)$	&	$0.195_{(-31)(-38)}^{(+38)(+58)}$	\\
	&	\multirow{2}{*}{$(0,0,2)$}	&	$n=1$	&	$ 0.0349(10)(17)$	&	$-0.0091(08)(14)$	&	$-0.082_{(-06)(-10)}^{(+07)(+13)}$	\\
	&		&	$n=2$	&	$ 0.0800(11)(30)$	&	$ 0.0299(10)(26)$	&	$0.23_{(-04)(-08)}^{(+04)(+17)}$	\\ \hline
\multirow{4}{*}{$48^3\times 96$}	&	\multirow{2}{*}{$(0,0,0)$}	&	$n=1$	&	$-0.0087(12)(13)$	&	$-0.0072(10)(11)$	&	$-0.082_{(-06)(-07)}^{(+07)(+08)}$	\\
	&		&	$n=2$	&	$ 0.0115(13)(14)$	&	$ 0.0096(11)(12)$	&	$0.026_{(-19)(-19)}^{(+22)(+27)}$	\\
	&	\multirow{2}{*}{$(0,0,2)$}	&	$n=1$	&	$ 0.0120(13)(16)$	&	$-0.0071(11)(13)$	&	$-0.081_{(-07)(-08)}^{(+08)(+10)}$	\\
	&		&	$n=2$	&	$ 0.0321(13)(17)$	&	$ 0.0099(11)(15)$	&	$0.033_{(-20)(-26)}^{(+27)(+37)}$	\\
\end{tabular}
\end{ruledtabular}
\end{table}
\begin{table}[hbt!]
\caption{The values of the energy shift $\Delta E$, the c.m.\ momentum $k^{*2}$, and $k^*\cot\delta$ for the $NN \; (\3s1)$ channel.}
\begin{ruledtabular}
\begin{tabular}{cccrrr}
Ensemble & Boost vector & State & \multicolumn{1}{c}{$\Delta E$ [l.u.]} & \multicolumn{1}{c}{$k^{*2}$ [l.u.]} & \multicolumn{1}{c}{$k^*\cot\delta$ [l.u.]} \\
\hline
\multirow{4}{*}{$24^3\times 64$}	&	\multirow{2}{*}{$(0,0,0)$}	&	$n=1$	&	$-0.0140(18)(19)$	&	$-0.0101(13)(14)$	&	$-0.058_{(-14)(-13)}^{(+17)(+20)}$	\\
	&		&	$n=2$	&	$ 0.0860(30)(26)$	&	$ 0.0643(22)(20)$	&	$1.3_{(-0.5)(-0.3)}^{(+1.4)(+2.8)}$	\\
	&	\multirow{2}{*}{$(0,0,2)$}	&	$n=1$	&	$ 0.0819(18)(27)$	&	$-0.0074(14)(20)$	&	$-0.023_{(-22)(-29)}^{(+31)(+55)}$	\\
	&		&	$n=2$	&	$ 0.1744(25)(36)$	&	$ 0.0658(20)(29)$	&	\multicolumn{1}{c}{-}	\\ \hline
\multirow{4}{*}{$32^3\times 96$}	&	\multirow{2}{*}{$(0,0,0)$}	&	$n=1$	&	$-0.0090(15)(12)$	&	$-0.0065(11)(08)$	&	$-0.056_{(-12)(-09)}^{(+15)(+13)}$	\\
	&		&	$n=2$	&	$ 0.0442(16)(12)$	&	$ 0.0326(12)(10)$	&	$0.340_{(-70)(-45)}^{(+97)(+98)}$	\\
	&	\multirow{2}{*}{$(0,0,2)$}	&	$n=1$	&	$ 0.0434(18)(12)$	&	$-0.0066(13)(08)$	&	$-0.057_{(-14)(-08)}^{(+20)(+14)}$	\\
	&		&	$n=2$	&	$ 0.0952(20)(23)$	&	$ 0.0328(15)(18)$	&	$0.44_{(-13)(-12)}^{(+30)(+75)}$	\\ \hline
\multirow{4}{*}{$48^3\times 96$}	&	\multirow{2}{*}{$(0,0,0)$}	&	$n=1$	&	$-0.0083(27)(15)$	&	$-0.0060(19)(11)$	&	$-0.073_{(-13)(-07)}^{(+18)(+12)}$	\\
	&		&	$n=2$	&	$ 0.0167(30)(17)$	&	$ 0.0122(22)(12)$	&	$0.09_{(-06)(-03)}^{(+13)(+11)}$	\\
	&	\multirow{2}{*}{$(0,0,2)$}	&	$n=1$	&	$ 0.0149(29)(15)$	&	$-0.0063(21)(11)$	&	$-0.076_{(-14)(-07)}^{(+19)(+12)}$	\\
	&		&	$n=2$	&	$ 0.0393(31)(16)$	&	$ 0.0117(23)(12)$	&	$0.08_{(-06)(-02)}^{(+12)(+09)}$	\\
\end{tabular}
\end{ruledtabular}
\end{table}
\begin{table}[hbt!]
\caption{The values of the energy shift $\Delta E$, the c.m.\ momentum $k^{*2}$, and $k^*\cot\delta$ for the $\Sigma N \; (\3s1)$ channel.}
\begin{ruledtabular}
\begin{tabular}{cccrrr}
Ensemble & Boost vector & State & \multicolumn{1}{c}{$\Delta E$ [l.u.]} & \multicolumn{1}{c}{$k^{*2}$ [l.u.]} & \multicolumn{1}{c}{$k^*\cot\delta$ [l.u.]} \\
\hline
\multirow{4}{*}{$24^3\times 64$}	&	\multirow{2}{*}{$(0,0,0)$}	&	$n=1$	&	$ 0.0012(57)(41)$	&	$ 0.0009(44)(31)$	&	\multicolumn{1}{c}{-}	\\
	&		&	$n=2$	&	$ 0.1325(28)(27)$	&	$ 0.1050(24)(22)$	&	$0.102_{(-40)(-36)}^{(+42)(+40)}$	\\
	&	\multirow{2}{*}{$(0,0,2)$}	&	$n=1$	&	$ 0.0816(57)(88)$	&	$-0.0048(46)(69)$	&	\multicolumn{1}{c}{-}	\\
	&		&	$n=2$	&	$ 0.2047(84)(60)$	&	$ 0.0975(71)(50)$	&	$0.03_{(-12)(-10)}^{(+13)(+09)}$	\\ \hline
\multirow{4}{*}{$32^3\times 96$}	&	\multirow{2}{*}{$(0,0,0)$}	&	$n=1$	&	$ 0.0096(32)(33)$	&	$ 0.0073(24)(25)$	&	$-0.107_{(-37)(-53)}^{(+27)(+26)}$	\\
	&		&	$n=2$	&	$ 0.0742(22)(13)$	&	$ 0.0578(18)(10)$	&	$0.047_{(-38)(-18)}^{(+40)(+24)}$	\\
	&	\multirow{2}{*}{$(0,0,2)$}	&	$n=1$	&	$ 0.0609(38)(16)$	&	$ 0.0088(30)(11)$	&	$-0.087_{(-40)(-17)}^{(+29)(+11)}$	\\
	&		&	$n=2$	&	$ 0.1252(46)(25)$	&	$ 0.0604(38)(19)$	&	$0.15_{(-10)(-05)}^{(+14)(+08)}$	\\ \hline
\multirow{4}{*}{$48^3\times 96$}	&	\multirow{2}{*}{$(0,0,0)$}	&	$n=1$	&	$ 0.0087(32)(25)$	&	$ 0.0066(24)(18)$	&	$-0.021_{(-33)(-28)}^{(+37)(+31)}$	\\
	&		&	$n=2$	&	$ 0.0352(34)(27)$	&	$ 0.0270(26)(19)$	&	$0.08_{(-09)(-06)}^{(+14)(+13)}$	\\
	&	\multirow{2}{*}{$(0,0,2)$}	&	$n=1$	&	$ 0.0291(35)(37)$	&	$ 0.0051(27)(29)$	&	$-0.041_{(-45)(-91)}^{(+39)(+44)}$	\\
	&		&	$n=2$	&	$ 0.0527(38)(52)$	&	$ 0.0235(30)(41)$	&	$-0.03_{(-14)(-65)}^{(+11)(+18)}$	\\
\end{tabular}
\end{ruledtabular}
\end{table}
\begin{table}[hbt!]
\caption{The values of the energy shift $\Delta E$, the c.m.\ momentum $k^{*2}$, and $k^*\cot\delta$ for the $\Xi \Xi \; (\3s1)$ channel.}
\label{tab:eshift_xx3s1}
\begin{ruledtabular}
\begin{tabular}{cccrrr}
Ensemble & Boost vector & State & \multicolumn{1}{c}{$\Delta E$ [l.u.]} & \multicolumn{1}{c}{$k^{*2}$ [l.u.]} & \multicolumn{1}{c}{$k^*\cot\delta$ [l.u.]} \\
\hline
\multirow{4}{*}{$24^3\times 64$}	&	\multirow{2}{*}{$(0,0,0)$}	&	$n=1$	&	$ 0.0012(11)(09)$	&	$ 0.0010(10)(08)$	&	\multicolumn{1}{c}{-}	\\
	&		&	$n=2$	&	$ 0.0944(28)(16)$	&	$ 0.0812(25)(14)$	&	$-0.38_{(-12)(-08)}^{(+09)(+05)}$	\\
	&	\multirow{2}{*}{$(0,0,2)$}	&	$n=1$	&	$ 0.0797(13)(28)$	&	$-0.0003(12)(24)$	&	\multicolumn{1}{c}{-}	\\
	&		&	$n=2$	&	$ 0.1712(19)(29)$	&	$ 0.0820(18)(26)$	&	$-0.29_{(-06)(-10)}^{(+05)(+07)}$	\\ \hline
\multirow{4}{*}{$32^3\times 96$}	&	\multirow{2}{*}{$(0,0,0)$}	&	$n=1$	&	$ 0.0024(15)(14)$	&	$ 0.0020(12)(11)$	&	$-0.27_{(-30)(-95)}^{(+08)(+06)}$	\\
	&		&	$n=2$	&	$ 0.0535(14)(13)$	&	$ 0.0455(12)(12)$	&	$-0.297_{(-77)(-89)}^{(+57)(+53)}$	\\
	&	\multirow{2}{*}{$(0,0,2)$}	&	$n=1$	&	$ 0.0471(13)(15)$	&	$ 0.0014(11)(13)$	&	\multicolumn{1}{c}{-}	\\
	&		&	$n=2$	&	$ 0.0976(14)(13)$	&	$ 0.0454(12)(11)$	&	$-0.262_{(-73)(-72)}^{(+52)(+44)}$	\\ \hline
\multirow{4}{*}{$48^3\times 96$}	&	\multirow{2}{*}{$(0,0,0)$}	&	$n=1$	&	$ 0.0014(17)(29)$	&	$ 0.0012(14)(23)$	&	\multicolumn{1}{c}{-}	\\
	&		&	$n=2$	&	$ 0.0195(67)(25)$	&	$ 0.0164(56)(21)$	&	\multicolumn{1}{c}{-}	\\
	&	\multirow{2}{*}{$(0,0,2)$}	&	$n=1$	&	$ 0.0145(78)(06)$	&	$-0.0049(66)(13)$	&	\multicolumn{1}{c}{-}	\\
	&		&	$n=2$	&	$ 0.0430(19)(45)$	&	$ 0.0192(16)(39)$	&	\multicolumn{1}{c}{-}	\\
\end{tabular}
\end{ruledtabular}
\end{table}
\begin{table}[hbt!]
\caption{The values of the energy shift $\Delta E$, the c.m.\ momentum $k^{*2}$, and $k^*\cot\delta$ for the $\Xi N \; (\3s1)$ channel.}
\label{tab:eshift_fin}
\begin{ruledtabular}
\begin{tabular}{cccrrr}
Ensemble & Boost vector & State & \multicolumn{1}{c}{$\Delta E$ [l.u.]} & \multicolumn{1}{c}{$k^{*2}$ [l.u.]} & \multicolumn{1}{c}{$k^*\cot\delta$ [l.u.]} \\
\hline
\multirow{4}{*}{$24^3\times 64$}	&	\multirow{2}{*}{$(0,0,0)$}	&	$n=1$	&	$-0.0186(16)(14)$	&	$-0.0144(12)(11)$	&	$-0.097_{(-08)(-07)}^{(+09)(+09)}$	\\
	&		&	$n=2$	&	$ 0.0602(21)(18)$	&	$ 0.0478(17)(15)$	&	$0.165_{(-26)(-21)}^{(+29)(+27)}$	\\
	&	\multirow{2}{*}{$(0,0,2)$}	&	$n=1$	&	$ 0.0651(19)(35)$	&	$-0.0166(15)(28)$	&	$-0.113_{(-08)(-14)}^{(+09)(+19)}$	\\
	&		&	$n=2$	&	$ 0.1436(23)(60)$	&	$ 0.0486(19)(51)$	&	\multicolumn{1}{c}{-}	\\ \hline
\multirow{4}{*}{$32^3\times 96$}	&	\multirow{2}{*}{$(0,0,0)$}	&	$n=1$	&	$-0.0161(13)(23)$	&	$-0.0124(10)(17)$	&	$-0.104_{(-06)(-09)}^{(+06)(+11)}$	\\
	&		&	$n=2$	&	$ 0.0302(14)(25)$	&	$ 0.0237(11)(20)$	&	$0.067_{(-16)(-27)}^{(+17)(+35)}$	\\
	&	\multirow{2}{*}{$(0,0,2)$}	&	$n=1$	&	$ 0.0331(13)(16)$	&	$-0.0124(10)(13)$	&	$-0.104_{(-06)(-07)}^{(+06)(+08)}$	\\
	&		&	$n=2$	&	$ 0.0801(14)(17)$	&	$ 0.0255(11)(14)$	&	$0.122_{(-26)(-29)}^{(+30)(+45)}$	\\ \hline
\multirow{4}{*}{$48^3\times 96$}	&	\multirow{2}{*}{$(0,0,0)$}	&	$n=1$	&	$-0.0151(15)(25)$	&	$-0.0117(11)(19)$	&	$-0.107_{(-05)(-09)}^{(+06)(+10)}$	\\
	&		&	$n=2$	&	$ 0.0053(49)(31)$	&	$ 0.0041(38)(24)$	&	\multicolumn{1}{c}{-}	\\
	&	\multirow{2}{*}{$(0,0,2)$}	&	$n=1$	&	$ 0.0068(16)(27)$	&	$-0.0118(13)(21)$	&	$-0.108_{(-06)(-09)}^{(+06)(+11)}$	\\
	&		&	$n=2$	&	$ 0.0296(18)(22)$	&	$ 0.0062(14)(17)$	&	$-0.026_{(-20)(-25)}^{(+21)(+27)}$	\\
\end{tabular}
\end{ruledtabular}
\end{table}

\end{document}